\documentclass[longauth]{aa}
\usepackage{epsfig,natbib,url,twoopt}
\usepackage{graphicx}
\usepackage{txfonts}
\usepackage{longtable}
\usepackage{supertabular}
\usepackage{lscape}
\usepackage{rotating}

\usepackage[labelfont=footnotesize,textfont=footnotesize]{caption}  
\usepackage{amstext}
\usepackage{color}

\begin{document}

\title{Limits on the LyC signal from $z\sim3$ sources with secure redshift and HST coverage in the E-CDFS field
  \thanks{Based on data obtained with the European Southern Observatory
Very Large Telescope, Paranal, Chile, under Programs 170.A--0788, 074.A--0709, 171.A--3045, 275.A--5060, and 185.A--0791.}  
}
%
\author{L. Guaita\inst{1}
  \and L. Pentericci\inst{1}
  \and A. Grazian\inst{1}
   \and E. Vanzella\inst{2} 
\and M. Nonino\inst{3}
\and M. Giavalisco \inst{1,5}
\and G. Zamorani\inst{2}
\and A. Bongiorno\inst{1}
\and P.~Cassata\inst{7}
\and M. Castellano \inst{1}
\and B.~Garilli\inst{8}
\and E. Gawiser\inst{4}
\and V.~Le Brun\inst{6}
\and O.~Le F\`evre\inst{6}
\and B.~C.~Lemaux \inst{6}
\and D.~Maccagni\inst{8}
\and E. Merlin \inst{1}
\and P. Santini\inst{1}
\and L. A. M.~Tasca\inst{6}
\and R.~Thomas\inst{6,7}
\and E.~Zucca\inst{2}
\and S. De Barros \inst{2}
\and N. P. Hathi \inst{6}
\and R. Amorin\inst{1}
\and S. Bardelli\inst{2}
\and A. Fontana \inst{1}
}
\offprints{Lucia Guaita, \email{lucia.guaita@oa-roma.inaf.it}}
\institute{INAF - Osservatorio Astronomico di Roma, Via Frascati 33, 00040 Monteporzio (RM), Italy
  \and INAF - Osservatorio Astronomico di Bologna, via Ranzani 1, I-40127 Bologna, Italy
\and INAF - Osservatorio Astronomico di Trieste, via G.B. Tiepolo 11, I-34143 Trieste, Italy
  \and Department of Physics and Astronomy, Rutgers, The State University of New Jersey, Piscataway, NJ 08854, USA
\and Department of Astronomy, University of Massachusetts, 710 North Pleasant Street, Amherst, MA 01003
\and Aix Marseille Universit\'e, CNRS, LAM (Laboratoire d'Astrophysique de Marseille) UMR 7326, 13388, Marseille, France
\and Instituto de Fisica y Astronom\'ia, Facultad de Ciencias, Universidad de Valpara\'iso, Gran Breta$\rm{\tilde{n}}$a 1111, Playa Ancha, Valpara\'iso Chile
\and INAF--IASF, via Bassini 15, I-20133, Milano, Italy
}
\date{Accepted date: January 5th, 2016}
\abstract
{Determining the strength of the Lyman continuum (LyC) and the fraction of LyC escape have implications for the properties of the emitting sources at any redshift, but also for the re-ionization of the Universe at $z > 6$. 
}
{We aim to measure the LyC signal from a sample of sources in the Chandra deep field south. We collect star-forming galaxies (SFGs) and active galactic nuclei (AGN) with accurate spectroscopic redshifts, for which
Hubble Space Telescope (HST) coverage and multi-wavelength photometry are available.}
{We selected a sample of about 200 sources at $z\sim3$. 
Taking advantage of HST resolution, we applied a careful cleaning procedure and rejected sources showing nearby clumps with different colours, which could be lower-z interlopers. Our clean sample
consisted of 86 SFGs 
(including 19 narrow-band selected Ly$\alpha$ emitters) and 8 AGN (including 6 detected in X-rays). We measured the LyC flux from aperture photometry in four narrow-band filters covering wavelengths below a 912 {\AA} rest frame ($3.11<z<3.53$). We estimated the ratio between ionizing (LyC flux) and 1400 {\AA} non-ionizing emissions for AGN and galaxies. 
}
{
By running population synthesis models, we assume an average intrinsic L$_{\nu}$(1400 {\AA})/L$_{\nu}$(900 {\AA}) ratio of 5 as the representative value for our sample. With this value and an average treatment of the lines of sight of the inter-galactic
medium,  we estimate the LyC escape fraction relative to the intrinsic value (fesc$^{rel}$(LyC)).
We do not directly detect ionizing radiation from any individual SFG, but we are able to set a 1(2)$\sigma$ upper limit of fesc$^{rel}$(LyC) $<12(24)$\%. 
This result is consistent with other non-detections published in the literature. 

No meaningful limits can be calculated for the sub-sample of Ly$\alpha$ emitters. 
We obtain one significant direct detection for an AGN at $z=3.46$, with fesc$^{rel}$(LyC) $=(72\pm18)$ \%.
}
{
Our upper limit on fesc$^{rel}$(LyC) implies that the SFGs studied here do not present either the physical properties or the geometric conditions suitable for efficient LyC-photon escape. 
}


\titlerunning{LyC signal}
\authorrunning{L. Guaita}
\maketitle 
%

\section{Introduction}     \label{sec:introduction}

The radiation at wavelengths shorter than 912 {\AA} (Lyman continuum, LyC) is produced by massive OB-type stars in young star clusters \citep[e.g.][]{Avedisova1979} and by active galactic nuclei (AGN). At $z>2.5$, it is redshifted into the optical light spectral region and, in principle, it could be detected from the ground. 

Stellar models \citep[e.g.][]{Bruzual:2003,Leitherer1999} predict that the intrinsic ratio between ionizing and non-ionizing radiation 
from star-forming galaxies is less than 0.2-0.5 (depending on the stellar population age, metallicity, AGN-galaxy fraction). 
As a result of its energy, 
the LyC is very likely to be absorbed by neutral hydrogen (HI) and dust in the inter-stellar medium (ISM) or by the HI in the inter-galactic medium (IGM). 
At $z>>4$, the IGM is completely opaque to this radiation, and the chance of detection from Earth becomes negligible \citep[e.g.][]{Madau:1995,Inoue2014}. Therefore, only the LyC signal coming from $2.5<z<4$ sources can be detected from the ground. 

Theoretically, the production, propagation, and escape of LyC photons are related to the physical properties of the galaxies.
Firstly, the production of LyC radiation implies the presence of young, 
massive stars, 
and therefore of on-going star formation. Because of the fast recombination timescale of the HI atoms, previous episodes of star formation have no significant impact on the production of ionizing photons eventually escaping the galaxy \citep[e.g.][]{Paardekooper2015}. 
Secondly, the propagation of LyC photons within the ISM is favoured by a negligible amount of dust and low column density of HI (N(HI) $\leq10^{18}$ cm$^{-2}$) in a 10-pc scale region around the emitting star clusters. This could be the case for galaxies embedded in dark-matter halos with masses less than $10^8$ M$_{\odot}$ \citep{Yajima2011,Wise2014,Paardekooper2015}. However, even galaxies residing in more massive halos can have lines of sight favourable to the propagation and the escape of LyC photons \citep{Gnedin2008,Roy2014}. Supernova explosions could have cleared 
 their ISM, and star-formation episodes could occur in their outskirts. 
In addition, 'runaway' OB stars up to 1 kpc away from the initial-origin regions are proposed to significantly contribute to the amount of LyC photons finally emitted into the IGM \citep{Conroy2012}. 
Thirdly, LyC photons emitted into the IGM can affect the galaxy environment, changing the ratio of neutral vs ionized gas, eventually fuelling the ISM \citep[e.g.][for a study of ionized-metal outflows and inflows]{Martin2012}. 
Simulations at intermediate redshift have shown that the LyC escape fraction (fesc(LyC)) steeply decreases as the dark-matter halo mass (M$_{h}$) increases at $3<z<6$ \citep[e.g.][]{Yajima2011} and that the median fesc(LyC) also changes with redshift at $z=4-6$ \citep[e.g.,][]{CenKimm2015}.
%
%
It is worth stressing that while some authors find that the LyC escape fraction decreases with the increase in the halo mass 
\citep[see also][]{Ferrara2013}, other works find the opposite trend: 
 fesc(LyC) is found to range from a few percent \citep[e.g.][]{Gnedin2008} up to $20-30$\% \citep[e.g.][]{Mitra2013} or even higher \citep[e.g.][]{WiseCen2009}. 

Searches for LyC emission from galaxies have been very difficult so far.
In the local Universe, two galaxies have been observed to be ionizing-radiation emitters. One is Haro11, a blue compact and metal poor (oxygen abundance 12+log(O/H)=7.9) galaxy at $z=0.02$, characterized by M(HI) $<10^8$ M$_{\odot}$ \citep{Bergvall2006}. 
It is composed of three main young ($<40$ Myr old) star-formation clumps, probably experiencing a merging event. \citet{Leitet2011} estimated that $\sim17$\% of the intrinsic LyC flux escapes Haro11 \citep[see also][for the method]{Bergvall2013}. A comparable value was estimated by \citet{Leitet2013} for a second blue galaxy, Tol 1247-232, which is also characterized by low metallicity and low dust content. 

At $z\sim0.24$, \citet{Borthakur2014} discovered a similarly
blue galaxy, J0921+4509, that could have
a LyC escape fraction of $(21\pm5)$\%. 
J0921+4509 is characterized by a moderately high stellar mass ($10^{10.8}$ M$_{\odot}$), metallicity (12+log(O/H)=8.67), and a moderate star-formation rate, SFR $=55$ M$_{\odot}$ yr$^{-1}$. However, the analyses of hard X-ray and radio emissions of this galaxy 
did not exclude the possible contribution by an AGN \citep[][]{Jia2011,Alexandroff2012}.

At $0.3\leq z\leq3$, not one undisputed detection of LyC from a pure star-forming galaxy exists. \citet{Bridge2010} found one LyC leaking galaxy with an active galactic nucleus at $z\sim0.7$ ($\sim15$\% of the intrinsic LyC), 
but no LyC signal from pure star-forming galaxies. \citet{Siana2010} searched for LyC emission in starburst galaxies at $z\sim1.3$, but found no emission from any of them. 
\citet{Iwata2009}, \citet{Nestor2013}, and \citet{ Mostardi2013} compiled samples of a few LyC emitter candidates among $z\sim3$ UV-continuum-selected star-forming galaxies and narrow-band-selected Lyman alpha emitters. A LyC signal from individual sources was estimated by measuring the flux in narrow bands, covering the $\sim$900 {\AA} rest frame. However, the contamination from nearby lower-$z$ neighbours was not considered on the object-by-object basis in these samples, and this could have led to false detections \citep{Siana2015, Mostardi2015}.
\citet{Vanzella2010c}, \citet{Boutsia2011}, and \citet{Grazian2015b} analysed large samples of $z\sim3$ star-forming galaxies with secure redshift, without providing any significant detection. 
A promising LyC detection comes from Ion2, a galaxy at $z\sim3.218$, studied in \citet{Vanzella2015,deBarros2015}, Vanzella et al. in prep.

The total number of sources investigated in these high-redshift studies is quite high, but not all of the analysed star-forming galaxies have homogeneous archival data. 
In particular, only some of them have Hubble Space Telescope (HST) coverage and multi-wavelength photometry, which is essential to identify nearby low-redshift contaminants. As shown by \citet{Vanzella2012}, one of the main limitations of LyC studies is the difficulty of identifying low-redshift contaminants, which can be responsible for a false LyC signal. \citet{Siana2015} reached the same conclusion after following up five LyC emitter candidates that were previously photometrically selected through narrow-band imaging \citep[see also][]{Mostardi2015}. 


The direct detection of LyC photons also has implications for understanding the origin of the re-ionization of the Universe at $z>6$ and which objects keep it ionized at $z\sim6$. It is still uncertain if galaxies or AGN were the main drivers of the re-ionization. The uncertainty depends on the fact that we still have not observed exactly how ionizing radiation escapes from individual galaxies. 
%
However, the average ionization rate of the Universe has been estimated by studying the Ly$\alpha$ forest on the line of sight of bright sources and the physical properties of the high-redshift IGM \citep[e.g.][]{HaardtMadau2012, Becker2013}. Depending on the redshift, either galaxies or AGN have been considered the main sources of the ionization.

A Lyman continuum from an AGN is observed in many cases \citep[e.g.][]{Cowie2009,Stevans2014,Worseck2014,Lusso2015,Tilton2015}, but the number of bright AGN significantly decreases at $z>3$ \citep[e.g.,][]{Fan2006}. However, \citet{Giallongo2015} evaluated the luminosity function of faint AGN at $z=4-6.5$. They demonstrated that the AGN population in their sample could produce ionization rates comparable with those required to keep the IGM as ionized as the observed Ly$\alpha$ forest of similar-redshift quasar spectra. Despite this, the majority of the studies still support the idea that low-mass star-forming galaxies (below the current
survey detection limits) are the most probable sources of the re-ionization 
\citep{Duncan2015, Dayal2015b,Robertson2015}. A few rare cases of more massive galaxies with active phenomena of feedback could also contribute, and we may have the chance to find galaxies like them by exploring large samples at $z<4$ \citep[see also][]{Chardin2015}. 

Finally, it is important to consider the fact that the physical properties of galaxies/AGN and their LyC escape fraction at the epoch of re-ionization in turn could be different from those at $2<z<4$. \citet{Duncan2015}, for instance, showed that star-forming galaxies at $z>6$ could be bluer and therefore able to produce a larger number of ionizing photons than at intermediate redshift.

%
In this work, we start from sources with secure redshift, since photometric-redshift samples always contain some interlopers. This can be achieved by starting with large spectroscopic samples, such as VUDS \citep[the VIMOS Ultra Deep Survey,][]{LeFevre2015} and the ESO-GOODS master catalogue; then we analyse images at the resolution of HST to eliminate close-by possible lower-$z$ contaminants, and  take advantage of multi-wavelength photometry and HST images to 
correlate the LyC signal with physical and morphological properties \citep[see also][]{Bog2010}. 
The scope is to measure either a direct LyC signal or a reliable LyC upper limit for our sample of star-forming galaxies, free of contamination from low-$z$ interlopers.
%
%

The paper is organized as follows. In Sect. 2 we present the data and the selection of our sample of star-forming galaxies and AGN, in Sect. 3 we describe the method used to estimate the LyC signal, in Sect. 4 we collect the archival properties of our sample, in Sects. 5 and 6 we show our results and discuss them, and we summarize our work in Sect. 7. Throughout the paper we use AB magnitudes.

\section{Data and sample selection}     \label{sec:Data}

%
We measured the LyC flux in narrow-band filters that cover the rest-frame wavelength range of $860<\lambda<912$ {\AA}. 
We started from a sample of photometrically selected star-forming galaxies that are spectroscopically confirmed. This ensures that we avoid the contamination that is often present when only photometric redshifts and/or narrow-band selections are available.
With this, we built a sample of star-forming galaxies and AGN with accurate redshift, and we required the availability of at least two HST broad bands to be able to identify 
low-$z$ interlopers by eye. 

\subsection{Narrow-band filters} \label{sec:NB}

\begin{figure} 
\centering
\includegraphics[width=80mm]{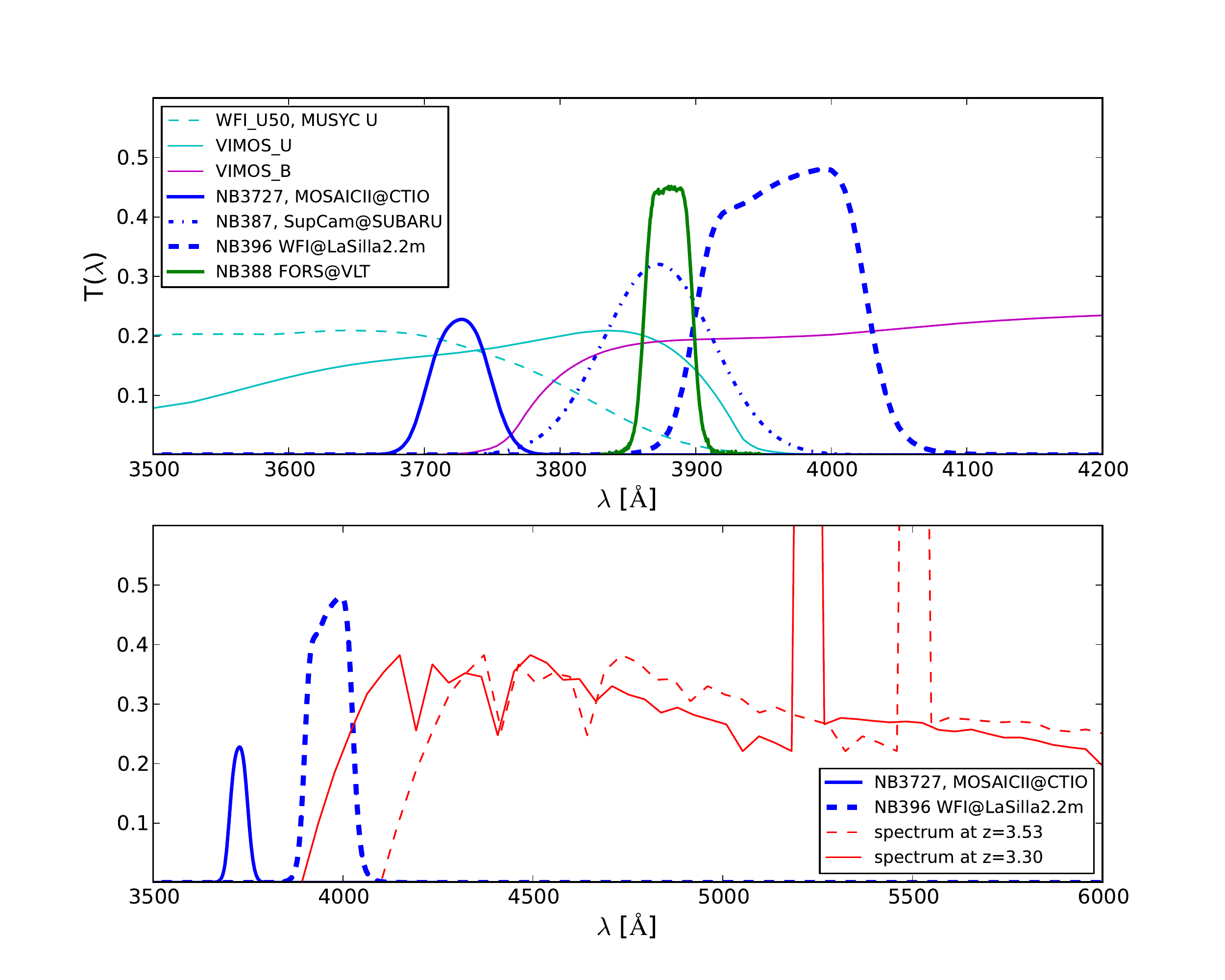}
\caption{$Upper      panel$: Narrow-band filter transmission curves, convolved with the instrument characteristics. 
They are not affected by the red leak in the optical wavelength coverage. $Lower panel$: NB3727 and NB396 together with two synthetic spectra of star-forming galaxies, obtained by running BC03 models at sub-solar metallicity and extracting a stellar population of 10 Myr. The solid (dashed) red line corresponds to a spectrum redshifted to $z=3.30(3.53)$.}
\label{NBcurve}
\end{figure}

The narrow-band filters we considered for measuring the LyC flux are NB3727, NB387, NB388, and NB396. They cover the wavelength range between the $U$ and $B$ bands (Table \ref{tab:NB} and Fig. \ref{NBcurve}). For the images obtained in these filters, we measured a seeing PSF between 0.7'' and 1.4'', and 3$\sigma$ detection limit between 26.7 and 29.1 magnitudes. The PSF and detection limits 
were obtained by following the prescription described in \citet{Gawiser:2006a} and \citet{Guaita2010}. 

The four narrow-band observations were originally designed for different surveys, as explained below 
%
%
\begin{table*} 
\centering
\caption{Properties of the narrow bands covering Lyman continuum signal at $860<\lambda<912$ {\AA}}
\label{tab:NB}
\scalebox{1.1}{
\begin{tabular}{|c|c|c|c|c|c|c|}
\hline\hline 
NBfilter & INSTR/TEL  & 3$\sigma$det lim &  3$\sigma$det lim &  Aper diam('') & fractional signal of PS &  PSF \\
(FWHM)&  & &  &  &  &   \\
&   &  (1'' radius) &  (PSF/2 radius) &  highest S/N of PS& within Aper diam&  \\
\hline
\hline
NB3727 & MOSAICII/CTIO4m &  26.19 & 26.71 &  1.4'' & 0.390 &  1.4'' \\
(50{\AA}) &  & &  &  &  &    \\
\hline
NB-L-387& SupCam/Subaru & 26.90 & 28.20 & 0.8'' & 0.374 &  0.74'' \\
(110{\AA})&  & &  &  &  &    \\
\hline
NB388 & FORS1/VLT &  27.64 & 29.08 &  0.8'' & 0.338 &  0.80'' \\
(37{\AA}) &  & &  &  &  &    \\
\hline
NB396& WFI/LaSilla2.2m &  26.53 & 27.43 &  1.0'' & 0.398 &  0.94'' \\
(129{\AA}) &  & &  &  &  &    \\
\hline
%
%
%
\end{tabular}
}
\end{table*}

The NB3727 image was obtained with the MOSAICII instrument at the Cerro Tololo Interamerican Observatory (CTIO) for a total exposure time of 36 hours (36 exposures of 1 hour each); it was reduced and analysed by  \citet{Guaita2010}. It covers a total area of 31.5' x 31.5'. The E-CDFS was imaged with NB3727 to detect Lyman alpha  (Ly$\alpha$) emitters at $z\simeq2.1$.  This narrow band samples the LyC for sources at $3.11<z<3.30$.

For the other three narrow bands, raw images were recovered from telescope archives and reduced as explained in Sect. \ref{sec:Red}. 

The NB387 image was obtained with SuprimeCam at Subaru telescope for a total exposure time of 8.5 hours; NB387 is the narrow-band filter used in \citet{Nakajima2012a} to probe Ly$\alpha$ emission lines at $z\sim2.2$. We focused on the central area of NB387 image, where astrometry is of better quality (see Sec. \ref{sec:Red}). This narrow band samples the LyC for sources at $3.33<z<3.43$.

The NB388 image was obtained with the FORS1 instrument at the VLT for a total exposure time of 17 hours; the NB388 filter was also used by \citet{Hayes2010}  to probe Ly$\alpha$ emission lines at $z\sim2.2$. It samples the LyC for sources at $3.28<z<3.49$. The image area is about 12 arcmin$^2$.

The NB396 image was obtained with the WFI instrument 
at the ESO 2.2 meter telescope at La Silla for a total exposure time of 13 hours; this filter covers the Ly$\alpha$ emission line at $z\sim2.25$ \citep[like in][]{Nilsson:2009}. We focused on the central area of the NB396 image (see Sect. \ref{sec:Red}). This narrow band samples the LyC for sources at $3.41<z<3.53$.

\begin{figure} 
\centering
\includegraphics[width=90mm]{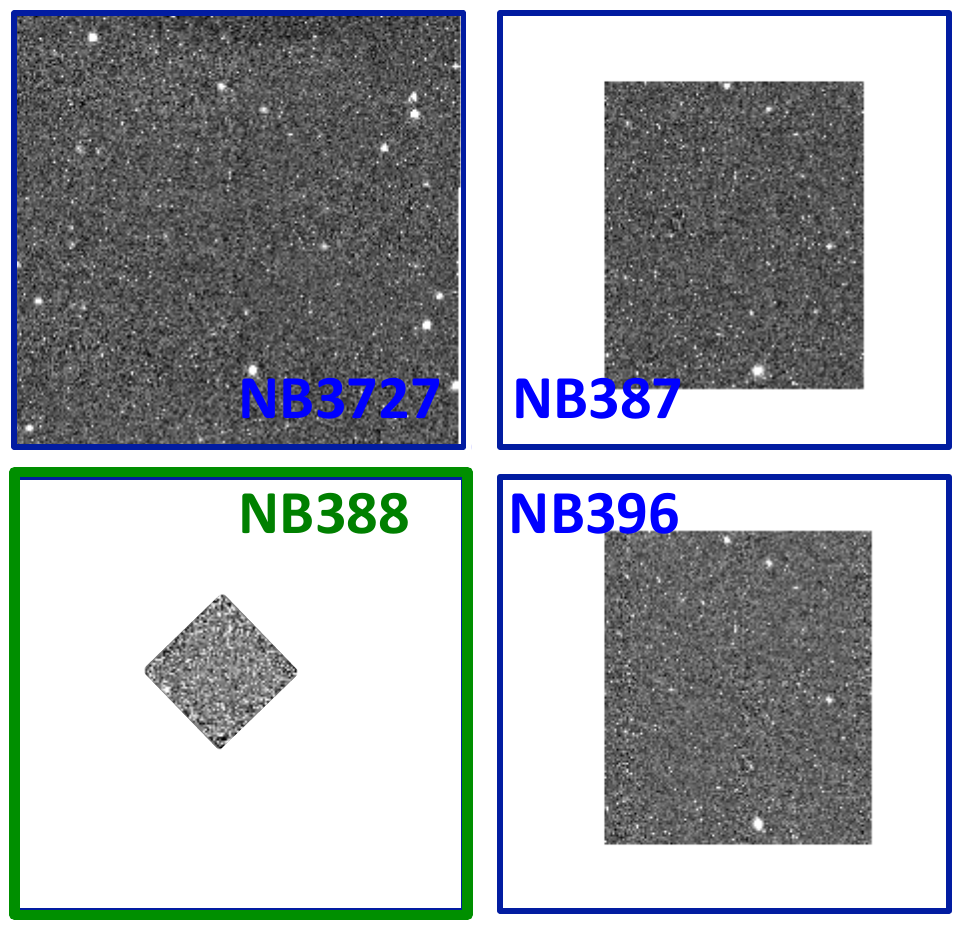}
\caption{
Area in E-CDFS covered by the four narrow-band images. The image obtained with NB3727 covers the entire E-CDFS area of 30'x30', the other images a part of it. The blue and green squares indicate the area of E-CDFS and are coloured as the transmission curves in Fig. \ref{NBcurve}. 
}
\label{NBarea}
\end{figure}

The area covered by the four narrow-band images is shown in Fig. \ref{NBarea}.
%
%
%



\subsubsection{Reduction of archival narrow-band images}  \label{sec:Red}

The raw images in the NB387, NB388, and NB396 bands were processed as described by \citet{Nonino2009}. The only difference is the final stacking of the individual frames because we here adopted the SWARP software \citep{Bertin2002}. Astrometry calibration was performed by considering a catalogue of sources detected in the VIMOS $B$-band image, which is available as part of CANDELS.

Flux calibration was performed by using a set of stars from the CANDELS catalogue \citep{Guo2013}, together with a set of star templates \citep{Pickles1998, Gunn1983}. 
The star templates were convolved with the narrow-band filter transmission curves to obtain their narrow-band magnitudes. The spectral energy distributions of CANDELS stars were fitted with the templates. From the best fit, we inferred the $\text{true}$ narrow-band magnitudes of the stars. We measured observed narrow-band magnitudes of CANDELS stars by running the IRAF task $\sf{PHOT}$ (aperture radius equal to 5'') on our narrow-band images. The narrow-band zero points was then calculated as the difference between the $\text{true}$ and the observed narrow-band magnitudes. We applied a Galactic correction of the order of 0.02-0.03 magnitudes \citep{SF2011} for the reddest and bluest filter, respectively, at E-CDFS coordinates. The zero-point error is about 10\%.

\subsection{Spectroscopic sample and HST coverage} 

A large number of measured redshifts is available in the extended Chandra Deep Field South (E-CDFS). We considered redshifts of continuum-selected star-forming galaxies from the ESO GOODS/E-CDFS master catalogue\footnote{http://www.eso.org/sci/activities/garching/projects/goods/\\MasterSpectroscopy.html}, including the compilations of ESO-GOODS/FORS2 \citep[][and references therein]{Vanzella2008} and ESO-GOODS/VIMOS \citep[][and references therein]{Popesso2009,Balestra2010}. 
We included redshifts 
from the GOODS-MUSIC \citep[Multi-wavelength southern infrared catalog,][]{Grazian2006,Santini2009}, CANDELS \citep[cosmic assembly near-infrared deep extragalactic legacy survey,][]{Grazian2015a} surveys, and the most recent compilation used in this paper. 
We also incorporated data from VUDS \citep[VIMOS Ultra-Deep Survey,][]{LeFevre2015} using the VUDS Data Release 1 (Tasca et al., in preparation), which encompasses about one third of the final sample, and a few redshifts from the test run of VANDELS\footnote{http://vandels.inaf.it} (a deep VIMOS survey of the CANDELS UDS and CDFS fields), a recently approved public survey (PI L. Pentericci and R. McLure). 
In all cases we only considered the most secure redshifts. For example, for VUDS we used the reliability flags 3 and 4, corresponding to a probability greater than 95\% for the redshift to be correct, supplemented by flags 2 and 9, corresponding to a probability of about 80\% for the redshift to be correct, only when the photometric redshift agrees with the measured spectroscopic redshift \citep{LeFevre2015}.
%
We finally added narrow-band selected Ly$\alpha$ emitters, spectroscopically confirmed during MUSYC \citep[Multi-wavelength survey by Yale-Chile,][]{Gawiser:2006a} follow up.
Throughout the paper, we use the acronym SFGs for continuum-selected star-forming galaxies and LAEs for Ly$\alpha$ emitters.

The catalogues by \citet{Xue2011} and \citet{Fiore2012} were used to identify X-ray detections and to isolate AGN.  


We compiled an initial sample of about 200 sources at $3.11<z<3.53$ in E-CDFS. 
A few ($\sim 15$) of the sources were previously studied by \citet{Vanzella2010c, Vanzella2012a, Vanzella2012b}, including the LyC candidate, Ion2 \citep{Vanzella2015}, which is at the redshift covered by our shallowest narrow-band image, however. 

GOODS-S is part of CANDELS\footnote{http://candels.ucolick.org/} and thus every source has multi-wavelength photometry, from VIMOS $U$ and $B$, 
 through HST F435W to F160W, through HUGS (Hawk-I UDS and GOODS survey) 
$Ks$ to Spitzer IRAC 8.0$\mu$m bands. 

The HST data in E-CDFS were obtained within GEMS \citep[Galaxy Evolution from Morphology and Spectral energy distributions,][]{Rix2004} in HST/ ACS F606W 
and F850LP. 
Multi-wavelength photometry is also available for the extended area, and it includes HST and ground-based fluxes \citep{Cardamone:2010,Hsu2014}. 

In Table \ref{tab:NUM} we list the number of galaxies initially selected in the corresponding redshift bin for each narrow-band filter, chosen according to the narrow-band filter transmission curves; the number of galaxies within the area covered by each narrow band; and the number of galaxies in $\text{clean}$ regions (see Sect. \ref{sec:Cleaning}).

\section{Measuring the LyC signal: method}     \label{sec:Method}

In this section, we present the method used to estimate the LyC escape fraction.
\begin{table}[]
\caption{Number of star-forming galaxies with robust redshift}
\label{tab:NUM}
\scalebox{0.8}{
\begin{tabular}{|c|c|c|c|c|}
\hline
\hline
NBfilter & z range & N. gal within & N. gal within & N. gal in  \\
 CDF-S &  & z range$^a$ & NB area & clean regions \\
\hline
NB3727$^b$ & $3.11<z<3.30$ & 88 & 88 & 45 \\ 
NB387 & $3.33<z<3.43$ & 29 & 21 & 7 \\ 
NB388 & $3.28<z<3.49$ & 68 & 22 & 9 \\ 
NB396 & $3.41<z<3.53$ & 59 & 47 &  25\\   
\hline
\hline
\end{tabular}
}
\tablefoot{$^a$ Some of the sources corresponding to different NB ranges overlap. \\
$^b$ The number includes continuum-selected star-forming galaxies and narrow-band selected LAEs (19 in $clean$ regions).  
%
%
%
}
\end{table}

\citet{Vanzella2010a} discussed in detail the role of foreground contamination in estimating the LyC radiation from galaxies at $z\geq3$. 
To reduce the contamination from nearby sources, we performed a $\text{cleaning}$ procedure (Sect. \ref{sec:Cleaning}). 
We measured LyC fluxes 
only for sources that passed the three steps of the procedure,
and we adopted as non-ionizing flux the HST/ACS F606W band flux from the CANDELS and GEMS catalogues. This band covers the UV rest-frame wavelength of 1400 {\AA} at the redshift considered here.
The LyC flux was measured by running Source Extractor \citep[SExtractor,][]{bertin1996} on the narrow bands at pre-defined positions. SExtractor configuration parameters for background, source extraction, and optimal photometry were obtained following \citet{Guaita2010}. Optimal aperture diameters were defined as those that allowed the highest signal-to-noise ratio for point sources, and the corresponding fractional-to-total signal was calculated from the curve of growth of bright stars. 

Narrow-band photometry can be performed with two methods. In the first method, we can optimize SExtractor detection and build narrow-band detection catalogues by adopting optimized configuration parameters. 
We can then search for matches 
between our targets and all the sources detected in narrow band. This method was used in the literature to take into account a possible offset (observed up to 10 kpc) between ionizing and non-ionizing emissions \citep[e.g.][]{Mostardi2013}. These offsets could be observable if special channels for the escape of LyC photons are opened in an off-centre star-formation clump. But the method is sensitive to the contamination by nearby sources, which could be located exactly in the position of a matched object. By applying it before cleaning the sample, low-$z$ sources within the ground-based PSF area are indeed identified as matches. 
  
In the second method, which we adopted here, we measure narrow-band flux at the position of the source with secure redshift \citep[e.g.][]{Vanzella2010c,Boutsia2011}, by running SExtractor 
in double mode. As a detection image we create a narrow-band image by injecting artificial bright stars (IRAF $\sf{artdata.mkobject}$) at the position of the sources. The original narrow-band images are then used for aperture photometry. We chose the optimized apertures (Table \ref{tab:NB}) 
that have the advantage to be small enough to additionally reduce the contamination from nearby sources. 

The optimized-aperture flux was translated into the total flux
by dividing it by the fractional signal (Table \ref{tab:NB}). This total flux is the LyC flux we refer to throughout the paper. The observed flux ratio 
for every source was then calculated as the ratio between LyC and HST/ACS F606W band 
flux densities, 
%
$\frac{f_{\nu}(900)}{f_{\nu}(1400)}$.

The ratio was corrected for the effect of the IGM by dividing it by the IGM transmissivity, exp(-$\tau_{IGM,z}$).
From the observed flux ratio, we can estimate the fesc(LyC) relative to the intrinsic value, 

\begin{equation} \label{eq:fescrel}
fesc^{rel}(LyC) = L_{\nu}(1400)/L_{\nu}(900)\frac{      f_{\nu}(900)/f_{\nu}(1400)       }{exp(-\tau_{IGM,z})}
\label{eq2}
,\end{equation}

%
%
%

%

where the unknowns are the intrinsic L$_{\nu}$(1400)/L$_{\nu}$(900) and exp(-$\tau_{IGM,z}$) (Sect. \ref{sec:exp}).

In the following subsections, we describe  the $\text{cleaning}$ procedure in detail and derive the unknowns of Eq. \ref{eq2}.

\subsection{Cleaning procedure}  \label{sec:Cleaning}

We applied a procedure to identify the final sample of galaxies for which
we performed measurements of LyC escape fraction by removing  the galaxies with possible projected contamination on 
an object-by-object basis. 
We started this procedure with the total sample of 178 galaxies selected in 
Sect. 2.2. 
Of these, 35\% of the SFGs are only covered by GEMS imaging and 65\% fall within the area covered by CANDELS; 70\% of the LAEs fall outside the CANDELS area. The $\text{cleaning}$ procedure was performed in the following steps. 

\begin{figure*} 
\centering
\includegraphics[width=120mm]{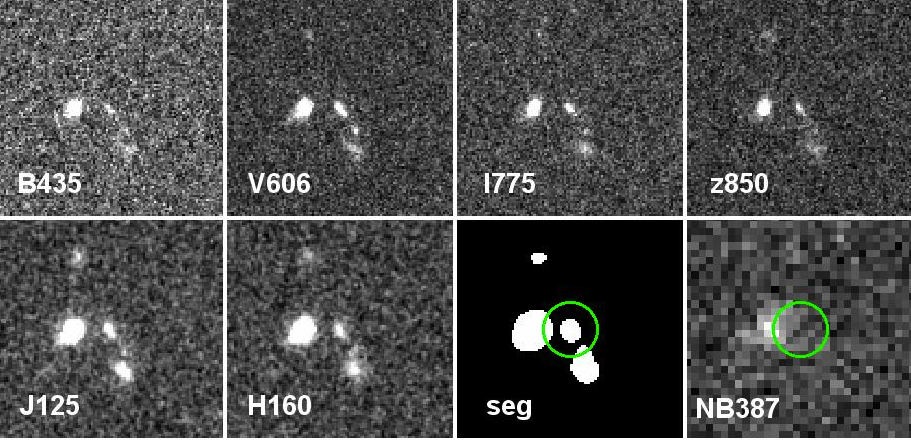} 
\caption{Example of a source rejected in the first step of our cleaning procedure, object ID\_CANDELS=5164 ($z=3.403$). Every stamp covers 6''x6''.  From the top left we show the following filter cut-outs: B435, V606, I775, z850, J125, H160, CANDELS detection segmentation map, and NB387. The green circle has a radius equal to the PSF in NB387. Object ID\_CANDELS=5197 (on the left, zphot\_CANDELS $=2.00\pm0.05$) would be blended with ID\_CANDELS=5164 and ID\_CANDELS=5166 (object at the bottom right, zphot\_CANDELS $=0.47\pm0.13$) in a ground-based image with PSF as large as that of NB387. In the NB387 stamp we see emission coming mainly from the lower-$z$ source on the left-hand side of our target.
%
%
}
\label{excludedstep1}
\end{figure*}

%
We first checked that no other source in the CANDELS/GEMS photometric catalogue is centroided within the size of narrow-band PSF full width at half maximum around the centroid of the candidates. In total, this first step of the cleaning removed 70 of the 178 in our sample. 
An example of a source removed during this step is shown in Fig \ref{excludedstep1}. 

%

Because the success or failure of the first step for our purposes depends sensitively 
on the deblending that is set to generate the various catalogues, the second step of our process is the direct inspection of the HST images. For galaxies that allow it, we inspect cut-outs in all of the available CANDELS HST bands
\footnote{Throughout the paper, we abbreviate the HST ACS/F606W and F850LP bands as V606 and z850, respectively, and the ACS/F435W and F814W as B435 and I814, respectively, for simplicity.}
For the remainder of our sample, we inspected the HST/ACS V606 and HST/ACS F850LP from GEMS. The purpose of this inspection is to separate isolated, single-component sources from those with multiple components. The former are retained
in our sample without further scrutiny
, the latter are flagged and are analysed in the third stage of our cleaning process.

The final stage of our cleaning procedure begins with generating colour images of each
multi-component source selected in the previous step and examining each sub-component. 
We use the IRAF task $\sf{IMSTACK}$ to generate colour images in 
BVi, BVz, BIz, VIz, and V-z 
for the galaxies that were imaged as part of CANDELS. For sources outside the CANDELS area, only V-z images were generated.
An individual knot with appreciably different colours with respect to the main galaxy component is likely to be an interloper, which
means that generating these
images allows us to reject these contaminated sources from the sample \citep{Vanzella2012, Siana2015}. However, the situation is slightly more complex as 
sub-components (hereafter clumps) belonging to the same galaxy can be redder or bluer than the main component of the galaxy, depending on the properties of their stellar populations, dust, and the presence and strength of emission lines falling within the broad-band filters. For this reason, we performed an SED fitting of individual clumps for eight star-forming galaxies for which the analysis was possible (i.e. within the 
CANDELS area, see Appendix \ref{sec:smart} for details). For each clump we
obtained the best-fit photometric redshift and checked whether the SED was 
consistent with the spectroscopic redshift coming from the integrated 
spectrum of the dominant source. We find that clumps with colours very similar ($\Delta$(B-V) $<$ 0.1, $\Delta$(V-z) $<$ 0.1) to that of the main 
component are always consistent with being at the spectroscopic redshift. 
However, we also find three cases where, despite relatively large colour offsets 
($\Delta$(B-V) $\sim$ 0.4 and $\Delta$(V-z) $\sim$ 0.2), the SEDs of the clumps are 
consistent with the clumps being at the same redshift as their main component.
Three of the eight multi-component objects were removed from our sample because they are surrounded by different redshift neighbours. This proves the importance of having complete HST photometry to study LyC candidates. 

For sources that are only covered by GEMS data, we were unable
to perform the complete SED fitting analysis. Therefore, we relied on colour criteria to estimate the contamination
for these sources ($\Delta$(V-z) $<$ 0.2). The analysis presented in Appendix \ref{sec:smart} suggests that relying
on colours alone may be a too conservative approach because it tends 
to overestimate the number of contaminated sources. Despite this, we decided to adopt it because the purity of our sample is our primary concern. The presence of contaminants is the
most worrying aspect in the search for LyC emitters \citep{Vanzella2012,Mostardi2015}.

\begin{figure*} 
\centering
\includegraphics[width=40mm]{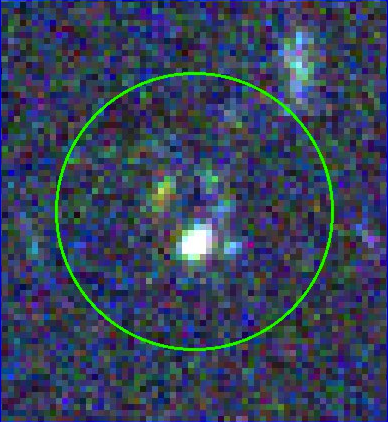} 
\includegraphics[width=120mm]{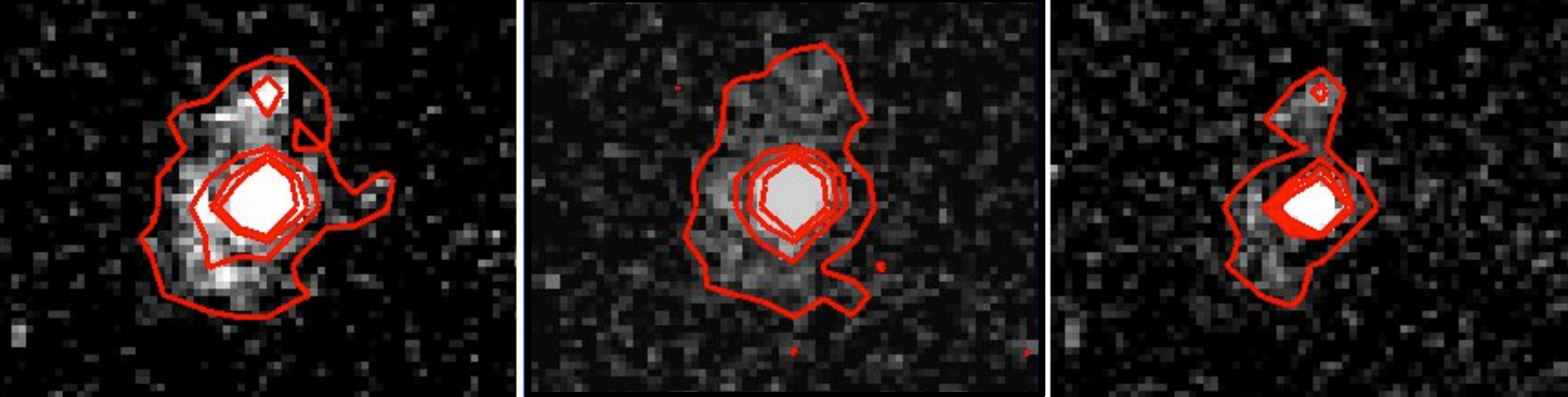}
\caption{Examples of sources rejected in the third step of our cleaning procedure. $Left$: BIz colour cut-outs of a narrow-band selected LAE (ID\_CANDELS=23527, $z=3.116$). The green circle has a radius equal to the NB3727 PSF full width at half maximum. The stamp covers about 4''x4.5''. See the appendix for the details on each clump. 
$Right$: Object ID\_ECDFS=42881 ($z=3.16$). It is found in the
GEMS catalogue. From the left V, z, and V-z, stamps taken from GEMS images. The stamps cover an area of about 2''x1.5''. 
The red contours represent levels equal to 4,7, and 10 times the value of background. This AGN is excluded from our sample because a source lies on top of it. 
}
\label{SFGLAEcolExAGN}
\end{figure*}

\begin{figure} 
\centering
\includegraphics[width=90mm]{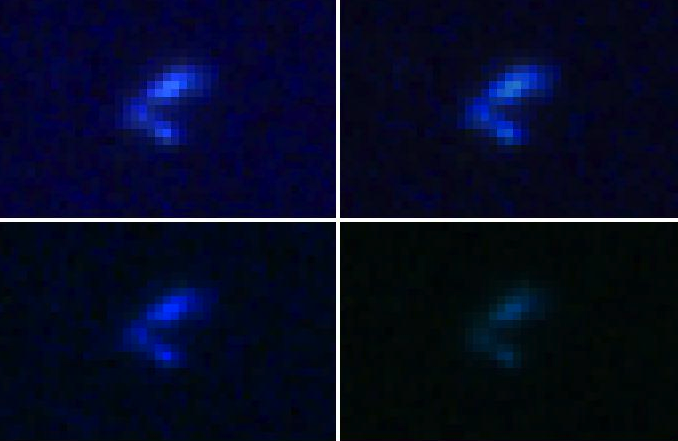}
\caption{Example of a source kept in the third step of our cleaning procedure, object ID\_CANDELS=6839 ($z=3.2078$). 
From the top left BVi, BVz, BIz, VIz 
stamps taken from CANDELS mosaic images. The stamps cover an area of about 2''x1.5''. See the appendix for the details on each clump.}
\label{keptstep2}
\end{figure}

After imposing the colour cuts discussed above, we excluded an
additional 22 sources. Figure \ref{SFGLAEcolExAGN} shows an example of a narrow-band-selected LAE rejected at this step. The V- and z-band images and a V-z colour image obtained from GEMS imaging for an 
AGN excluded during this step are also shown in Fig. \ref{SFGLAEcolExAGN}. It is clear from the inspection of the images that there is another
component in addition to the main source. This second source contaminates
the narrow-band (PSF full width at half maximum equal to 1.4$\arcsec$)
flux of the main source, which was classified as AGN according to the 
criteria reported in \citet{Hsu2014}. Figure \ref{keptstep2} shows an example of a
multi-component source kept in the sample because it is composed of two clumps with similar colours (see also Appendix \ref{sec:smart}). 
%
%
%

The number of objects left in our sample from the cleaning procedure is 86 galaxies (their distribution in CANDELS/GEMS areas is listed in Tables 4 and 5). From here on, we refer to this sample as the $\text{clean}$ sample. 
For sources at a redshift sampled by
more than one narrow band, we adopt the deepest narrow-band image to attempt to measure possible LyC signal. We also 
study eight AGN in clean regions (see Sect. \ref{sub:AGN}).
\subsection{Intrinsic L$_{\nu}$(1400)/L$_{\nu}$(900) ratio and intergalactic-medium transmissivity} 
\label{sec:exp}

We now proceed to estimate the intrinsic L$_{\nu}$(1400)/L$_{\nu}$(900) value and the intergalactic-medium transmissivity, exp(-$\tau_{IGM,z}$), at $3.11<z<3.53$.

The effective wavelength of the V606 filter corresponds to $\lambda=1400$ {\AA} for the redshift covered on average by our narrow-band filters.
The intrinsic L$_{\nu}$(1400)/L$_{\nu}$(900) of galaxies depends on initial mass function (IMF), star-formation history (SFH), metallicity (Z), stellar-population age, and star properties. 
For AGN, it depends on the hardness and the obscuration of their spectrum. For mixed SFG/AGN populations, it depends on the relative contribution of the two components and on all the other effect listed above.
\begin{table}[]
\caption{Intrinsic non-ioniziang versus ionizing radiation ratio}
\label{tab:intrLyCesc}
\scalebox{0.65}{
\begin{tabular}{|c|c|c|c|c|c|} 
\hline
\hline
IMF & SFH  & Z  & age & L$_{\nu}$(1400)/L$_{\nu}$(900)&  Evol Track, Stel Atm\\
&  & Z$_{\odot}$  & Myr &  &  \\
\hline
BC03& &   & &  & Padova94, KBFA\\
\hline
Chab & cSFR  & 0.02  & 1 &   1.76&  \\    
Chab & cSFR  & 0.02  & 10 &   3.41&  \\
Chab & cSFR  & 0.02  & 100 &   5.97&  \\
Chab & cSFR  & 0.02  & 1000 &  7.06&  \\
\hline
Chab & $\tau=0.1$Gyr  & 0.02  & 100 &  7.10&  \\
Chab & $\tau=-0.1$Gyr  & 0.02 & 100 &   5.27&  \\
\hline
Chab & cSFR  & 1  & 1 &   1.94&  \\
Chab & cSFR  & 1  & 10 &   4.28&  \\
Chab & cSFR  & 1  & 100 &   6.43&  \\
Chab & cSFR  & 1  & 1000 &   6.86&  \\
\hline
Salp & cSFR  & 1  & 100 &   6.71&  \\
\hline
\hline
BPASS\_v2& &   & &  & single+binary, BaSeL PoWR \\
\hline
Salp & cSFR  & 1  & 100 &  4.62 &  \\
\hline
\hline
Starburst99& &   & &  &  \\
\hline
Salp & cSFR  & 1  & 100 & 5.24 &  Padova94,PH\\
Salp & cSFR  & $\sim$1  & 100 &  4.62 &  GenevaV00,PH\\
Salp & cSFR  & $\sim$1  & 100 &  3.28 &  GenevaV40,PH\\
Salp & cSFR  & $\sim$1  & 100 &  3.63&  70\%ROT+30\%noROT,PH\\
\hline
\hline
Stel+AGN&  &   & &  & \\
\hline
&  &   & &  2.76& AGN-dominated \\ 
&  &   & &  7.04& obscured AGN-dominated \\ 
&  &   & & 2.66 & AGN-dominated \\ 
&  &   & & 2.35 & theoretical AGN$^{(1)}$ \\ 
\hline
\end{tabular}
}
\tablefoot{Intrinsic L$_{\nu}$(1400)/L$_{\nu}$(900) ratios obtained from different stellar population models: the BC03, assuming different initial mass functions (Salpeter, Salp, and Chabrier, Chab), star-formation histories (constant, cSFR, exponentially declining with $\tau=0.1$Gyr, exponentially rising with $\tau=-0.1$Gyr), metallicties (Z$_{\odot}$ and 0.02$\times$Z$_{\odot}$), and stellar-population ages; the BPASS 
 \citep[Binary Population and Spectral Synthesis code][]{Stanway2015}; the Starburst99 
 models. We consider a range of wavelengths around 900 {\AA} and 1400 {\AA}, comparable to that covered by the narrow-band and the HST/ACS F606W filters in the rest frame, to estimate the average L$_{\nu}$.
As described in \citet{Bruzual:2003}, BC03 provides the spectral energy distribution of stars obtained from a comprehensive library of theoretical model atmospheres (KBFA in the table). The library consists of \citet{Kurucz1996} spectra for O-K stars, \citet{Bessell1991} and \citet{Fluks1994} spectra for M giants, and \citet{Allard1995} spectra for M dwarfs. In BPASS the stellar evolutionary tracks contain a contribution from isolated stars and binary systems. Also, the stellar atmosphere models are selected from the BaSeL v3.1 library \citep{Westera2002}, supplemented by Wolf-Rayet stellar atmosphere models from the Potsdam PoWR group \citep{Hamann2003}. With Starburst99, it is possible to generate SEDs assuming a bunch of stellar evolutionary tracks, including stars with and without rotation \citep[GenevaV40 and GenevaV00 respectively,][]{Leitherer2014}, and stellar atmospheres \citep[the combination of model atmosphere from][is the recommended option]{Pauldrach1998,Hillier1998}. The 70\%ROT+30\%noROT entry indicates a model that is a combination of GenevaV40 for the 70\% and GenevaV00 for the 30\% \citep{Levesque2012}. The GenevaV40 and GenevaV00 tracks were released for metallicity equal to Z=0.014 \citep[$\sim$Z$_{\odot}$,][]{Eldridge2012}. The change in the intrinsic ratio due to the different evolutionary tracks and stellar atmospheres is shown for a Salpeter IMF, Z$_{\odot}$, cSFR, and 100 Myr old stellar population.
In the bottom part of the table, we report the ratios calculated from the best fit templates of AGN \citep{Bongiorno2012}, in which the SEDs are dominated by the radiation coming from an (un-)obscured AGN. $^{(1)}$ The intrinsic L$_{\nu}$(1400)/L$_{\nu}$(900) ratio for an unobscured TypeII AGN can be as low as 2.35 \citep{R06}.
In the estimation of the ratios we take into account the HI absorption occurred in star atmospheres, but we neglect that within the inter-stellar and circum-galactic medium. 
}
\end{table}

Table \ref{tab:intrLyCesc} summarizes the intrinsic ratios between non-ionizing and ionizing flux densities for a variety of galaxy physical properties and stellar population models. Since the emission in the rest-frame UV can strongly depend on the stellar evolutionary tracks and on the stellar atmosphere assumptions, 
we explore the intrinsic ratio adopting the BC03 \citep{Bruzual:2003}, the Starburst99\footnote{http://www.stsci.edu/science/starburst99/docs/parameters.html} \citep{Leitherer1999}, and the BPASS\footnote{http://www.bpass.org.uk/} \citep[binary population and spectral synthesis,][]{Stanway2015} codes. 

Stars with rotations that are located in binary systems can produce more ionizing photons than is expected for non-rotational, isolated stars \citep[e.g.][]{Levesque2012,Eldridge2012}. Therefore, they can in principle reduce the intrinsic L$_{\nu}$(1400)/L$_{\nu}$(900) value.

Evolutionary tracks of stars with rotation are implemented in Starburst99. However, as explained in \citet{Leitherer2014}, those models consider rotational velocities that might be too extreme. 
A more realistic experiment might be obtained by considering 70\% of evolutionary tracks with high rotational velocity and 30\% of evolutionary tracks without rotation \citep[70\%ROT+30\%noROT in the table,][]{Levesque2012}. 
%
%
%
%
We estimated the intrinsic ratios by varying 
evolutionary tracks \citep[from the Padova1994 tracks to the Geneva tracks with the GenevaV40 and without the GenevaV00 implementation of stellar rotations,][]{Leitherer2014}, and we also performed some tests by varying metallicity, IMF, and stellar atmospheres. 

As pointed out in \citet{Leitherer2014}, 50\% of the massive stars are in binary systems and experience evolutionary processes that  might alter the expected number of ionizing photons. Binary evolution also extends the period of ionizing-photon production, mainly at low metallicities, and so it is worth considering them in the studies of the re-ionization of the Universe. As described in \citet[][and references therein]{Stanway2015}, 
the increase in the production of ionizing photons that is due to binary systems is significantly lower than that produced by rotation models, such as the GenevaV40. We followed \citet{Nestor2013} and compared the previous intrinsic ratio estimations with those from BPASS version v2.0 (Salpeter IMF, constant SFR, Z$_{\odot}$, 100 Myr age). 

Our tests show that the strongest change in the intrinsic L$_{\nu}$(1400)/L$_{\nu}$(900) ratio is due to the age parameter. Variations with metallicity are only lower than 10\% in BC03. Different star-formation histories create variations of the order of 20-30\%.

Including binary stars in BPASS produces changes in the intrinsic ratio of up to 60\% with respect to BC03 for ages of 100 Myr.  Within Starburst99, changing the evolutionary tracks (Padova94 vs GenevaV00) produces variations of 20\% at the same age, while including rotation models reduces the intrinsic ratio of 30(70\%ROT+30\%noROT) to 40(GenevaV40) per cent. The variation caused by the IMF is negligible within BC03 and is up to 20\% in general. Finally, the change of stellar atmosphere models causes L$_{\nu}$(1400)/L$_{\nu}$(900) to increase or decrease by less than 20\% within Starburst99 and between BC03 and Starburst99. 

In conclusion, as a result of different evolutionary tracks and stellar-population models, the intrinsic ratio can present a 60\% uncertainty. In particular, for a Salpeter IMF, constant SFR, solar metallicity, and 100 Myr age, it can vary from 3 to 6, while for ages of a few hundreds of Myr, it can increase up to 8. We would like to stress that the intrinsic ratio is a multiplicative constant in the expression of the fesc$^{rel}$(LyC). By changing its value, it is possible to simply re-scale the fesc$^{rel}$(LyC), depending on the assumptions, to compare with other works. For this reason, we assumed an intermediate value of 5 as representative for our entire sample of star-forming galaxies. 
The value of 5 is strictly related to the assumptions of a Salpeter IMF and constant SFR, which is most frequently used in the literature and is the easiest to be compared with.

We also calculated the intrinsic ratios corresponding to the physical parameters listed in Tables 4 and 5. To estimate the galaxy physical parameters in the tables, we considered the BC03 models (Chabrier IMF) because they are among those that better reproduce the majority of the galaxy properties both in the UV and in the NIR \citep[][and references therein]{Zibetti2013}. They well reproduce the entire SED of intermediate-mass galaxies, like those studied here \citep{Santini2015}. 
In addition to this, with the BC03 code we are able to consider constant, rising, and declining star-formation histories, which include quite a wide variety of histories (they also mimic more than one single population). 

Under the assumption of a BC03 declining tau model, the range of intrinsic L$_{\nu}$(1400)/L$_{\nu}$(900) ratios is 4-16 for the galaxies (SFGs and LAEs) studied here, characterized by a median age of 300 Myr.

We refer to \citet{Bongiorno2012} to infer the expected L$_{\nu}$(1400)/L$_{\nu}$(900) ratio for AGN. They presented the observed SEDs of un-obscured and obscured AGN and also showed how typical templates vary as
a result of the host galaxy obscuration (their Figs. 4 and 2).   
The largest possible AGN contribution in the LyC regime produces a L$_{\nu}$(1400)/L$_{\nu}$(900) = 2.35. A typical AGN-dominated source can have a ratio of 2.6. Since the observed values already contain the IGM contribution, we adopt 2.35 throughout the paper. 
Sources in which the galaxy emission dominates that from the AGN are expected to present L$_{\nu}$(1400)/L$_{\nu}$(900) ratios similar to those calculated for galaxy templates. 

\begin{figure} 
\centering
\includegraphics[width=90mm]{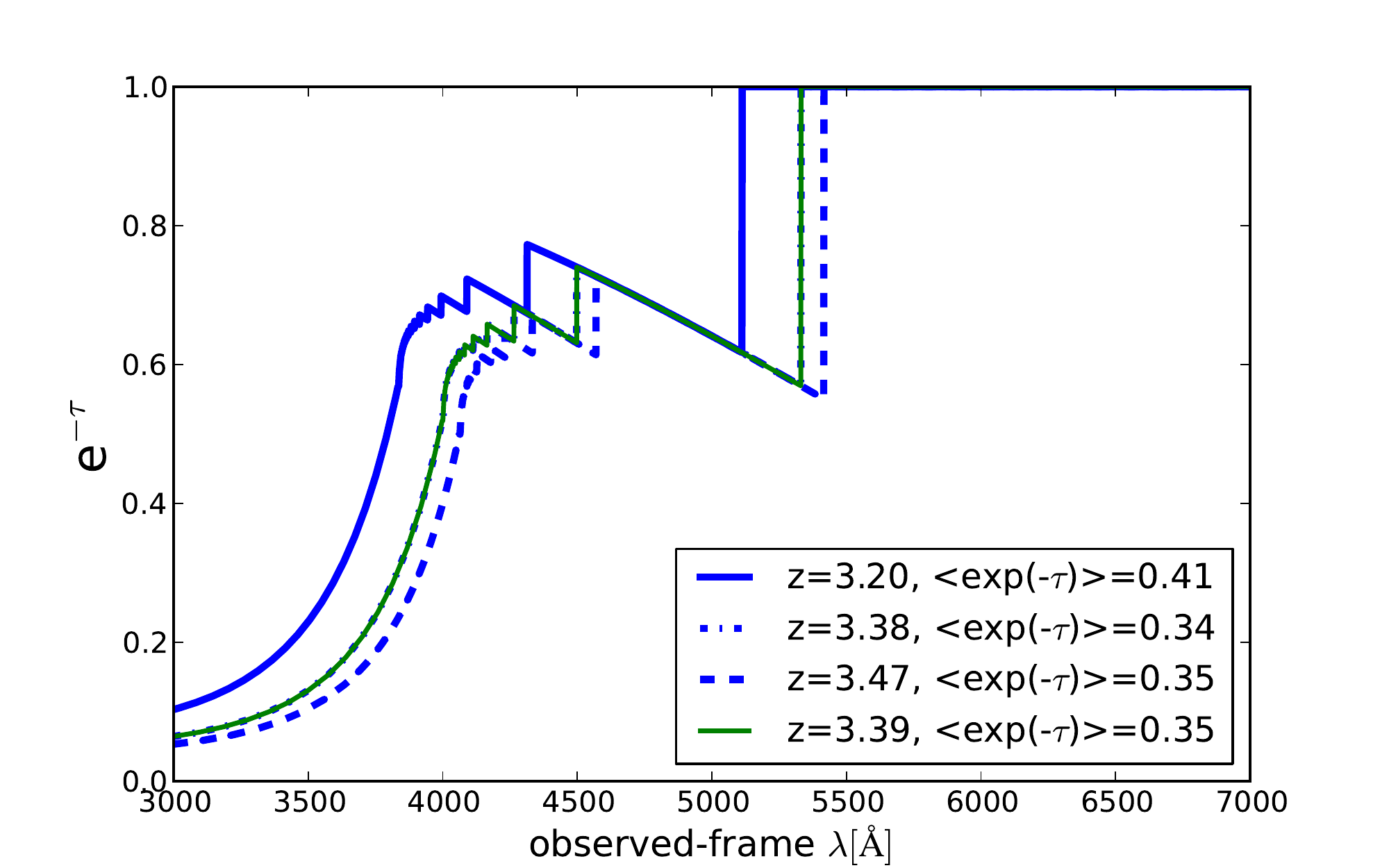}
\caption{Mean IGM transmissivity as a function of wavelength at the redshift corresponding to the centre of our narrow-bands \citep{Inoue2014}. Solid, dotted-dashed, dashed blue, and green curves indicate the average redshifts sampled by NB3727, NB387, NB396, and NB388, respectively.}
\label{InoueIGM}
\end{figure}

By following the analytical prescription by \citet{Inoue2014}, we determine $\tau_{IGM,z}$ (Fig. \ref{InoueIGM}). In the range of redshift studied here, exp(-$\tau_{IGM,z}) \simeq 0.36$. Even though we considered narrow-band filters, we derived the average transmissivity by applying the equation

\begin{equation}
<exp(-\tau_{IGM})> = \frac{\int{  exp(-\tau_{IGM}) T(\lambda)    d\lambda }} {\int {T(\lambda) d\lambda} }
\label{eqtau}
,\end{equation}

where T($\lambda$) represents the narrow-band filter transmission curve. Measuring LyC for a number of sources allowed us to use a mean value for $<$exp(-$\tau_{IGM})>, $
%
such as that provided by the analytical prescription by \citet{Inoue2014}. We refer to this paper for understanding the limitations and uncertainty related to the application of this prescription. See also \citet{Thomas2014} as an example of a study of IGM variability for sources at the same redshift. 

Finally, we performed a calculation to demonstrate that we could expect some direct LyC measurement from the sources in our sample, given their UV-continuum magnitude (V606) and the depth of our narrow-band filters. 

\begin{figure} 
\centering
\includegraphics[width=100mm]{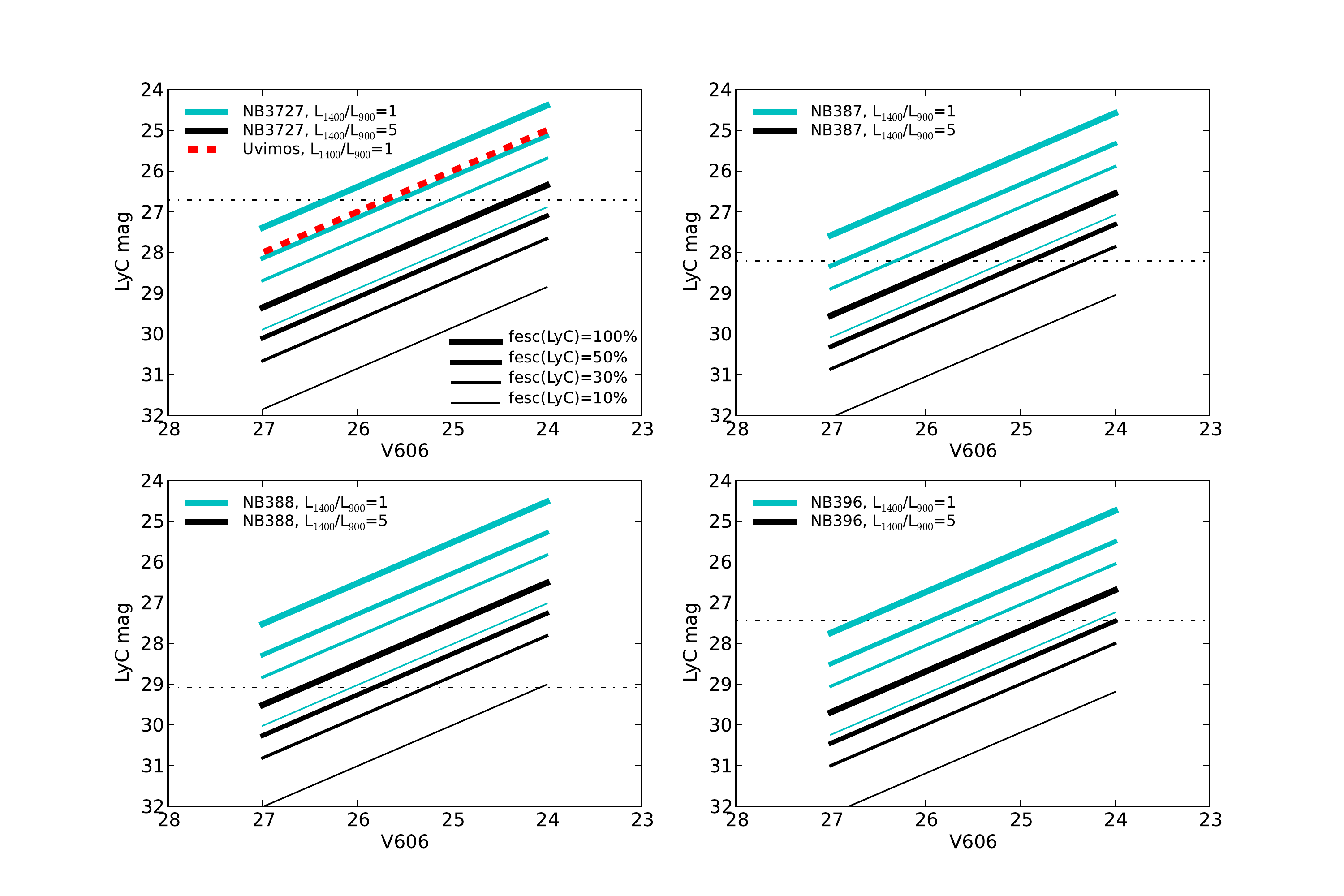}
\caption{Expected LyC mag as a function of the magnitude in the V606 band. For each narrow-band, we consider the redshift corresponding to 890 {\AA} at the centre of the filter. Thick (thin) lines correspond to the calculations under the assumption of fesc(LyC)=100\% (10\%), while the cyan (black) lines indicate L$_{\nu}$(1400)/L$_{\nu}$(900) = 1 (5). Horizontal dotted-dashed lines are located at the 3$\sigma$ detection limit of each narrow-band filter. The red dashed line indicates what is expected if the LyC mag is measured in the VIMOS $U$ broad-band.}
\label{LyCexpected}
\end{figure}

In this exercise, we considered a LyC escape fraction of 10-100\%, a galaxy template characterized by an intrinsic L$_{\nu}$(1400)/L$_{\nu}$(900) value between 1 and 5, and an average treatment of the IGM lines of sight. We estimated narrow-band magnitudes (LyC mag) as a function of V606, as shown in Fig. \ref{LyCexpected}. 
Given the depth of the available data, we expect more significant detections or upper limits from
the NB388 filter.
For a LyC emitter at $z=3.4$, characterized by an fesc(LyC)=100\% 
and V606=25 (the average magnitude of our sample of continuum-selected SFGs), we expect a LyC (NB388) mag = 25.5-27.5 (depending on the assumptions on the intrinsic L$_{\nu}$(1400)/L$_{\nu}$(900) value). This is brighter than the 3$\sigma$ detection limit of the narrow-band image and therefore in principle detectable. For fesc(LyC)=30\%, we expect a LyC (NB388) mag = 26-28, wich
is more difficult to be detected directly. We might also expect a LyC (NB387) mag = 25.6-27.6 (26.3-28.3) for an fesc(LyC)=100\% (30\%).
%
For the sources with LyC emission covered by NB396, an fesc(LyC)=100\% and V606=25, we expect a LyC (NB396) mag = 25.7-27.7, barely above the narrow-band detection limit. On the other hand, for a galaxy at $z=3.2$ we expect a LyC (NB3727) mag = 25.4-27.4, which is even below the NB3727 3$\sigma$ detection limit. Galaxies
that are characterized by an SED such as the one assumed in this exercise would show a LyC signal in NB3727 (fesc(LyC) $>$ 30\%) if they were brighter than V606=24. The same calculation performed with the VIMOS $U$ broad band would produce a LyC mag up to 0.6 magnitude fainter than the value calculated for NB3727. This demonstrates the advantage of using narrow instead of broad bands in the LyC measurements. Of course the $U$-band estimation can change if the broad band is significantly deeper than the narrow band. 
\section{Physical properties of the clean sample}  \label{sec:prop}

%
We cross-correlated our sources in the clean sample with the E-CDFS multi-wavelength photometric catalogue \citep{Guo2013,Hsu2014}. In this way, we associated  a multi-wavelength SED from the optical $U$ to the Spitzer bands with every source, thanks to SED fitting, physical parameters, such as stellar mass (M$^*$), star-formation rate (SFR), dust reddening (E(B-V), and stellar-population age. 

\subsection{Star-forming galaxies}

We adopted the physical parameters quoted in \citet{Santini2015} for the galaxies entering the CANDELS area. 
For the sources in E-CDFS, we estimated M$^*$, SFR, E(B-V), and stellar-population age starting from the \citet{Hsu2014} photometry, the spectroscopic redshift, and with the same code as in  \citet{Santini2015}, adopted for CANDELS. 
%
%
%

In Tables \ref{tab:propSFG} and \ref{tab:propLAE}, we report the physical parameters considered here. For reference, we report the IDs from the CANDELS and \citet{Hsu2014} catalogues for every source. 
It is important to note that our sample covers quite a wide range in UV absolute magnitude ($-21.5\lesssim$ M1400 $\lesssim-18.5$), stellar mass (10$^{8}$ M$_{\odot}\lesssim$ M$^{*} \lesssim 5 \times 10^{10}$), star-formation rate ($1\lesssim$ SFR $\lesssim50$ M$_{\odot}$ yr$^{-1}$), specific star-formation rate ($10^{-9.5} \lesssim$ sSFR  $\lesssim 10^{-7}$ yr$^{-1}$), dust reddening ($0\lesssim $ E(B-V) $\lesssim0.2$), and stellar population age ($7\lesssim$ log(age/yr)  $\lesssim 9.3$). Some of the LAEs are too faint in the continuum to enter the CANDELS/E-CDFS photometric catalogue. Therefore, we have a stellar mass estimation for only 11 of the 19 LAEs.

\begin{figure} 
\centering
\includegraphics[width=100mm]{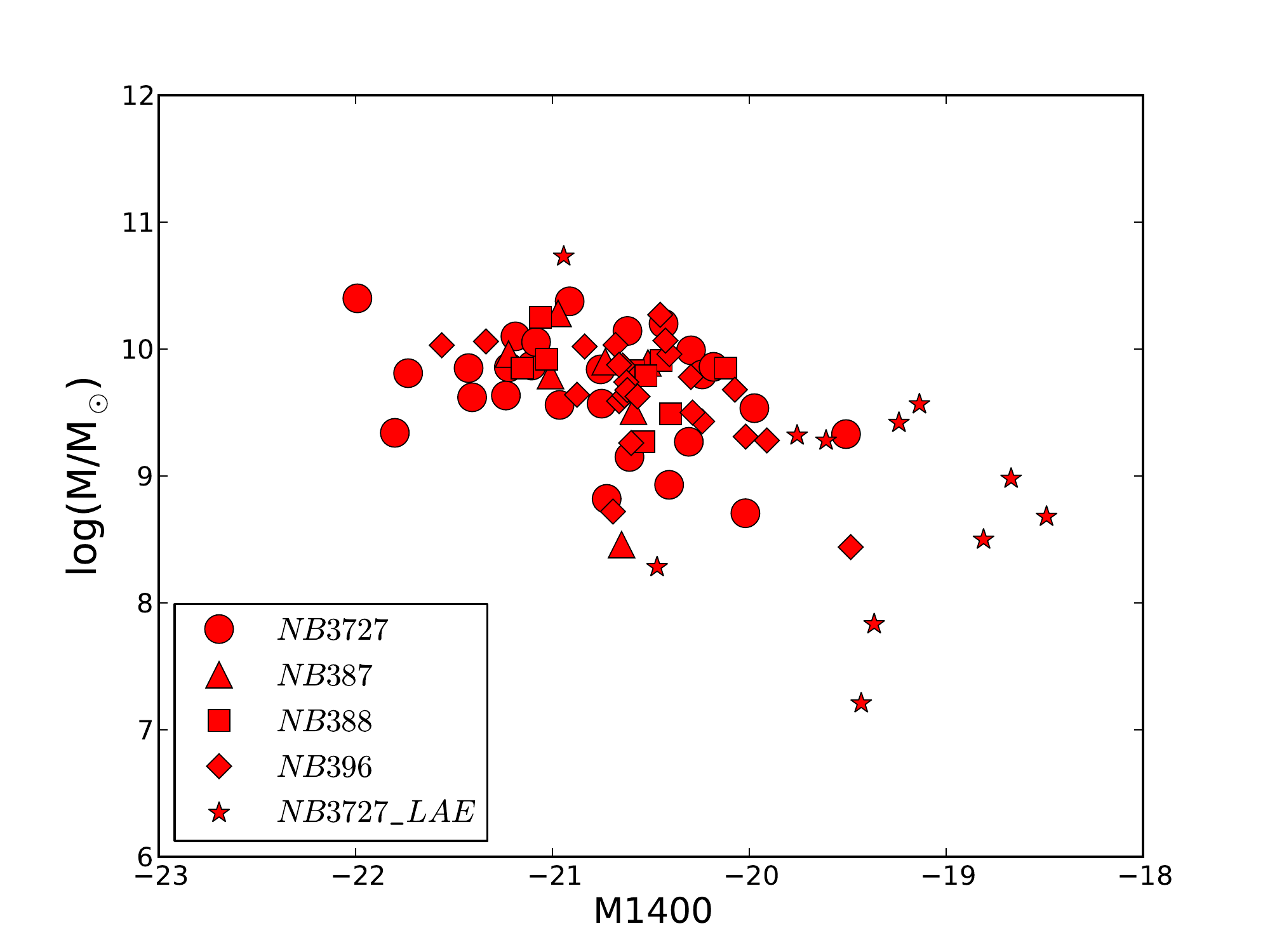}
\caption{Stellar mass vs rest-frame UV absolute magnitude. The symbols indicate SFGs in the range of redshift covered by NB3727 (red circles), NB387 (red triangles), NB388 (red squares), NB396 (red diamonds), and LAEs (red stars). The LAEs found in GEMS catalogue but not in the E-CDFS catalogue from \citet{Hsu2014} are not shown in this plot.}
\label{M1400mass}
\end{figure}

Figure \ref{M1400mass} shows stellar mass as a function of rest-frame UV magnitude. The sub-sample of Ly$\alpha$ emitters occupies the region with the lowest mass and the faintest magnitude. We refer to Fig. 7 in \citet{Grazian2015a} to compare our sample with all the sources in GOODS-S. In a similar redshift range, our sample occupies the bright-magnitude and high-mass tail of the general distribution. This is expected because the sources are all spectroscopically confirmed.

\begin{table*}[]
\centering
\caption{Description of the properties of the star-forming galaxies}
\label{tab:propSFG}
\scalebox{0.7}{
\begin{tabular}{|c|c|c|c|c|c|c|c|c|c|c|c|c|c|c|c|c|} 
\hline
\hline
(1)& (2) &  (3) &(4)  &(5)   &(6)  & (7) & (8) & (9) & (10) & (11) & (12)& (13) & (14) &(15) & (16) &(17)\\ 
ID\_ECDFS &RA &dec &zspec & log(M$^*$/M$_{\odot}$) & SFR & E(B-V)  & log(age/yr) & $\tau$ & Z &M$^*_{median}$ & V & errV  & $\frac{f_{900}}{f_{1400}}$ & err$\frac{f_{900}}{f_{1400}}$ & 2$\sigma$NB/fV &    $\frac{L_{\nu}(1400)}{L_{\nu}(900)}$ \\ 
 & deg & deg   & & &M$_{\odot}$ yr$^{-1}$  & &  & Gyr& Z$_{\odot}$& 1E+9M$_{\odot}$ &  &  &  & & & \\ 
\hline
NB3727&  &  &  &  &  &  &  &  &  & & & & & & & \\ 
\hline
23052 & 53.0090103 & -27.7084465 & 3.263  & 10.20 & 13.99 & 0.15 & 8.45 &  0.1 & 1.0 & 16.50$\pm$2.59 & 25.39 & 0.06& 0.108 & 0.234 & 0.504 & 9 \\ 
18320$^{(a)}$ &   53.0135155 & -27.7552338    & 3.213 &9.62 & 6.79 & 0.00 & 8.35 &   0.1  & 1.0   &  4.85$\pm$0.64 & 24.43 & 0.02 & -0.034 & 0.105 & 0.208 &9 \\ 
21398 &  53.0145454  & -27.7279186 &3.132& 9.87 & 9.00 & 0.06 & 8.40 &   0.1  & 0.2   &    7.33$\pm$1.29 & 24.75 & 0.03& 0.071 & 0.132 & 0.281 & 10\\ 
6839 & 53.0616188 &-27.8462505 & 3.208 &9.81 & 13.88 & 0.06 & 8.30 &   0.1  & 0.02   &  5.77$\pm$1.35 & 24.11 & 0.02 & 0.108 & 0.064 & 0.155 & 9\\ 
13340 & 53.0808716 &-27.7912197 & 3.274&  9.84 & 8.47 & 0.06 & 8.40 &   0.1  & 0.2   & 8.80$\pm$2.16 & 25.07 & 0.03& 0.124 & 0.183 & 0.375 &10\\ 
26236 & 53.1133347 & -27.6977139 & 3.168&  9.56 & 6.76 & 0.03 & 8.70 &   0.6  & 0.02   &    2.72$\pm$0.75 & 24.89 & 0.03 & -0.054 & 0.159 & 0.317 &7\\ 
5096 &  53.1165733 & -27.8634548& 3.287 &9.33 & 0.58 & 0.00 & 8.60 &   0.1  & 0.02   & 1.93$\pm$0.39 & 26.31 & 0.08& -0.312 & 0.544 & 1.177 &16 \\ 
697 & 53.1581573 & -27.9257145 & 3.192& 9.15 & 7.55 & 0.06 & 8.40 &   15.0  & 0.02   & 1.14$\pm$0.36 & 25.24 & 0.02& 0.135 & 0.242 & 0.437 & 7\\ 
24050 &53.1656418 &-27.6657372 & 3.271 & 9.57 & 4.57 & 0.03 & 8.40 &   0.1  & 0.02   & 2.82$\pm$0.94 & 25.07 & 0.03 & -0.442 & 0.209 & 0.377 &10\\ 
7259 &53.1716728 &-27.8424492 & 3.162 &9.85 & 14.98 & 0.06 & 8.30 &   0.1  & 1.0   & 6.37$\pm$1.76 & 24.43 & 0.02& -0.018 & 0.101 & 0.207 &9\\ 
8231 &  53.2197647 &-27.8338223 & 3.143 &   10.10 &  14.02 & 0.06 &    8.70   &   0.3  & 1.0   & 9.19$\pm$2.04 & 24.67 & 0.03& -0.088 & 0.131 & 0.259 & 7\\ 
16430 & 53.1937943 &-27.7712898 & 3.188 &9.99 & 13.67 & 0.10 & 8.65  &   0.3  & 1.0   &   9.17$\pm$2.41 & 25.55 & 0.05& 0.031 & 0.264 & 0.584  & 7\\ 
17885 &  53.140419 &-27.7592926& 3.236 & 9.80 & 3.87 & 0.03 & 8.50 & 0.1&1.0 & 6.52$\pm$1.82 & 25.59 & 0.04& -0.250 & 0.283 & 0.608 & 10\\ 
16526 &53.0275917 &-27.7702847 & 3.149 &  9.86 & 3.02 & 0.03 & 8.55 &   0.1  & 1.0   &  7.54$\pm$1.83 & 25.67 & 0.05& -0.073 & 0.322 & 0.654 & 10\\ 
18454 &  53.0399857 &-27.7540779 & 3.163 &      8.82 &             69.68 & 0.15 &    7.00     &   15.0  & 0.2   &       0.08$\pm$0.02& 25.13 & 0.04& 0.137 & 0.196 & 0.396 & 7\\ 
75666 &52.9956665 &-27.7398472 & 3.208 & 10.06 & 12.97 & 0.06 & 8.70 &  0.3   & 0.02   &  -- & 24.76 & 0.02 & 0.005 & 0.136 & 0.281 &8 \\ 
59083 & 53.0544014 &-27.9023724 & 3.232 &  9.86 & 9.07 & 0.00 & 9.00 &   3.0  & 1.0   &  -- & 24.61 & 0.01& 0.085 & 0.110 & 0.246 &7 \\ 
52315 & 53.1323662   & -27.9521828 & 3.238 & 9.34 & 231.40 & 0.15 & 7.00 & 15.0 &1.0 &-- & 24.03 & 0.02& -0.028 & 0.086 & 0.144 & 4\\ 
56382 &53.1925201 &-27.9246674  & 3.173 &9.53 & 12.04 & 0.06 & 8.60 &15.0 & 0.02&-- & 25.87 & 0.05& -0.065 & 0.436 & 0.788 &7\\ 
62873 & 53.2538567 &-27.8633366 &3.132 &  10.40 & 23.28 & 0.03 & 8.90 &0.6 & 0.02&-- & 23.87 & 0.01& 0.027 & 0.061 & 0.124 & 8\\ 
37590 & 53.019684  &	-28.045127 &3.173 & 10.14 & 8.71 & 0.06 & 8.50 & 0.1 &0.02 &-- & 25.23 & 0.03& -0.065 & 0.213 & 0.435 &12\\ 
46323 & 53.206498	&-27.991631 &3.153 &9.63 & 17.34 & 0.03 & 8.50 & 2.0 &0.2 &-- & 24.62 & 0.02& 0.063 & 0.125 & 0.247 &7\\ 
46783 &53.211483	&-27.989133  &3.192 & 10.38 & 29.01 & 0.06 & 8.90 & 1.0 & 1.0 &-- & 24.93 & 0.03 & 0.069 & 0.178 & 0.330 &7\\ 
91256 & 53.206902&-27.622635 &3.278 &9.27 & 18.23 & 0.10 & 8.10 & 2.0 & 0.2 &-- & 25.51 & 0.04& -0.073 & 0.335 & 0.565 &7\\ 
82659 & 53.027626&	-27.678058 &3.151 &8.71 & 14.09 & 0.10 & 7.60 & 0.3&1.0 &-- & 25.83 & 0.04& 0.454 & 0.362 & 0.759 &6\\ 
79422 & 52.970582 &-27.704338 &3.283 & 8.93 & 9.84 & 0.03 & 8.00 & 0.6 & 1.0 & -- & 25.41 & 0.02& -0.370 & 0.234 & 0.515 & 7\\ 
\hline
NB387&  &  &  &  &  &  &  &  &  & & & & & & & \\ 
\hline
24919 & 53.0953941& -27.6875267 &3.356 & 9.51 & 3.97 & 0.03 & 8.40 & 0.1& 0.02&2.48$\pm$0.63 & 25.21 & 0.04& -0.071 & 0.050 & 0.113 &10 \\ 
25151$^{(a)}$ &  53.1433334 &-27.690094 &3.419 &9.79 & 218.10 & 0.30 & 7.50 & 2.0 & 0.2 &10.30$\pm$3.02 & 24.78 & 0.03& -0.024 & 0.033 & 0.075  & 5\\ 
2584 & 53.1451225 &-27.8903446& 3.369 &10.28 & 82.52 & 0.20 & 8.50 &15.0 & 1.0&19.80$\pm$6.24 & 24.83 & 0.02& -0.007 & 0.035 & 0.079 & 7\\ 
26325$^{(a)}$ & 53.1785698 &-27.7017498 &3.412 & 9.96 & 19.10 & 0.06 & 8.55 & 0.3&0.2 & 9.13$\pm$1.74 & 24.57 & 0.02& -0.028 & 0.029 & 0.062 &8 \\ 
18882 &53.0257111 &-27.7505207 &3.403 & 9.89 & 6.74 & 0.03 & 8.45 & 0.1 & 1.0 &5.73$\pm$1.24 & 25.28 & 0.03& 0.046 & 0.052 & 0.119 & 9\\ 
14354 &53.0312309 &-27.7852192 &3.429 &  9.90 & 9.71 & 0.06 & 8.40 & 0.1 & 0.2 & 9.12$\pm$2.56 & 25.05 & 0.04& -0.093 & 0.051 & 0.097  &10 \\ 
69899 &53.2544556 &-27.7927914  &3.334 &8.46 & 30.26 & 0.06 & 7.00 & 15.0 & 0.2 &-- & 25.16 & 0.02& 0.021 & 0.054 & 0.107 & 4\\ 
\hline
NB388&  &  &  &  &  &  &  &  &  & & & & & & & \\ 
\hline
20417 & 53.1206131 &-27.7365837&  3.368 &9.92 & 7.32 & 0.00 & 8.75 & 0.3& 1.0& 9.08$\pm$1.52 & 24.77 & 0.02& -0.016 & 0.019 & 0.038 & 8\\ 
23842 & 53.1390076 & -27.7152863 &3.420 &9.85 & 11.58 & 0.03 & 8.35 & 0.1 & 1.0 & 6.47$\pm$1.18 & 24.63 & 0.02& 0.007 & 0.020 & 0.034 &9\\ 
16346 &  53.1487122 &-27.7729473& 3.316 & 9.83 & 4.30 & 0.03 & 8.50 & 0.1 &0.02 &0.01$\pm$0.00& 25.22 & 0.01& -0.025 & 0.028 & 0.058 & 12\\ 
11608 &53.2189064 &-27.8042641 & 3.472 & 9.91 & 10.63 & 0.06 & 9.05 & 15.0 & 0.2 & 7.04$\pm$1.44 & 25.33 & 0.05& 0.016 & 0.049 & 0.064 & 7\\ 
13190 & 53.2286491 &-27.7923336&3.327 &  9.79 & 3.83 & 0.03 & 8.50 & 0.1 & 0.02 & 4.30$\pm$2.14 & 25.28 & 0.09& -0.027 & 0.064 & 0.062 &12 \\ 
15220 & 53.1332321 &-27.7797909& 3.469 & 9.49 & 5.03 & 0.03 & 8.35 & 0.1 & 0.2 & 3.68$\pm$0.65 & 25.38 & 0.01& 0.014 & 0.032 & 0.067 & 9\\ 
19760 & 53.1512985 &-27.7429085& 3.413 & 10.25 & 19.67 & 0.06 & 8.70 & 0.3 & 1.0 & 14.00$\pm$3.41 & 24.73 & 0.02& -0.019 & 0.018 & 0.037 & 8\\ 
13734 & 53.0485344& -27.7886524& 3.309 &  9.27 & 3.20 & 0.00 & 8.60 & 0.3 &0.02 & 1.49$\pm$0.49 & 25.28 & 0.04& 0.019 & 0.061 & 0.061 &8 \\ 
20533 &53.1780052 &-27.7354355 & 3.478 &9.85 & 6.57 & 0.10 & 8.90 & 0.6& 0.02&5.29$\pm$0.97& 25.65 & 0.05 & -0.035 & 0.062 & 0.087  & 8\\ 
\hline
NB396&  &  &  &  &  &  &  &  &  & & & & & & & \\ 
\hline
20583 & 53.0307884 &-27.7348957 & 3.498 & 9.43 & 2.35 & 0.00 & 8.45 & 0.1 & 0.2 &3.76$\pm$0.97 & 25.528 & 0.049& -0.021 & 0.160 & 0.309 &12\\ 
5764 & 53.0401802 & -27.8563824 & 3.466 &  9.64 & 5.80 & 0.00 & 8.80 & 0.6 & 0.02 &3.02$\pm$0.83 & 24.900 & 0.015& 0.156 & 0.101 & 0.173 & 8\\ 
15347 & 53.0550652 &-27.7784977& 3.521 &  9.68 & 1.32 & 0.00 & 8.60 & 0.1& 0.02&4.50$\pm$1.25 & 25.689 & 0.057& 0.095 & 0.174 & 0.358 &16 \\ 
4717 & 53.0617599 & -27.8679447 & 3.473 &10.06 & 20.11 & 0.06 & 8.60 & 0.3 & 0.02 &12.00$\pm$4.38 & 24.436 & 0.025& 0.057 & 0.067 & 0.113 & 8\\ 
19613 & 53.0936394 &-27.7440109&3.494 & 9.96 & 2.48 & 0.03 & 8.60 & 0.1 & 0.02 & 8.54$\pm$0.96 & 25.362 & 0.039& -0.003 & 0.129 & 0.265 & 16\\ 
2324 &53.1360855 & -27.8937855 & 3.490 &  10.02 & 9.13 & 0.03 & 8.45 & 0.1& 1.0 &7.62$\pm$1.87 & 24.933 & 0.023& -0.073 & 0.086 & 0.178 & 9\\ 
3325 & 53.1867943 & -27.8814468& 3.473 & 9.87 & 11.60 & 0.06 & 8.90 & 2.0 & 0.02& 5.27$\pm$1.74 & 25.131 & 0.025& -0.070 & 0.104 & 0.214 & 7\\ 
27939 &53.2074356 &-27.8796673&  3.473 & 8.44 & 2.04 & 0.03 & 8.00 & 0.1 & 0.02 & 0.33$\pm$0.17 & 26.289 & 0.097& -0.020 & 0.313 & 0.622 &7 \\ 
3768 & 53.2158508 &-27.8768501& 3.468 &  9.74 & 9.23 & 0.03 & 8.60 & 0.3 & 1.0 &4.48$\pm$1.03 & 25.149 & 0.038& 0.131 & 0.106 & 0.218 & 7\\ 
3037 &  53.2237396 & -27.8844109 & 3.520 &9.26 & 6.18 & 0.03 & 8.20 & 0.1 &0.2 & 1.90$\pm$0.49 & 25.162 & 0.035& 0.059 & 0.107 & 0.220 &7 \\ 
3906 &53.2173729 & -27.8750782 & 3.466 &  9.59 & 4.76 & 0.03 & 8.40 & 0.1 & 0.02 & 3.24$\pm$0.86 & 25.114 & 0.029& 0.097 & 0.104 & 0.211 & 10\\ 
4022 & 53.2294273 & -27.8742199  &  3.464 &   10.27 & 17.10 & 0.10 & 9.00 & 1.0 & 0.02 &8.83$\pm$4.53 & 25.323 & 0.052 & 0.003 & 0.129 & 0.255 & 8\\ 
9692 &  53.0268402 &-27.8209667 & 3.470 &9.31 & 2.52 & 0.03 & 8.40 & 0.1 & 0.02 & 2.23$\pm$1.15 & 25.756 & 0.049& 0.338 & 0.215 & 0.380 & 10\\ 
7835 &53.0394936 &-27.8370953 &3.456 &  9.50 & 305.30 & 0.35 & 7.05 & 15.0 & 0.2 & 5.03$\pm$1.71 & 25.489 & 0.067& -0.201 & 0.167 & 0.298 & 4\\ 
20745 & 53.0459061 &-27.7336273  & 3.498 &   9.28 & 2.36 & 0.03 & 8.40 & 0.1 & 0.02 & 1.46$\pm$0.33 & 25.857 & 0.070& 0.304 & 0.205 & 0.418 & 10\\ 
73615& 52.9943237 &-27.7592602 &3.446 &  9.63 & 7.48 & 0.03 & 8.60 & 0.3 & 0.02 &-- & 25.146 & 0.025& 0.086 & 0.106 & 0.217 & 8\\ 
59311 & 53.0645256 & -27.9011326& 3.461 & 9.68 & 5.78 & 0.00 & 8.40 & 0.1 & 0.2 &-- & 25.157 & 0.017& -0.043 & 0.127 & 0.219 &10 \\ 
89369 &  53.1025162 &-27.6338024& 3.508 & 8.72 & 24.27 & 0.10 & 7.35 & 0.1 & 1.0 & -- & 25.073 & 0.017& -0.047 & 0.113 & 0.203 & 5\\ 
90129 & 53.1687737 &-27.6299114& 3.502 & 10.07 & 18.85 & 0.03 & 8.35 & 0.1 & 1.0 & -- & 25.341 & 0.018& -0.069 & 0.125 & 0.260 &9\\ 
61551 & 53.2748108 &-27.8745422& 3.466 &10.03 & 2.96 & 0.03 & 8.60 &0.1 & 0.02 &-- & 25.096 & 0.023& -0.163 & 0.140 & 0.207 &16\\ 
49017  & 52.993400 & -27.972620& 3.490 &   9.78 & 11.31 & 0.06 & 8.70 & 0.6 & 0.02 & -- & 25.473 & 0.029& -0.248 & 0.141 & 0.293 &7\\ 
70198 & 53.002103&	-27.790144 &3.470 & 9.88 & 28.22 & 0.10 & 8.50 & 1.0 & 0.02 & -- & 25.113 & 0.039& 0.052 & 0.120 & 0.210 &7\\ 
70236 & 53.238829 &-27.792083	  &3.489 &10.03 & 17.41 & 0.03 & 8.35 & 0.1 & 0.2 & -- & 24.207 & 0.012& 0.035 & 0.068 & 0.091 &9\\ 
78739 & 53.251029 & -27.711819 &3.475 & 10.38 & 15.05 & 0.03 & 9.11 & 1.0 & 0.2 & --& 25.073 & 0.018& -0.171 & 0.103 & 0.203 &7\\ 
83725 & 53.035204 &-27.670960 &3.484 &9.63 & 9.05 & 0.03 & 8.30 & 0.1 & 0.02 & --& 25.203 & 0.019& -0.090 & 0.118 & 0.229 &9\\ 
\hline
\end{tabular}
}
\tablefoot{(1) CANDELS/E-CDFS catalog identification number; 
(2)(3) right ascension and declination; 
(4) spectroscopic redshift; (5)(6)(7)(8)(9)(10) stellar mass, SFR, dust reddening, parametrized by E(B-V), stellar-population age obtained by assuming exponentially declining $\tau$ models, $\tau$ value in Gyr, and metallicity in solar units. The ratio between $\tau$ and stellar-population age is such that the models resemble an instantaneous burst as well as a constant SFR. The SED fit is run fixing the spectroscopic redshift; (11) median mass, taken from the compilation of \citet{Santini2015}. It is reported only for sources in CANDELS; (12)(13) V-band AB magnitude taken from CANDELS/GEMS catalogs and its error; (14)(15) ratio between narrow-band and V-band flux densities and its error; (16) ratio between the 2$\sigma$ background noise of the narrow-band image and the source V-band flux density; (17) intrinsic L$_{\nu}$(1400)/L$_{\nu}$(900) ratio for the age, $\tau$, metallicity in columns (8)(9)(10), and adopting BC03 models.\\ 
$^{(a)}$ sources already studied in \citet{Vanzella2010c}; ID\_ECDFS=18320 is Ion2, the source also studied in \citet{Vanzella2015} as a LyC emitter candidate. The sources identified with a 530 million number come from VUDS. 
}
\end{table*}

\begin{table*}[]
\centering
\caption{Description of the properties of the narrow-band selected Ly$\alpha$ emitters}
\label{tab:propLAE}
\scalebox{0.7}{
\begin{tabular}{|c|c|c|c|c|c|c|c|c|c|c|c|c|c|c|c|c|} 
\hline
\hline
(1)& (2) & (3)& (4)& (5)  &(6)   &(7)  & (8) & (9) & (10) & (11) & (12) & (13) &(14) & (15) & (16) & (17) \\ 
ID\_ECDFS  & RA & dec & zspec & log(M$^*$/M$_{\odot}$)& SFR & E(B-V) & log(age/yr) &$\tau$ & Z & M$^*_{median}$ & V & errV  & $\frac{f_{900}}{f_{1400}}$ & err$\frac{f_{900}}{f_{1400}}$ & 2$\sigma$NB/fV & $\frac{L_{\nu}(1400)}{L_{\nu}(900)}$ \\ 
 &deg   &deg & &  &  M$_{\odot}$ yr$^{-1}$  &  & & Gyr & Z$_{\odot}$ & M$_{\odot}$1E+9 & & & & &  & \\ 
\hline
NB3727&  &  &  &  &  &  &  &  &  & & & & & & & \\ 
\hline 
20433 &  53.065781 & -27.736179 &3.119 & 9.32 & 3.06 & 0.06 & 9.00 & 15.0 & 0.2 & 2.19$\pm$0.52 &26.11 & 0.05 & 0.052 & 0.466 & 0.975 & 7\\ 
28279 & 53.229096 & -27.868610 &3.134 &8.98 &  0.72 & 0.03 & 8.50 & 0.1 & 0.02 & 0.67$\pm$0.30 &27.16  & 0.18 & 0.522 & 1.240& 2.563 &12 \\ 
19293 & 53.053879 &-27.747627 & 3.117 &  10.73 & 61.91 & 0.25 & 9.00 & 2.0 & 1.0 & 45.90$\pm$6.67 & 24.92 & 0.05 & -0.106 & 0.161 & 0.327 &7\\ 
34808  & 53.180985 &-27.673071 &3.130 & 8.68  & 0.87  & 0.00 & 8.50 &  0.1 & 0.02 & 0.37$\pm$0.08& 27.37 & 0.21 & -0.790 & 1.691 & 3.131 & 12\\ 
63466 &  52.978683 & -27.855867 & 3.115 &8.28 & 20.37 & 0.10 & 7.00 & 15.0 & 1.0 & -- & 25.40 & 0.02 & 0.0001 & 0.217 & 0.507 & 4\\ 
84199 &53.194806 & -27.666645 &3.112 & 9.57 & 5.77 & 0.15 & 8.75 & 0.6 & 0.2 & -- & 26.73 & 0.05 & -0.150 & 0.962 & 1.730 &8\\ 
52172 &  52.971719 & -27.952267& 3.115 &     9.28 & 2.48 & 0.00 & 9.06 & 15.0 & 2.5 &-- & 26.25 & 0.04 & 0.340 & 0.517 & 1.117 &7\\ 
74439 & 52.949147 &-27.750925 &  3.107 &  9.42 & 1.49 & 0.00 & 9.25 & 2.0 & 0.2 &-- & 26.63 & 0.05 & -0.897 & 0.782 & 1.574 & 7\\ 
-$^{(a)}$ & 53.078858 &-27.644502 &3.122 & -- & -- &  -- & -- & --& -- & -- & 26.88 & 0.07 & -1.082 & 0.950 & 1.991 &--\\ 
46913 &  52.959872 & -27.986139 & 3.123 &7.83 & 2.45 & 0.00 & 7.50 & 2.0 & 1.0 & -- & 26.50 & 0.04 & 0.460 & 0.724 & 1.396 & 6\\ 
-$^{(a)}$ & 53.236134 & -27.830346 &3.102 &-- & -- &  -- & -- & --& -- & -- & 26.82 & 0.06 & -0.747 & 0.994 & 1.874 &--\\ 
-$^{(a)}$ & 53.278931 & -27.707732& 3.111 &-- & -- &-- & -- & --& -- & -- &  26.91 & 0.07 & 0.270 & 0.967 & 2.044 &--\\ 
-$^{(a)}$ & 52.999277 &-27.829559& 3.120 &  -- &  -- & -- & -- & -- & -- & -- & 26.07 & 0.04 & -0.475 & 0.457 & 0.940 &--\\ 
-$^{(a)}$ &53.367375 &-28.055773 & 3.122 &-- & -- & -- & -- & -- &-- & -- & 27.35 & 0.07 & -0.326 & 1.553 & 3.070 &--\\ 
-$^{(a)}$ &53.043862 &-27.988235 &3.122 &  -- & -- & -- & -- & -- & -- & -- &27.02 & 0.05 & -0.100 & 1.236 & 2.269 &--\\ 
-$^{(a)}$ & 53.131038 &-27.726939 &3.133 &-- & -- & -- & -- & -- & -- & -- &26.99 & 0.11 &  -0.985 & 1.111 & 2.196 &--\\ 
75954 & 53.348875 &-27.735881&3.114 &8.50 & 1.38  &0.00  & 8.50 & 15.0 & 0.2 & - & 27.05 & 0.09 &  -1.311 & 1.121 & 2.322 &7\\ 
-$^{(a)}$ &53.351875 &-27.742758 &3.113 & -- & -- & -- & -- & -- & -- & -- &27.61 & 0.13 & -2.904 & 1.895 & 3.903 &--\\ 
\hline
\end{tabular}
}
\tablefoot{(1) Identification number from CANDELS/E-CDFS catalog; (2)(3) right ascension and declination in degrees; 
(4) spectroscopic redshift; (5)(6)(7)(8)(9)(10) stellar mass, SFR, dust reddening, parametrized by E(B-V), stellar-population age obtained by assuming exponentially declining $\tau$ models, $\tau$ value in Gyr, and metallicity in solar units. 
The SED fit is run fixing the spectroscopic redshift; (11) median mass, taken from the compilation of \citet{Santini2015}. It is reported only for sources in CANDELS;  (12)(13) V-band AB magnitude taken from CANDELS/GEMS catalogs and its error; (14)(15) ratio between narrow-band and V-band flux densities and its error; (16) ratio between the 2$\sigma$ background noise of the narrow-band image and the source V-band flux density; (17) intrinsic L$_{\nu}$(1400)/L$_{\nu}$(900) ratio for the age, $\tau$, metallicity in columns (8)(9)(10), and adopting BC03 models. \\ 
$^{(a)}$ sources detected in GEMS, but without counterpart in the photometric CDF-S catalog.\\
}
\end{table*}

\subsection{Active galactic nuclei}
\label{sub:AGN}


We matched our initial sample with AGN catalogues. We chose the \citet{Xue2011} X-ray catalogue 
and the \citet{Hsu2014} AGN photometric list.
In Table \ref{tab:propAGN} we present the AGN matches. 
Most them have a counterpart in X-ray \citep{Xue2011, Fiore2012}. In the table 
we show their absolute magnitudes, the spectroscopic redshift, the LyC flux and the relative LyC escape fraction. 
The relative LyC escape fraction is obtained from the equations in Sect. \ref{sec:Results}, assuming an intrinsic L$_{\nu}$(1400)/L$_{\nu}$(900)$^{AGN}=2.35$
 \citep[][and Table \ref{tab:intrLyCesc}]{Bongiorno2012} and the spectroscopic redshift to account for the IGM opacity.
 We also report the
hardness ratio for the AGN in the \citet{Xue2011} catalogue. 
Some of these AGN may be obscured and/or affected by broad absorption region, as were those discussed by \citet{Civano2012} in the COSMOS field. 
Moreover, the presence of intervening damped Lyman-alpha or Lyman limit systems close to (or associated with) the AGN can absorb the LyC radiation and decrease their measured fesc(LyC). A dedicated work on the connection between obscuration and LyC emission is ongoing.

\begin{table}[]
\centering
\caption{Description of the properties of the AGN}
\label{tab:propAGN}
\scalebox{0.8}{
\begin{tabular}{|c|c|c|c|c|c|c|} 
\hline
\hline
(1) & (2) &(3)  &(4)  &(5)   & (6) & (7)\\
ID & ID\_ECDFS & zspec & M1400  &f(LyC) & fesc$^{rel}$(LyC) & HR \\
 &  &  &   & $\mu$J& &\\
\hline
NB3727&  &  &  &  &  &\\
\hline
433$^{(1)}$ & 20936 & 3.256 & -22.59& $<0.061$ & $<0.23$ &-0.04\\ 
718$^{(1)}$ & 14587 & 3.193 & -21.28 & $<0.051$ & $<0.52$&-0.26\\ 
819$^{(2)}$ & 48101 & 3.195 &-20.79 & $<0.062$& $<1.00$&\\ 
9$^{(1,2)}$ & 66174 &3.153 & -22.09 & $0.108\pm0.055$ & $0.46\pm0.24$&0.20\\    
\hline
NB387&  &  &  &  &  & \\
\hline
94$^{(2)}$ & 48250 & 3.381 & -21.29 & $<0.019$ & $<0.23$ &\\ 
\hline
NB396&  &  &  &  &  &\\
\hline
78$^{(1)}$ &19824& 3.462 & -21.93 & $0.111\pm0.028$& $0.72\pm0.18$&\\ 
748$^{(1,2)}$ & 3360& 3.471& -21.79&$<0.028$ & $<0.23$& -0.40\\ 
816$^{(1)}$ & 3372 & 3.470&-20.88& $<0.029$& $<0.52$& -0.24\\  
\hline
\end{tabular}
}
\tablefoot{Spectroscopic and photometric properties for eight AGN in clean regions. (1) AGN identification number; (2) CANDELS/E-CDFS catalog identification number; (3) spectroscopic redshift; (4) absolute magnitude;  (5) flux measured in NB within optimized apertures, turned into total flux. The upper limits correspond to 1$\sigma$; (6) relative LyC escape fraction obtained from the equations in Sec. \ref{sec:Results}; (7) hardness ration from \citet{Xue2011}, as the ratio between H-S and H+S, where H and S are the hard and soft Xray fluxes in counts.
It is worth noting that objects 3360, 14587, and 19824 are detected in the VIMOS $U$ band.
$^{(1)}$ Xray detected. ID78 is listed in \citet{Vanzella2010c} as an AGN due to the presence of NV and CIV in its spectrum. It also appears in the list of faint AGN by \citet{Fiore2012}, for which they reported F(0.5-2 keV)=9.5E-17 erg sec$^{-1}$ cm$^{-2}$ and L(2-10 keV)=43.6 erg sec$^{-1}$.  $^{(2)}$ Within the AGN list by \citet{Hsu2014}.}
\end{table}

\section{Results}     \label{sec:Results}

\begin{figure*} 
\centering
\includegraphics[width=200mm]{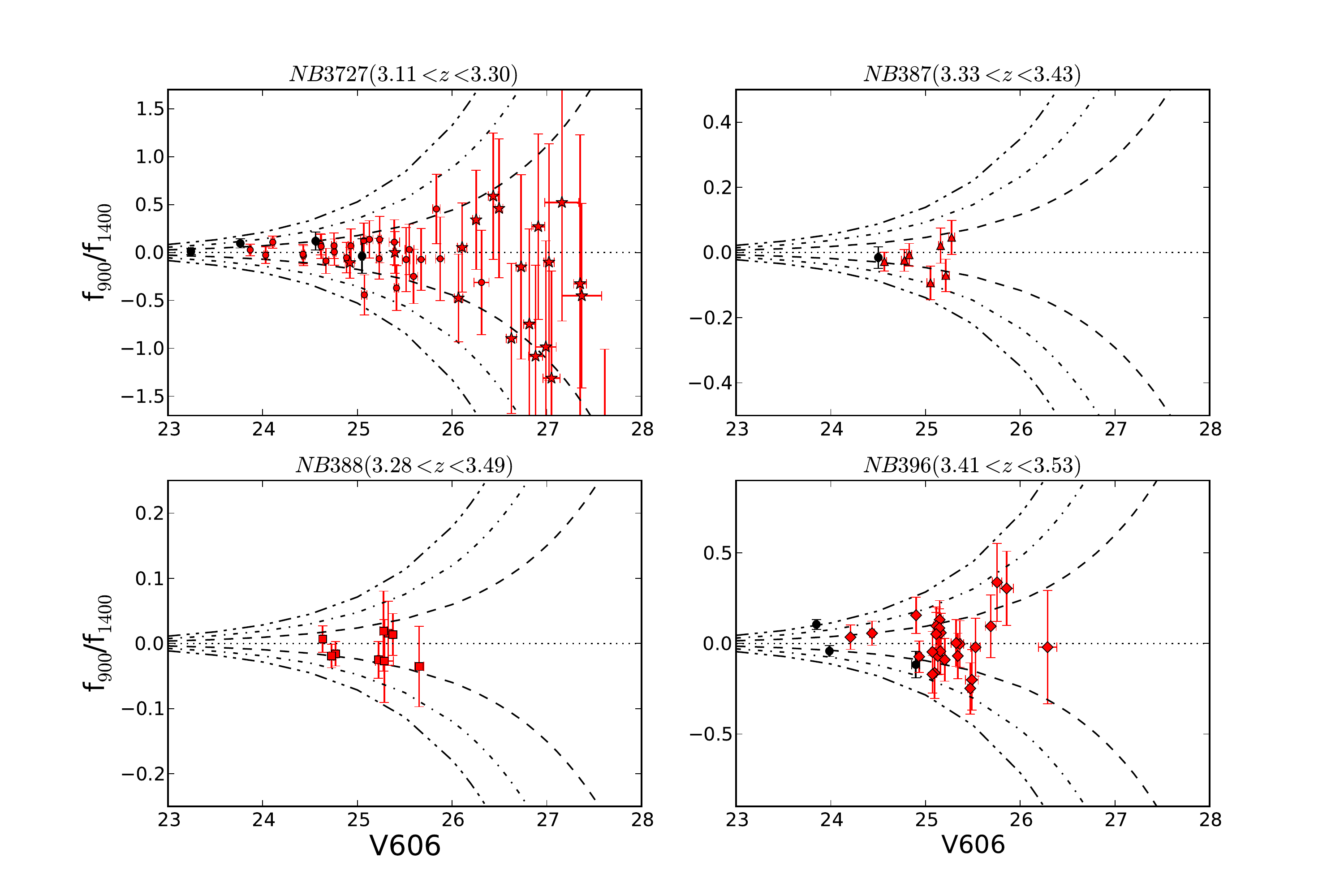}
\caption{Observed flux ratio, f$_{\nu}$(900)/f$_{\nu}$(1400), as a function of V606 band magnitude. From the top left to bottom right: spectroscopically confirmed sources entering the range of redshift covered by NB3727, NB387, NB388, and NB396. Dashed (dot-dashed)(dot-dot-dashed) curves indicate the ratios between 1 (2)(3)$\sigma$ background rms values 
and V-band flux. The symbol coding is the same as in Fig. \ref{M1400mass}. 
Black dots represent the eight AGN in our sample.}
\label{ratio_allNB}
\end{figure*}

\begin{figure}
\centering
\includegraphics[width=90mm]{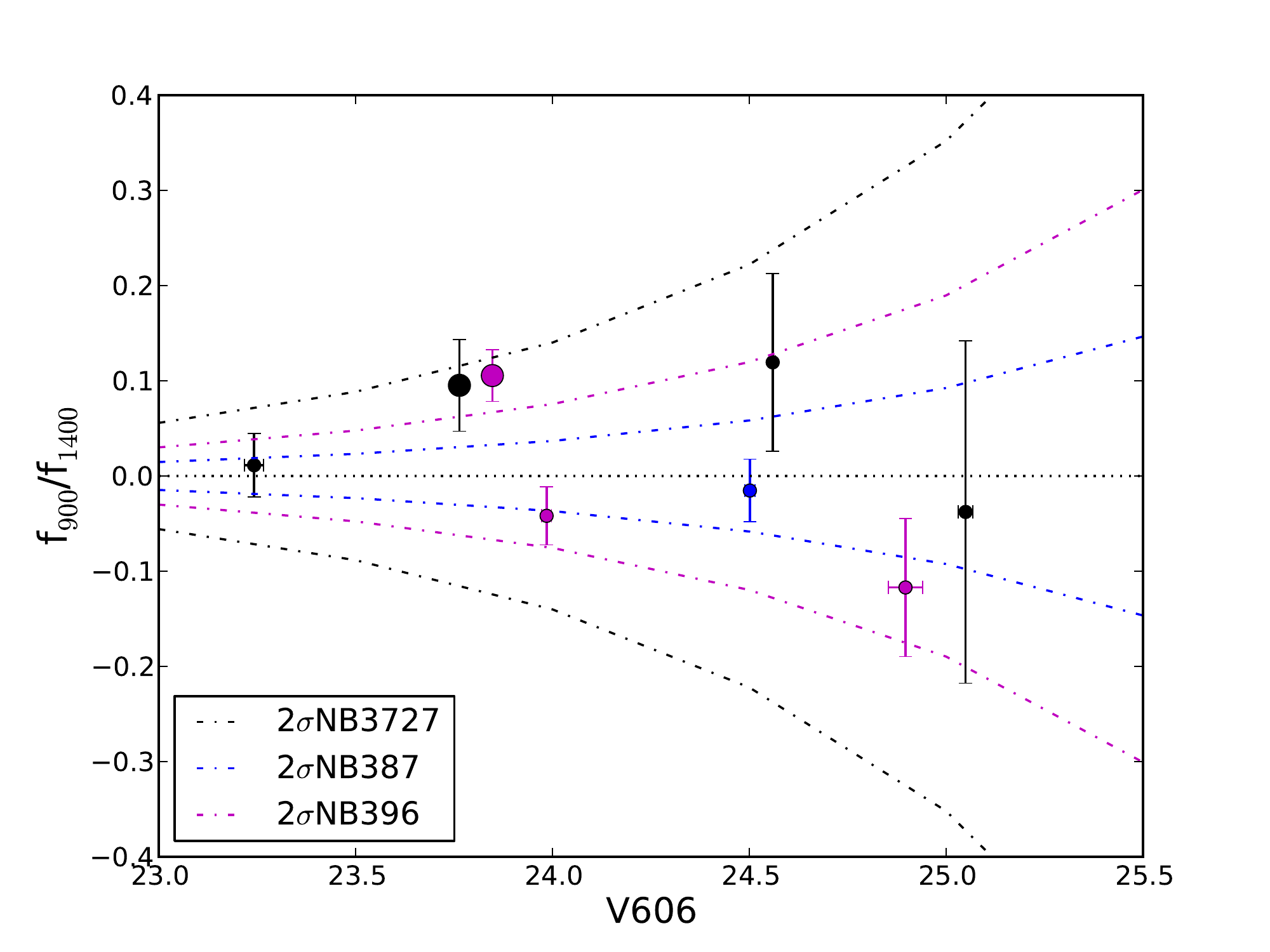}
\caption{Observed flux ratio, f$_{\nu}$(900)/f$_{\nu}$(1400), as a function of $V$ band magnitude for the eight AGN studied here. Black, blue, and magenta filled circles correspond to the AGN for which the LyC flux is measured in NB3727, NB387, and NB388 respectively. The dot-dashed lines indicate the 2$\sigma$ background level corresponding to NB3727 (black), NB387 (blue), and NB388 (magenta). The two largest symbols indicate the AGN for which the flux ratio is not just an upper limit in Table \ref{tab:propAGN}.}
\label{ratio_onlyAGN}
\end{figure}

%
In Tables \ref{tab:propSFG} and \ref{tab:propLAE} we present the observed f$_{\nu}$(900)/f$_{\nu}$(1400) flux ratios for SFGs and LAEs, and the 1$\sigma$ photometric errors on the measurements.
Figure \ref{ratio_allNB} shows the observed f$_{\nu}$(900)/f$_{\nu}$(1400) flux ratio 
as a function of V606 band magnitude for each filter.  In NB3727 we measure LyC signal from continuum-selected star-forming galaxies, narrow-band selected Ly$\alpha$ emitters, 
 and AGN
. In NB387 and NB396 for SFGs and AGN, and in NB388 only for SFGs. The measured LyC fluxes of all the candidates are within the 2$\sigma$ background level
, except for one AGN. 
This source, identified by ID=78, 
is detected in NB396 at more than 3$\sigma$. 
In addition, the AGN identified by ID=9 is detected at $\sim$2$\sigma$ in NB3727 (Fig. \ref{ratio_onlyAGN}). 

Previous works in the literature have shown that LAEs are characterized by a stronger LyC signal than continuum-selected star-forming galaxies. 
For instance, \citet{Mostardi2013} found an observed f$_{\nu}$(900)/f$_{\nu}$(1400) of 0.14 
for LAEs, larger than the value they obtained for LBGs of 0.01 \citep[see also the recent work by][]{Micheva2015}. We do not find any significant detection for any of the LAEs. One of the reasons might be that the NB3727 image is the shallowest, characterized by the largest PSF among the four narrow-band filters. The result is also partially due to the small number of sources and their faint average V606 magnitude (27 against 25 for the SFGs, see calculation at the end of Sect. \ref{sec:exp}). 

We conclude that our whole population of star-forming galaxies does not show any direct LyC signal. Moreover, the 1$\sigma$ photometric errors on the observed f$_{\nu}$(900)/f$_{\nu}$(1400) flux ratios 
depend on the depth of the narrow bands in which the ratios are measured.
Since the LyC measurements come from four different narrow bands with different detection limits, the contribution of the sample as a whole to the LyC emission can be estimated as a weighted mean (LyC(SFGs) = wmean $\pm$ $\sigma$wmean).
Because of the lack of individual detections, wmean is consistent with 0 and $\sigma$wmean provides an upper limit of the LyC signal of the whole sample. By definition, $\sigma$wmean = 1/$\sqrt{\sum_i^N (1/w_i^2)}$, where N is the number of sources considered in the calculation and the weights, w$_i$, are related to the individual photometric uncertainties (w$_i$ = 1/$\sigma_i^2$).

Considering that the V606 band covers the rest-frame UV ($\lambda \sim1400$ {\AA}) for the entire range of redshift probed here, we also calculated the weighted mean of the V606 fluxes in the estimation of the observed f$_{\nu}$(900)/f$_{\nu}$(1400) flux ratio for the whole sample of SFGs. 

Following the same approach, we estimate an upper limit of the observed flux ratio for the sub-samples of the galaxies in each narrow-band redshift range. 
%
%
%
%
%
%
%
%
%
%
By assuming L$_{\nu}$(1400)/L$_{\nu}$(900) $=5$ and exp(-$\tau_{IGM,z})$ (Sec. \ref{sec:exp}), we derive the following 1$\sigma$ upper limits on the fesc$^{rel}$(LyC) for each narrow-band sub-sample (Eq. \ref{eq2}), \\

$fesc^{rel}(LyC_{NB3727}) <44\%$  \\

$fesc^{rel}(LyC_{NB387}) <25\%$  \\

$fesc^{rel}(LyC_{NB388}) <17\%$  \\

$fesc^{rel}(LyC_{NB396}) <33\%.$  \\

These values reflect the different depths of the individual narrow bands. 
For the whole sample of 67 SFGs ($<z>=3.397$, $<$ exp(-$\tau_{IGM,z})>=0.36$, 
L$_{\nu}$(1400)/L$_{\nu}$(900) $=5$), we obtain the following the 1(2)$\sigma$ upper limit:\\ 

$fesc^{rel}(LyC) < 12(24)\%$.\\

As shown in Sect. 3, IGM transmissivity and intrinsic L$_{\nu}$(1400)/L$_{\nu}$(900) ratio correspond to multiplicative factors in the calculation of $fesc^{rel}$(LyC). In particular, L$_{\nu}$(1400)/L$_{\nu}$(900) $=5$ is obtained assuming constant SFR, Salpeter IMF, and age of a few hundred Myr in the galaxy SEDs. We point out that a change of $\sim$60\% in the intrinsic ratio is expected for a different assumption of stellar evolutionary models and a change
of $\sim$20\% for a different assumption of star-formation history. These changes would merely re-scale the $fesc^{rel}$(LyC) value. We also note that the highest ratios are obtained with the BC03 code. If we were to assume lower ratios, we would obtain even more stringent limits on $fesc^{rel}$(LyC).

\begin{figure} 
\centering
\includegraphics[width=90mm]{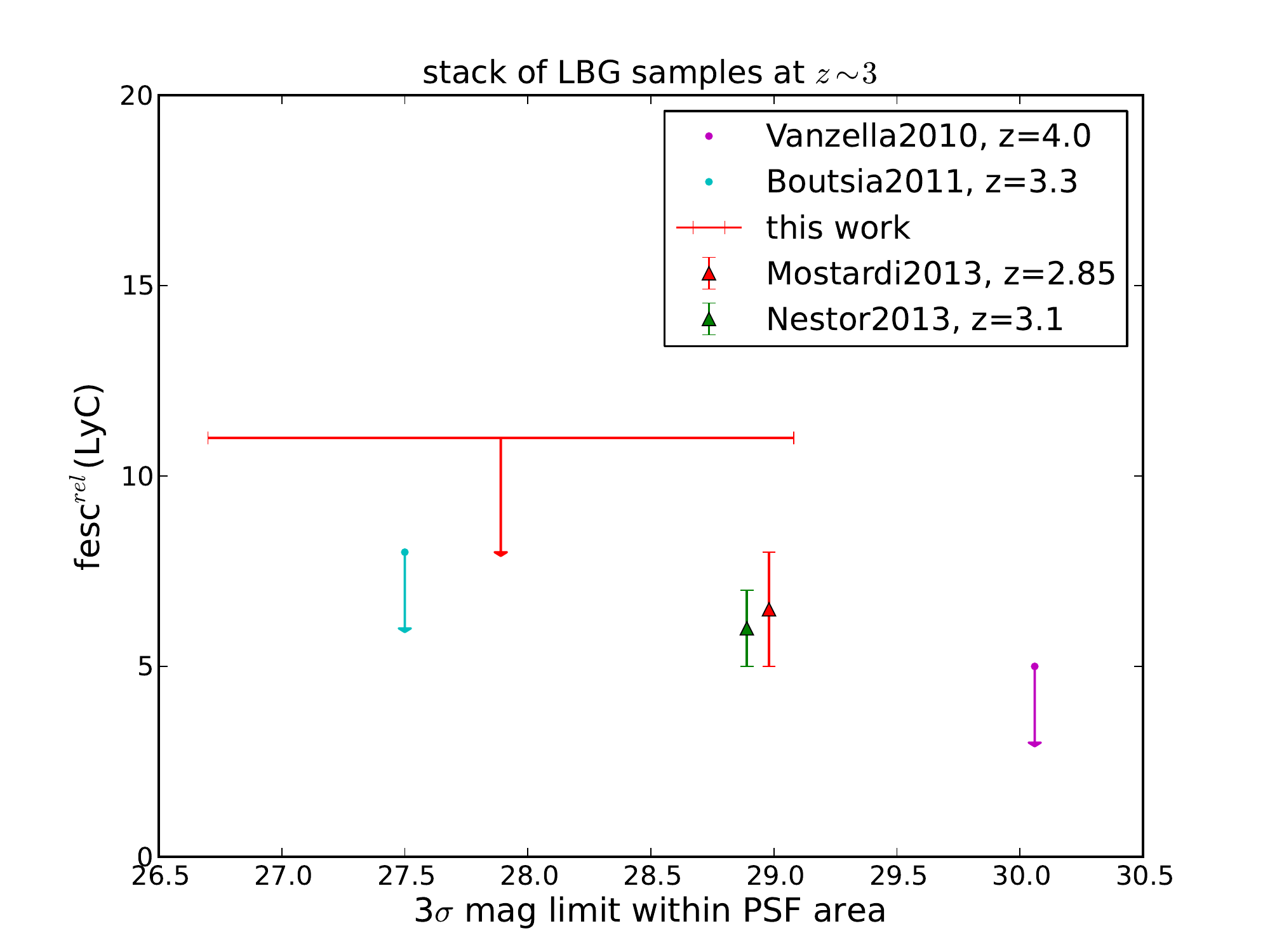}
\caption{Relative LyC escape fraction as a function of survey depth. We translate the fesc$^{rel}$(LyC) estimations in the literature into values corresponding to L$_{\nu}$(1400)/L$_{\nu}$(900) $=5$, where needed. The limits correspond to 1$\sigma$. The survey depths all correspond to the 3$\sigma$ detection limit in an area equal to the survey PSF seeing. Together with our upper limit (red arrow), we include the upper limits from the stack of \citet{Vanzella2010c} at $3.4<z<3.8$ (magenta arrow and dot), of \citet{Boutsia2011} at $z\simeq3.3$ (cyan arrow and dot), and the measurements from the stack of \citet{Nestor2013} at $z\simeq3.1$ (green triangle) and \citet{Mostardi2013} at $z\simeq2.85$ for Lyman break galaxies. 
}
\label{comparison}
\end{figure}

The upper limit we obtained is consistent with previous findings \citep[e.g.][]{Vanzella2010c,Boutsia2011,Nestor2013,Mostardi2013,Grazian2015b} for Lyman break galaxies at similar redshift. 
These limits result from a combination of LyC-image depth and sample size. They could be more stringent if deeper images and larger sample size were available. A quantitative example is presented in the next section. 
Figure \ref{comparison} reports our 1$\sigma$ upper limit together with measurements in the literature. 
The error bar on the magnitude for our work encompasses the depths of our four narrow bands. Despite the inhomogeneity of the available archival data of the samples in the literature, the upper limits and the measurements all agree with values of fesc$^{rel}$(LyC) lower than about 10\% for continuum-selected star-forming galaxies.

We checked our upper limit with an alternative approach, that of stacking the narrow-band images before measuring LyC fluxes. 
We stacked 26 NB3727, 
seven NB387, nine NB388, and 25 NB396 24''x24'' stamps of galaxies. We performed weighted average stacks by running the IRAF task $\sf{IMCOMBINE}$. The weights are obtained from the narrow-band image weight maps. 
Even if the stack is performed to increase the signal-to-noise
ratio, the LyC flux we estimate in the stack images is consistent with their background rms, which means that the final 1$\sigma$ upper limits are consistent with the 12\% reported above.


%

 In Table \ref{tab:propAGN} we report fesc$^{rel}$(LyC) for every individual AGN, where we adopt an intrinsic L$_{\nu}$(1400)/L$_{\nu}$(900) of 2.35 (Table \ref{tab:intrLyCesc}). The 1$\sigma$ NB photometric error is used as LyC flux for the
upper limits. The IGM transmission is calculated for each AGN redshift. For ID=78, we are able to perform a significant direct LyC measurement and obtain fesc$^{rel}$(LyC)=72 $\pm$ 18\%; for ID=9, barely detected at 2$\sigma$, we measure fesc$^{rel}$(LyC)=46$\pm$ 24\%.



\section{Discussion}     \label{sec:Discussion}
\subsection{Dirty sample before the cleaning procedure}
\label{dirty}     

As an example of the danger of reporting LyC measurements before applying a proper cleaning procedure, we stacked 
the initial sample of 145 SFGs (Table \ref{tab:NUM}). 
By assuming the same intrinsic L$_{\nu}$(1400)/L$_{\nu}$(900) value and transmissivity as for the cleaned sample, we would obtain fesc$^{rel}$(LyC) = $(33\pm8)$\%. To calculate this value, we considered the weighted mean of the individual-galaxy LyC measurements, which can be affected by foreground contaminations within the narrow-band PSF. 

This would be the value we would report if HST images were not available and the object-by-object cleaning were not possible. By using only the four sources with f(LyC) $>3 \times$ err\_f(LyC), 
we would obtain f$_{\nu}$(900)/f$_{\nu}$(1400)=0.29 and fesc$^{rel}$(LyC) = $(396\pm32)$ \%.

We can theoretically estimate the maximum value for the observed f$_{\nu}$(900)/f$_{\nu}$(1400) that can be expected, given reasonable assumptions on intrinsic LyC escape fraction and IGM transmissivity. For L$_{\nu}$(1400)/L$_{\nu}(900)=1.5$, exp(-$\tau_{IGM,z}$)=0.6 \citep[the most favourable values that allow LyC photons to reach us,][]{Inoue2008,Inoue2014} and assuming fesc$^{rel}$(LyC) $= 100$\%, we obtain

\begin{equation}
 f_{\nu}(900)/f_{\nu}(1400) = \frac{1 \times exp(-\tau_{IGM,z})}{L_{\nu}(1400)/L_{\nu}(900)}=0.4
\label{eq5}
.\end{equation}

For a very favourable line of sight with a transmissivity equal to 1 \citep[this has an almost null probability to occur, ][]{Inoue2008,Vanzella2015}, f$_{\nu}$(900)/f$_{\nu}$(1400) cannot be larger than 0.7. Based on our average assumptions, f$_{\nu}$(900)/f$_{\nu}$(1400) $\leq 0.08$ for fesc$^{rel}$(LyC) $\leq$ 100\%. The cases in which f$_{\nu}$(900)/f$_{\nu}$(1400) $>0.08(0.4)$ are probably (certainly) unreasonable LyC emitters, in which the observed LyC emission is likely to be produced by close-by contaminating sources. The four sources mentioned above were indeed all rejected after inspecting the images and the colours of neighbouring sources. 

This simple calculation demonstrates the importance of the cleaning procedure before quoting fesc$^{rel}$(LyC).

\subsection{Clean sample and LyC upper limits}     

For our sample of 67 SFGs, we estimate a 1(2)$\sigma$ upper limit in fesc$^{rel}$(LyC) of 12(24)\%.
More 
stringent limits on fesc$^{rel}$(LyC) can be obtained from a sample with hundreds of sources. 
This would also increase the chance of finding strong LyC emitters. The currently planned public spectroscopic survey VANDELS has the scope to provide deep spectra for many hundreds of bright ($H<24$) and faint ($H>27$) star-forming galaxies at $z\sim3$ and it will be possible to use it for this purpose. 
An estimation from \citet{CenKimm2015} suggested that a sample of at least 100 sources needs to be analysed to constrain the true escape fraction that is due to the solid angle effect. 

On the other hand, we could reach an upper limit of 5\% on the fesc$^{rel}$(LyC) with the same spectroscopic sample as in this analysis, but increasing the sensitivity of the narrow-band images. We estimate that we would need to reach a magnitude limit about 1.5 mag deeper than the current values (Table \ref{tab:NB}). This can be achieved with an integration about four times as long  as what is now available, which would be of the order of 40-100 hours.
%

\subsection{Physical properties and LyC escape fraction}     

One of our aims is to understand if and how physical properties of star-forming galaxies are related to the escape of LyC photons.

\begin{figure*}
\centering
\includegraphics[width=200mm]{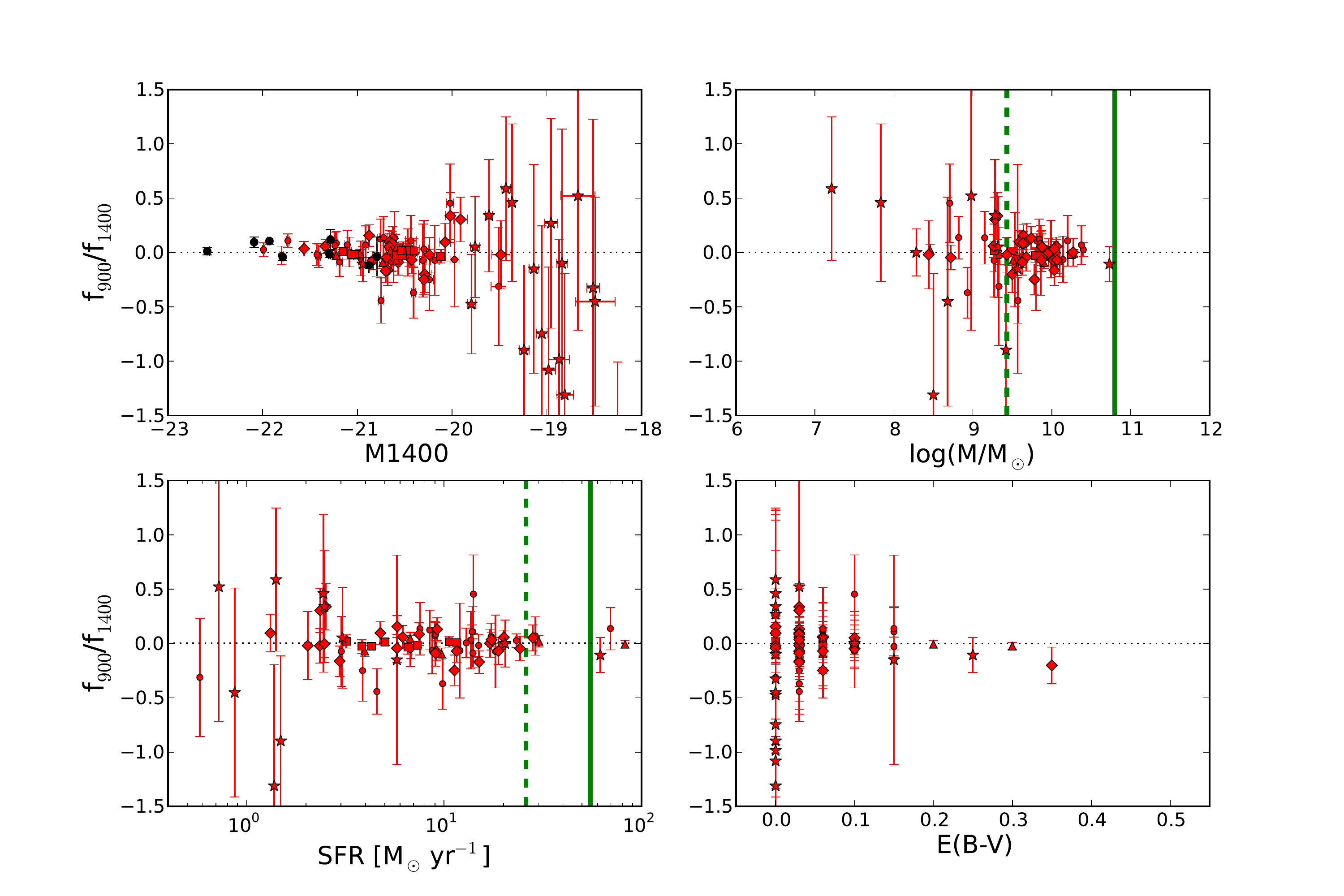}
\caption{Observed flux ratio, f$_{\nu}$(900)/f$_{\nu}$(1400), as a function
of physical parameters. From the top left to the bottom right the physical
parameters are rest-frame absolute magnitude (M1400),
Log(M$^{*}$/M$_{\odot}$), SFR, and dust reddening. The dashed lines
indicate the observed flux ratio equal to zero and are drawn to aid the eye.
The symbol coding is the same as in Fig. \ref{M1400mass}. The vertical dashed and
solid green lines indicate the approximate values of stellar mass and SFR
of two high-$z$ LyC-emitter candidates in the literature, $Ion1$ and $J0921+4509,$ respectively.}
\label{ratioPARAM}
\end{figure*}

In Fig. \ref{ratioPARAM} we show the observed f$_{\nu}$(900)/f$_{\nu}$(1400) flux ratios as a function of rest-frame absolute magnitude, stellar mass, SFR, and dust reddening as derived in Sect. \ref{sec:prop}. Since we do not find any direct detection, we do not expect any dependency between LyC signal and galaxy properties. Therefore, one possibility is that galaxies with properties similar to those of our sample are physically characterized by either low or negligible LyC escape fraction. 

The figure also shows the values of stellar mass and SFR for two LyC
emitters found in the literature. With this limited sample of confirmed LyC emitters we are unable to conclude whether stellar mass or SFR are the main physical parameters related to a high
 LyC escape. But it is noticeable that their star-formation rate is higher than that of most of our
sample and that the stellar mass of one of them is similar to that of many of the galaxies in our sample.  

%
%
%
Our study indicates that we need to enlarge the range of physical
properties explored to find the ones 
suitable to the LyC emission, we will be able to assess the role of physical parameters in the escape of LyC photons.

We can
place constraints on the prediction of models for the inter-stellar
medium properties, although we cannot
finally conclude. In particular, given that we
do not find any
individual object with high escape fraction, we can rule out geometries that
favour a high escape of LyC photons.
For example, \citet{Erik2013} discussed
ISM geometries such as an inter-stellar medium with
free channels, slightly oriented towards us, or an inter-stellar
medium with low HI column density that could allow a fesc(LyC) $>$ 50\%. They also
mentioned that population III stars (popIII) might be responsible for the
production of a significant amount of LyC photons.
The upper limits on fesc(LyC) we found in our work do not support the
expectations
for these geometries, or at least tell us that they are very rare at $z\sim3$.

\subsection{LyC escape fraction and the epoch of re-ionization}     

The study of the LyC escape fraction has implications for the properties and evolution of the galaxies themselves at any redshift, but also for the re-ionization of the Universe at $z>6$. In this context, we can adopt our upper limit to assess the possible role of the SFGs studied here in producing the UV background \citep{Nestor2013, Grazian2015b}. To do this, we estimated the average ionization rate of our sample, $\Gamma_{-12}$ \citep{FaG2008}, 

\begin{equation} \label{eq:gamma}
\Gamma_{-12} = \frac{10^{12} \rho_{900} \sigma_{HI} \Delta l (1+z)^3} {h_{p} (3-\alpha_{LyC})}  s^{-1}
,\end{equation}

where $\rho_{900}$ is the ionizing luminosity density, $\sigma_{HI}$ is the Thomson cross section, $\Delta l$ is the mean free path of ionizing photons, (1+z)$^3$ indicates the conversion from co-moving to proper density per volume, $h_{p}$ is the Planck constant, $\alpha_{LyC}$ is the power-law index of the spectrum in the LyC region (f$_{\nu} \propto \nu^{\alpha_{LyC}}$).

The value of $\rho_{900}$ can be estimated from the rest-frame non-ionizing luminosity density. 
We adopted $\rho_{900}=3.04\times$10$^{24}$ erg sec$^{-1}$ Hz$^{-1}$ Mpc$^{-3}$ \citep[computed from $\rho_{1400}$ as in][by integrating the luminosity function from \citet{ReddySteidel2009} 
 down to $R\leq25.5$, also applicable to our sample] 
{Grazian2015b}. As shown in \citet{Boutsia2011}, $\rho_{900}$ is independent of the assumed value of intrinsic L$_{\nu}$(1400)/L$_{\nu}$(900) ratio. Therefore, $\Gamma_{-12}$ is also independent of it. 

We calculated $\Delta l$ from the fit in Fig. 10 of \citet{Worseck2014}. By assuming  $\Delta l = C [(1+z)/5]^{nn}$, C=37, nn=-5.4, and $z=3.32$, we used $\Delta l =81.5$ proper Mpc. The uncertainty on $\Delta l$ is of the order of 9, given the uncertainties on the C and nn parameters quoted in \citet{Worseck2014}. A greater
(shorter) $\Delta l$ would increase (decrease) the $\Gamma_{-12}$ upper limit.

We derive $\Gamma_{-12}<0.58$ s$^{-1}$ for $\alpha_{LyC}=-0.5$ \citep{FaG2008} and $\Gamma_{-12}<0.34$ s$^{-1}$ for $\alpha_{LyC}=-3$ \citep{Mostardi2013}. These values are 
lower than 
the most recent measurements by \citet{Becker2013} for galaxies at $z\sim3$, 0.7s$^{-1}<\Gamma_{-12}<1.2$ s$^{-1}$, even assuming a longer mean free path, $\Delta l=91$. The authors performed measurements of the IGM properties at $z>2$, taking into account the IGM opacity to ionizing photons.
They found that the hydrogen photo-ionization rate is roughly constant at $2<z<5$ by taking into account cosmological radiative transfer effects (see their Fig. 11). 
To be able to obtain the ionization rate derived by them, we would need a direct measurement of the LyC escape fraction twice as large as the upper limit calculated for our sample, given our assumptions on $\Delta l$ and $\alpha_{LyC}$.

We would probably need to explore sources with physical properties and geometries of the interstellar medium different from those of the galaxies in our sample to ensure greater fesc$^{rel}$(LyC). We might also need to consider a broader redshift range to increase the chance of detecting strong LyC emitters.

\citet{Becker2013} concluded that the averaged ionization efficiency of star-forming galaxies should have globally increased with redshift to produce a fully ionized IGM by $z\sim6$. This creates a caveat in the extrapolation of the results on the LyC escape fraction at $z\sim3$ and the epoch of re-ionization.
 

\section{Summary}     \label{sec:Summary}

With the aim of directly measuring the LyC signal from sources at $z\sim3$, we started from a sample of about 200 spectroscopically
confirmed SFGs, LAEs, and AGN in E-CDFS (Fig. \ref{NBarea}). They were collected from public spectroscopic surveys, including the ESO/GOODS master catalogue, VUDS (for about one third of the initial sample), and VANDELS. The sample has HST coverage and multi-wavelength photometry. 

The key step in our study is the cleaning of the sample.  
The cleaning procedure includes the inspection of CANDELS/GEMS catalogue and the inspection of colour frames created from HST images. The purpose is to eliminate possible low-$z$ interlopers on an individual-source basis. 
After cleaning, our final sample was composed of 94 targets (Table \ref{tab:NUM}), including 67 SFGs, 19 LAEs, and 8 AGN. The sources in the clean regions occupy a certain range of physical properties (Fig. \ref{M1400mass}, Tables \ref{tab:propSFG}, \ref{tab:propLAE}, and \ref{tab:propAGN}), which could help us in identifying regions of the physical parameter space for typical LyC emitters and non-emitters.

%
We performed photometry in four narrow-band filters (Fig. \ref{NBcurve} and Table \ref{tab:NB}) within optimized apertures. These apertures allow the highest signal-to-noise ratio and are small enough to reduce any eventual remaining contamination from nearby sources. None of our star-forming galaxies is detected at more than 3$\sigma$ in our narrow bands. 

By using the V606 band flux from CANDELS and GEMS catalogues as the non-ionizing radiation flux, we calculated the relative LyC escape fraction.
For the sample of continuum-selected star-forming galaxies we could set a 1$\sigma$ upper limit of 12\% 
(Fig. \ref{ratio_allNB}, under the assumptions described in Sect. 3.2), which is consistent with previous results in the literature at the same redshift (Fig. \ref{comparison}).
For the 19 LAEs we were unable to set a meaningful limit. Our results would have been quite different if we had stacked all 145 SFGs in our spectroscopic sample without cleaning (fesc$^{rel}=$33\%, mainly coming from neighbours at different redshifts). We also discovered four sources in the entire sample
that alone produce an average relative escape fraction of more than 300\% \citep[see also][]{Grazian2015b}. In the four cases, the analysis of HST data clearly shows that the LyC signal is spurious and is due to lower redshift interlopers. This demonstrates the importance of applying a cleaning procedure before deriving LyC escape fractions.

Our results indicate that the galaxies in the range of physical parameters probed here have a typical LyC escape fraction that
is too low for individual detections in our narrow-band images (Fig. \ref{ratioPARAM}). 

As shown in \citet{CenKimm2015}, galaxies with relatively low SFRs could also present high fesc(LyC) because they could be in post-starburst phases in which the interstellar medium has been cleared out by the previous star-formation bursts.
However, we noted that two LyC emitter candidates found in the literature present significantly higher star-formation rate than most of our sources. This could indicate that 
the star-formation activity is an important ingredient for allowing the propagation in the ISM and the final escape of LyC photons along a specific line of sight.

The upper limit on the LyC escape fraction can be converted into an IGM hydrogen ionization rate. We derived $\Gamma_{-12}<0.34$ s$^{-1}$ \citep[assuming $\alpha_{LyC}=-3$, as in][]{Mostardi2013}. This value is lower than the measurements by \citet{Becker2013},
obtained for galaxies at $z\sim3$. The discrepancy could imply that sources with LyC emission fainter than our detection limits 
could be mainly responsible for the re-ionization of the intergalactic medium by $z=6$. Alternatively, it could imply that the physical properties of the re-ionization sources may be extremely rare in galaxies at $z\sim3$.

We found one individual significant detection on the relative LyC escape fraction of one AGN (out of eight) at $z=3.462$ (Table \ref{tab:propAGN}). This fesc$^{rel}$(LyC) could indicate that AGN are an interesting population to investigate in 
the study of the sources responsible for the re-ioniziation of the Universe also at $z>6$. In fact, AGN have recently been proposed as possible main drivers of the cosmic re-ionization; there are enough of them even at $z>2$ \citep[e.g.,][]{Giallongo2015}. 
We will investigate the significance and the importance of detecting AGN in LyC in a subsequent paper, where we will also study AGN properties
and obscuration in detail.

Larger samples with secure redshifts from spectroscopic surveys, in fields where deep HST imaging is available, are necessary to make progress. 
In the near future, surveys like VANDELS will add many (bright, but also faint) hundreds of star-forming galaxies and about 100 AGN, enabling a significant improvement on LyC measurements.

%



\begin{acknowledgements}
We acknowledge the BPASS, NeSI, BC03 and, Starburst99 codes,
which we used to perform the calculations in Table 3. 
We thank Konstantina Boutsia, Pratika Dayal, Daniel Schaerer, Tommaso Treu, Erik Zackrisson, and Stefano Zibetti for useful discussions.  
LG and LP acknowledge the support of the INAF-PRIN 105.01.98.01. 
The VUDS collaboration thanks the ESO staff for their continuous support for the survey, particularly the Paranal staff, who conduct the observations, and Marina Rejkuba and the ESO user support group in Garching. VUDS was supported by funding from the European Research Council Advanced Grant ERC--2010--AdG--268107--EARLY and by INAF Grants PRIN 2010, PRIN 2012 and PICS 2013.
The VUDS survey is based on data products made available at the CESAM data center, Laboratoire d'Astrophysique de Marseille, and in part on data products produced at TERAPIX and the Canadian Astronomy Data Centre as part of the Canada-France-Hawaii Telescope Legacy Survey, a collaborative project of NRC and CNRS. The VUDS survey partly uses observations obtained with MegaPrime/MegaCam, a joint project of CFHT and CEA/DAPNIA, at the Canada--France--Hawaii Telescope (CFHT) which is operated by the National Research Council (NRC) of Canada, the Institut National des Sciences de l'Univers of the Centre National de la Recherche Scientifique (CNRS) of France, and the University of Hawaii.
\end{acknowledgements}


\begin{thebibliography}{102}
\expandafter\ifx\csname natexlab\endcsname\relax\def\natexlab#1{#1}\fi

\bibitem[{{Alexandroff} {et~al.}(2012){Alexandroff}, {Overzier}, {Paragi},
  {Basu-Zych}, {Heckman}, {Kauffmann}, {Bourke}, {Lobanov}, {Ptak}, \&
  {Schiminovich}}]{Alexandroff2012}
{Alexandroff}, R., {Overzier}, R.~A., {Paragi}, Z., {et~al.} 2012, \mnras, 423,
  1325

\bibitem[{{Allard} \& {Hauschildt}(1995)}]{Allard1995}
{Allard}, F. \& {Hauschildt}, P.~H. 1995, \apj, 445, 433

\bibitem[{{Avedisova}(1979)}]{Avedisova1979}
{Avedisova}, V.~S. 1979, \sovast, 23, 544

\bibitem[{{Balestra} {et~al.}(2010){Balestra}, {Mainieri}, {Popesso},
  {Dickinson}, {Nonino}, {Rosati}, {Teimoorinia}, {Vanzella}, {Cristiani},
  {Cesarsky}, {Fosbury}, {Kuntschner}, \& {Rettura}}]{Balestra2010}
{Balestra}, I., {Mainieri}, V., {Popesso}, P., {et~al.} 2010, \aap, 512, A12

\bibitem[{{Becker} \& {Bolton}(2013)}]{Becker2013}
{Becker}, G.~D. \& {Bolton}, J.~S. 2013, \mnras, 436, 1023

\bibitem[{{Bergvall} {et~al.}(2013){Bergvall}, {Leitet}, {Zackrisson}, \&
  {Marquart}}]{Bergvall2013}
{Bergvall}, N., {Leitet}, E., {Zackrisson}, E., \& {Marquart}, T. 2013, \aap,
  554, A38

\bibitem[{{Bergvall} {et~al.}(2006){Bergvall}, {Zackrisson}, {Andersson},
  {Arnberg}, {Masegosa}, \& {{\"O}stlin}}]{Bergvall2006}
{Bergvall}, N., {Zackrisson}, E., {Andersson}, B.-G., {et~al.} 2006, \aap, 448,
  513

\bibitem[{{Bertin} \& {Arnouts}(1996)}]{bertin1996}
{Bertin}, E. \& {Arnouts}, S. 1996, \aaps, 117, 393

\bibitem[{{Bertin} {et~al.}(2002){Bertin}, {Mellier}, {Radovich}, {Missonnier},
  {Didelon}, \& {Morin}}]{Bertin2002}
{Bertin}, E., {Mellier}, Y., {Radovich}, M., {et~al.} 2002, in Astronomical
  Society of the Pacific Conference Series, Vol. 281, Astronomical Data
  Analysis Software and Systems XI, ed. D.~A. {Bohlender}, D.~{Durand}, \&
  T.~H. {Handley}, 228

\bibitem[{{Bessell} {et~al.}(1991){Bessell}, {Brett}, {Scholz}, \&
  {Wood}}]{Bessell1991}
{Bessell}, M.~S., {Brett}, J.~M., {Scholz}, M., \& {Wood}, P.~R. 1991, \aaps,
  89, 335

\bibitem[{{Bogosavljevi{\'c}}(2010)}]{Bog2010}
{Bogosavljevi{\'c}}, M. 2010, PhD thesis, California Institute of Technology

\bibitem[{{Bongiorno} {et~al.}(2012){Bongiorno}, {Merloni}, {Brusa},
  {Magnelli}, {Salvato}, {Mignoli}, {Zamorani}, {Fiore}, {Rosario}, {Mainieri},
  {Hao}, {Comastri}, {Vignali}, {Balestra}, {Bardelli}, {Berta}, {Civano},
  {Kampczyk}, {Le Floc'h}, {Lusso}, {Lutz}, {Pozzetti}, {Pozzi}, {Riguccini},
  {Shankar}, \& {Silverman}}]{Bongiorno2012}
{Bongiorno}, A., {Merloni}, A., {Brusa}, M., {et~al.} 2012, \mnras, 427, 3103

\bibitem[{{Borthakur} {et~al.}(2014){Borthakur}, {Heckman}, {Leitherer}, \&
  {Overzier}}]{Borthakur2014}
{Borthakur}, S., {Heckman}, T.~M., {Leitherer}, C., \& {Overzier}, R.~A. 2014,
  Science, 346, 216

\bibitem[{{Boutsia} {et~al.}(2011){Boutsia}, {Grazian}, {Giallongo}, {Fontana},
  {Pentericci}, {Castellano}, {Zamorani}, {Mignoli}, {Vanzella}, {Fiore},
  {Lilly}, {Gallozzi}, {Testa}, {Paris}, \& {Santini}}]{Boutsia2011}
{Boutsia}, K., {Grazian}, A., {Giallongo}, E., {et~al.} 2011, \apj, 736, 41

\bibitem[{{Bridge} {et~al.}(2010){Bridge}, {Teplitz}, {Siana}, {Scarlata},
  {Conselice}, {Ferguson}, {Brown}, {Salvato}, {Rudie}, {de Mello}, {Colbert},
  {Gardner}, {Giavalisco}, \& {Armus}}]{Bridge2010}
{Bridge}, C.~R., {Teplitz}, H.~I., {Siana}, B., {et~al.} 2010, \apj, 720, 465

\bibitem[{{Bruzual} \& {Charlot}(2003)}]{Bruzual:2003}
{Bruzual}, G. \& {Charlot}, S. 2003, \mnras, 344, 1000

\bibitem[{{Cardamone} {et~al.}(2010){Cardamone}, {van Dokkum}, {Urry},
  {Taniguchi}, {Gawiser}, {Brammer}, {Taylor}, {Damen}, {Treister}, {Cobb},
  {Bond}, {Schawinski}, {Lira}, {Murayama}, {Saito}, \&
  {Sumikawa}}]{Cardamone:2010}
{Cardamone}, C.~N., {van Dokkum}, P.~G., {Urry}, C.~M., {et~al.} 2010, \apjs,
  189, 270

\bibitem[{{Cen} \& {Kimm}(2015)}]{CenKimm2015}
{Cen}, R. \& {Kimm}, T. 2015, \apjl, 801, L25

\bibitem[{{Chardin} {et~al.}(2015){Chardin}, {Haehnelt}, {Aubert}, \&
  {Puchwein}}]{Chardin2015}
{Chardin}, J., {Haehnelt}, M.~G., {Aubert}, D., \& {Puchwein}, E. 2015, \mnras,
  453, 2943

\bibitem[{{Civano} {et~al.}(2012){Civano}, {Elvis}, {Brusa}, {Comastri},
  {Salvato}, {Zamorani}, {Aldcroft}, {Bongiorno}, {Capak}, {Cappelluti},
  {Cisternas}, {Fiore}, {Fruscione}, {Hao}, {Kartaltepe}, {Koekemoer}, {Gilli},
  {Impey}, {Lanzuisi}, {Lusso}, {Mainieri}, {Miyaji}, {Lilly}, {Masters},
  {Puccetti}, {Schawinski}, {Scoville}, {Silverman}, {Trump}, {Urry},
  {Vignali}, \& {Wright}}]{Civano2012}
{Civano}, F., {Elvis}, M., {Brusa}, M., {et~al.} 2012, \apjs, 201, 30

\bibitem[{{Conroy} \& {Kratter}(2012)}]{Conroy2012}
{Conroy}, C. \& {Kratter}, K.~M. 2012, \apj, 755, 123

\bibitem[{{Cowie} {et~al.}(2009){Cowie}, {Barger}, \& {Trouille}}]{Cowie2009}
{Cowie}, L.~L., {Barger}, A.~J., \& {Trouille}, L. 2009, \apj, 692, 1476

\bibitem[{{Dayal} {et~al.}(2015){Dayal}, {Choudhury}, {Bromm}, \&
  {Pacucci}}]{Dayal2015b}
{Dayal}, P., {Choudhury}, T.~R., {Bromm}, V., \& {Pacucci}, F. 2015, ArXiv
  1501.02823

\bibitem[{{de Barros} {et~al.}(2016){de Barros}, {Vanzella}, {Amor{\'{\i}}n},
  {Castellano}, {Siana}, {Grazian}, {Suh}, {Balestra}, {Vignali}, {Verhamme},
  {Zamorani}, {Mignoli}, {Hasinger}, {Comastri}, {Pentericci},
  {P{\'e}rez-Montero}, {Fontana}, {Giavalisco}, \& {Gilli}}]{deBarros2015}
{de Barros}, S., {Vanzella}, E., {Amor{\'{\i}}n}, R., {et~al.} 2016, \aap, 585,
  A51

\bibitem[{{Duncan} \& {Conselice}(2015)}]{Duncan2015}
{Duncan}, K. \& {Conselice}, C.~J. 2015, \mnras, 451, 2030

\bibitem[{{Eldridge}(2012)}]{Eldridge2012}
{Eldridge}, J.~J. 2012, \mnras, 422, 794

\bibitem[{{Fan} {et~al.}(2006){Fan}, {Strauss}, {Becker}, {White}, {Gunn},
  {Knapp}, {Richards}, {Schneider}, {Brinkmann}, \& {Fukugita}}]{Fan2006}
{Fan}, X., {Strauss}, M.~A., {Becker}, R.~H., {et~al.} 2006, \aj, 132, 117

\bibitem[{{Faucher-Gigu{\`e}re} {et~al.}(2008){Faucher-Gigu{\`e}re}, {Lidz},
  {Hernquist}, \& {Zaldarriaga}}]{FaG2008}
{Faucher-Gigu{\`e}re}, C.-A., {Lidz}, A., {Hernquist}, L., \& {Zaldarriaga}, M.
  2008, \apj, 688, 85

\bibitem[{{Ferrara} \& {Loeb}(2013)}]{Ferrara2013}
{Ferrara}, A. \& {Loeb}, A. 2013, \mnras, 431, 2826

\bibitem[{{Fiore} {et~al.}(2012){Fiore}, {Puccetti}, {Grazian}, {Menci},
  {Shankar}, {Santini}, {Piconcelli}, {Koekemoer}, {Fontana}, {Boutsia},
  {Castellano}, {Lamastra}, {Malacaria}, {Feruglio}, {Mathur}, {Miller}, \&
  {Pannella}}]{Fiore2012}
{Fiore}, F., {Puccetti}, S., {Grazian}, A., {et~al.} 2012, \aap, 537, A16

\bibitem[{{Fluks} {et~al.}(1994){Fluks}, {Plez}, {The}, {de Winter},
  {Westerlund}, \& {Steenman}}]{Fluks1994}
{Fluks}, M.~A., {Plez}, B., {The}, P.~S., {et~al.} 1994, \aaps, 105, 311

\bibitem[{{Fontana} {et~al.}(2000){Fontana}, {D'Odorico}, {Poli}, {Giallongo},
  {Arnouts}, {Cristiani}, {Moorwood}, \& {Saracco}}]{Fontana2000}
{Fontana}, A., {D'Odorico}, S., {Poli}, F., {et~al.} 2000, \aj, 120, 2206

\bibitem[{{Gawiser} {et~al.}(2006){Gawiser}, {van Dokkum}, {Herrera}, {Maza},
  {Castander}, {Infante}, {Lira}, {Quadri}, {Toner}, {Treister}, {Urry},
  {Altmann}, {Assef}, {Christlein}, {Coppi}, {Dur{\'a}n}, {Franx}, {Galaz},
  {Huerta}, {Liu}, {L{\'o}pez}, {M{\'e}ndez}, {Moore}, {Rubio}, {Ruiz}, {Toft},
  \& {Yi}}]{Gawiser:2006a}
{Gawiser}, E., {van Dokkum}, P.~G., {Herrera}, D., {et~al.} 2006, \apjs, 162, 1

\bibitem[{{Giallongo} {et~al.}(2015){Giallongo}, {Grazian}, {Fiore}, {Fontana},
  {Pentericci}, {Vanzella}, {Dickinson}, {Kocevski}, {Castellano}, {Cristiani},
  {Ferguson}, {Finkelstein}, {Grogin}, {Hathi}, {Koekemoer}, {Newman}, \&
  {Salvato}}]{Giallongo2015}
{Giallongo}, E., {Grazian}, A., {Fiore}, F., {et~al.} 2015, \aap, 578, A83

\bibitem[{{Gnedin} {et~al.}(2008){Gnedin}, {Kravtsov}, \& {Chen}}]{Gnedin2008}
{Gnedin}, N.~Y., {Kravtsov}, A.~V., \& {Chen}, H.-W. 2008, \apj, 672, 765

\bibitem[{{Grazian} {et~al.}(2006){Grazian}, {Fontana}, {de Santis}, {Nonino},
  {Salimbeni}, {Giallongo}, {Cristiani}, {Gallozzi}, \&
  {Vanzella}}]{Grazian2006}
{Grazian}, A., {Fontana}, A., {de Santis}, C., {et~al.} 2006, \aap, 449, 951

\bibitem[{{Grazian} {et~al.}(2015){Grazian}, {Fontana}, {Santini}, {Dunlop},
  {Ferguson}, {Castellano}, {Amorin}, {Ashby}, {Barro}, {Behroozi}, {Boutsia},
  {Caputi}, {Chary}, {Dekel}, {Dickinson}, {Faber}, {Fazio}, {Finkelstein},
  {Galametz}, {Giallongo}, {Giavalisco}, {Grogin}, {Guo}, {Kocevski},
  {Koekemoer}, {Koo}, {Lee}, {Lu}, {Merlin}, {Mobasher}, {Nonino}, {Papovich},
  {Paris}, {Pentericci}, {Reddy}, {Renzini}, {Salmon}, {Salvato}, {Sommariva},
  {Song}, \& {Vanzella}}]{Grazian2015a}
{Grazian}, A., {Fontana}, A., {Santini}, P., {et~al.} 2015, \aap, 575, A96

\bibitem[{{Grazian} {et~al.}(2016){Grazian}, {Giallongo}, {Gerbasi}, {Fiore},
  {Fontana}, {Le F{\`e}vre}, {Pentericci}, {Vanzella}, {Zamorani}, {Cassata},
  {Garilli}, {Le Brun}, {Maccagni}, {Tasca}, {Thomas}, {Zucca},
  {Amor{\'{\i}}n}, {Bardelli}, {Cassar{\`a}}, {Castellano}, {Cimatti},
  {Cucciati}, {Durkalec}, {Giavalisco}, {Hathi}, {Ilbert}, {Lemaux}, {Paltani},
  {Ribeiro}, {Schaerer}, {Scodeggio}, {Sommariva}, {Talia}, {Tresse},
  {Vergani}, {Bonchi}, {Boutsia}, {Capak}, {Charlot}, {Contini}, {de la Torre},
  {Dunlop}, {Fotopoulou}, {Guaita}, {Koekemoer}, {L{\'o}pez-Sanjuan},
  {Mellier}, {Merlin}, {Paris}, {Pforr}, {Pilo}, {Santini}, {Scoville},
  {Taniguchi}, \& {Wang}}]{Grazian2015b}
{Grazian}, A., {Giallongo}, E., {Gerbasi}, R., {et~al.} 2016, \aap, 585, A48

\bibitem[{{Guaita} {et~al.}(2010){Guaita}, {Gawiser}, {Padilla}, {Francke},
  {Bond}, {Gronwall}, {Ciardullo}, {Feldmeier}, {Sinawa}, {Blanc}, \&
  {Virani}}]{Guaita2010}
{Guaita}, L., {Gawiser}, E., {Padilla}, N., {et~al.} 2010, \apj, 714, 255

\bibitem[{{Gunn} \& {Stryker}(1983)}]{Gunn1983}
{Gunn}, J.~E. \& {Stryker}, L.~L. 1983, \apjs, 52, 121

\bibitem[{{Guo} {et~al.}(2013){Guo}, {Ferguson}, {Giavalisco}, {Barro},
  {Willner}, {Ashby}, {Dahlen}, {Donley}, {Faber}, {Fontana}, {Galametz},
  {Grazian}, {Huang}, {Kocevski}, {Koekemoer}, {Koo}, {McGrath}, {Peth},
  {Salvato}, {Wuyts}, {Castellano}, {Cooray}, {Dickinson}, {Dunlop}, {Fazio},
  {Gardner}, {Gawiser}, {Grogin}, {Hathi}, {Hsu}, {Lee}, {Lucas}, {Mobasher},
  {Nandra}, {Newman}, \& {van der Wel}}]{Guo2013}
{Guo}, Y., {Ferguson}, H.~C., {Giavalisco}, M., {et~al.} 2013, \apjs, 207, 24

\bibitem[{{Haardt} \& {Madau}(2012)}]{HaardtMadau2012}
{Haardt}, F. \& {Madau}, P. 2012, \apj, 746, 125

\bibitem[{{Hamann} \& {Gr{\"a}fener}(2003)}]{Hamann2003}
{Hamann}, W.-R. \& {Gr{\"a}fener}, G. 2003, \aap, 410, 993

\bibitem[{{Hayes} {et~al.}(2010){Hayes}, {{\"O}stlin}, {Schaerer}, {Mas-Hesse},
  {Leitherer}, {Atek}, {Kunth}, {Verhamme}, {de Barros}, \&
  {Melinder}}]{Hayes2010}
{Hayes}, M., {{\"O}stlin}, G., {Schaerer}, D., {et~al.} 2010, \nat, 464, 562

\bibitem[{{Hillier} \& {Miller}(1998)}]{Hillier1998}
{Hillier}, D.~J. \& {Miller}, D.~L. 1998, \apj, 496, 407

\bibitem[{{Hsu} {et~al.}(2014){Hsu}, {Salvato}, {Nandra}, {Brusa}, {Bender},
  {Buchner}, {Donley}, {Kocevski}, {Guo}, {Hathi}, {Rangel}, {Willner},
  {Brightman}, {Georgakakis}, {Budav{\'a}ri}, {Szalay}, {Ashby}, {Barro},
  {Dahlen}, {Faber}, {Ferguson}, {Galametz}, {Grazian}, {Grogin}, {Huang},
  {Koekemoer}, {Lucas}, {McGrath}, {Mobasher}, {Peth}, {Rosario}, \&
  {Trump}}]{Hsu2014}
{Hsu}, L.-T., {Salvato}, M., {Nandra}, K., {et~al.} 2014, \apj, 796, 60

\bibitem[{{Inoue} \& {Iwata}(2008)}]{Inoue2008}
{Inoue}, A.~K. \& {Iwata}, I. 2008, \mnras, 387, 1681

\bibitem[{{Inoue} {et~al.}(2014){Inoue}, {Shimizu}, {Iwata}, \&
  {Tanaka}}]{Inoue2014}
{Inoue}, A.~K., {Shimizu}, I., {Iwata}, I., \& {Tanaka}, M. 2014, \mnras, 442,
  1805

\bibitem[{{Iwata} {et~al.}(2009){Iwata}, {Inoue}, {Matsuda}, {Furusawa},
  {Hayashino}, {Kousai}, {Akiyama}, {Yamada}, {Burgarella}, \&
  {Deharveng}}]{Iwata2009}
{Iwata}, I., {Inoue}, A.~K., {Matsuda}, Y., {et~al.} 2009, \apj, 692, 1287

\bibitem[{{Jia} {et~al.}(2011){Jia}, {Ptak}, {Heckman}, {Overzier},
  {Hornschemeier}, \& {LaMassa}}]{Jia2011}
{Jia}, J., {Ptak}, A., {Heckman}, T.~M., {et~al.} 2011, \apj, 731, 55

\bibitem[{{Kurucz}(1996)}]{Kurucz1996}
{Kurucz}, R.~L. 1996, in Astronomical Society of the Pacific Conference Series,
  Vol. 108, M.A.S.S., Model Atmospheres and Spectrum Synthesis, ed. S.~J.
  {Adelman}, F.~{Kupka}, \& W.~W. {Weiss}, 2

\bibitem[{{Le F{\`e}vre} {et~al.}(2015){Le F{\`e}vre}, {Tasca}, {Cassata},
  {Garilli}, {Le Brun}, {Maccagni}, {Pentericci}, {Thomas}, {Vanzella},
  {Zamorani}, {Zucca}, {Amorin}, {Bardelli}, {Capak}, {Cassar{\`a}},
  {Castellano}, {Cimatti}, {Cuby}, {Cucciati}, {de la Torre}, {Durkalec},
  {Fontana}, {Giavalisco}, {Grazian}, {Hathi}, {Ilbert}, {Lemaux}, {Moreau},
  {Paltani}, {Ribeiro}, {Salvato}, {Schaerer}, {Scodeggio}, {Sommariva},
  {Talia}, {Taniguchi}, {Tresse}, {Vergani}, {Wang}, {Charlot}, {Contini},
  {Fotopoulou}, {L{\'o}pez-Sanjuan}, {Mellier}, \& {Scoville}}]{LeFevre2015}
{Le F{\`e}vre}, O., {Tasca}, L.~A.~M., {Cassata}, P., {et~al.} 2015, \aap, 576,
  A79

\bibitem[{{Leitet} {et~al.}(2013){Leitet}, {Bergvall}, {Hayes}, {Linn{\'e}}, \&
  {Zackrisson}}]{Leitet2013}
{Leitet}, E., {Bergvall}, N., {Hayes}, M., {Linn{\'e}}, S., \& {Zackrisson}, E.
  2013, \aap, 553, A106

\bibitem[{{Leitet} {et~al.}(2011){Leitet}, {Bergvall}, {Piskunov}, \&
  {Andersson}}]{Leitet2011}
{Leitet}, E., {Bergvall}, N., {Piskunov}, N., \& {Andersson}, B.-G. 2011, \aap,
  532, A107

\bibitem[{{Leitherer} {et~al.}(2014){Leitherer}, {Ekstr{\"o}m}, {Meynet},
  {Schaerer}, {Agienko}, \& {Levesque}}]{Leitherer2014}
{Leitherer}, C., {Ekstr{\"o}m}, S., {Meynet}, G., {et~al.} 2014, \apjs, 212, 14

\bibitem[{{Leitherer} {et~al.}(1999){Leitherer}, {Schaerer}, {Goldader},
  {Delgado}, {Robert}, {Kune}, {de Mello}, {Devost}, \&
  {Heckman}}]{Leitherer1999}
{Leitherer}, C., {Schaerer}, D., {Goldader}, J.~D., {et~al.} 1999, \apjs, 123,
  3

\bibitem[{{Levesque} {et~al.}(2012){Levesque}, {Leitherer}, {Ekstrom},
  {Meynet}, \& {Schaerer}}]{Levesque2012}
{Levesque}, E.~M., {Leitherer}, C., {Ekstrom}, S., {Meynet}, G., \& {Schaerer},
  D. 2012, \apj, 751, 67

\bibitem[{{Lusso} {et~al.}(2015){Lusso}, {Worseck}, {Hennawi}, {Prochaska},
  {Vignali}, {Stern}, \& {O'Meara}}]{Lusso2015}
{Lusso}, E., {Worseck}, G., {Hennawi}, J.~F., {et~al.} 2015, \mnras, 449, 4204

\bibitem[{{Madau}(1995)}]{Madau:1995}
{Madau}, P. 1995, \apj, 441, 18

\bibitem[{{Martin} {et~al.}(2012){Martin}, {Shapley}, {Coil}, {Kornei},
  {Bundy}, {Weiner}, {Noeske}, \& {Schiminovich}}]{Martin2012}
{Martin}, C.~L., {Shapley}, A.~E., {Coil}, A.~L., {et~al.} 2012, \apj, 760, 127

\bibitem[{{Merlin} {et~al.}(2015){Merlin}, {Fontana}, {Ferguson}, {Dunlop},
  {Elbaz}, {Bourne}, {Bruce}, {Buitrago}, {Castellano}, {Schreiber}, {Grazian},
  {McLure}, {Okumura}, {Shu}, {Wang}, {Amor{\'{\i}}n}, {Boutsia}, {Cappelluti},
  {Comastri}, {Derriere}, {Faber}, \& {Santini}}]{Merlin2015}
{Merlin}, E., {Fontana}, A., {Ferguson}, H.~C., {et~al.} 2015, \aap, 582, A15

\bibitem[{{Micheva} {et~al.}(2015){Micheva}, {Iwata}, {Inoue}, {Matsuda},
  {Yamada}, \& {Hayashino}}]{Micheva2015}
{Micheva}, G., {Iwata}, I., {Inoue}, A.~K., {et~al.} 2015, ArXiv 1509.03996

\bibitem[{{Mitra} {et~al.}(2013){Mitra}, {Ferrara}, \& {Choudhury}}]{Mitra2013}
{Mitra}, S., {Ferrara}, A., \& {Choudhury}, T.~R. 2013, \mnras, 428, L1

\bibitem[{{Mostardi} {et~al.}(2013){Mostardi}, {Shapley}, {Nestor}, {Steidel},
  {Reddy}, \& {Trainor}}]{Mostardi2013}
{Mostardi}, R.~E., {Shapley}, A.~E., {Nestor}, D.~B., {et~al.} 2013, \apj, 779,
  65

\bibitem[{{Mostardi} {et~al.}(2015){Mostardi}, {Shapley}, {Steidel}, {Trainor},
  {Reddy}, \& {Siana}}]{Mostardi2015}
{Mostardi}, R.~E., {Shapley}, A.~E., {Steidel}, C.~C., {et~al.} 2015, \apj,
  810, 107

\bibitem[{{Nakajima} {et~al.}(2012){Nakajima}, {Ouchi}, {Shimasaku}, {Ono},
  {Lee}, {Foucaud}, {Ly}, {Dale}, {Salim}, {Finn}, {Almaini}, \&
  {Okamura}}]{Nakajima2012a}
{Nakajima}, K., {Ouchi}, M., {Shimasaku}, K., {et~al.} 2012, \apj, 745, 12

\bibitem[{{Nestor} {et~al.}(2013){Nestor}, {Shapley}, {Kornei}, {Steidel}, \&
  {Siana}}]{Nestor2013}
{Nestor}, D.~B., {Shapley}, A.~E., {Kornei}, K.~A., {Steidel}, C.~C., \&
  {Siana}, B. 2013, \apj, 765, 47

\bibitem[{{Nilsson} {et~al.}(2009){Nilsson}, {Tapken}, {M{\o}ller},
  {Freudling}, {Fynbo}, {Meisenheimer}, {Laursen}, \&
  {{\"O}stlin}}]{Nilsson:2009}
{Nilsson}, K.~K., {Tapken}, C., {M{\o}ller}, P., {et~al.} 2009, \aap, 498, 13

\bibitem[{{Nonino} {et~al.}(2009){Nonino}, {Dickinson}, {Rosati}, {Grazian},
  {Reddy}, {Cristiani}, {Giavalisco}, {Kuntschner}, {Vanzella}, {Daddi},
  {Fosbury}, \& {Cesarsky}}]{Nonino2009}
{Nonino}, M., {Dickinson}, M., {Rosati}, P., {et~al.} 2009, \apjs, 183, 244

\bibitem[{{Paardekooper} {et~al.}(2015){Paardekooper}, {Khochfar}, \& {Dalla
  Vecchia}}]{Paardekooper2015}
{Paardekooper}, J.-P., {Khochfar}, S., \& {Dalla Vecchia}, C. 2015, \mnras,
  451, 2544

\bibitem[{{Pauldrach} {et~al.}(1998){Pauldrach}, {Lennon}, {Hoffmann},
  {Sellmaier}, {Kudritzki}, \& {Puls}}]{Pauldrach1998}
{Pauldrach}, A.~W.~A., {Lennon}, M., {Hoffmann}, T.~L., {et~al.} 1998, in
  Astronomical Society of the Pacific Conference Series, Vol. 131, Properties
  of Hot Luminous Stars, ed. I.~{Howarth}, 258

\bibitem[{{Pickles}(1998)}]{Pickles1998}
{Pickles}, A.~J. 1998, \pasp, 110, 863

\bibitem[{{Popesso} {et~al.}(2009){Popesso}, {Dickinson}, {Nonino}, {Vanzella},
  {Daddi}, {Fosbury}, {Kuntschner}, {Mainieri}, {Cristiani}, {Cesarsky},
  {Giavalisco}, {Renzini}, \& {GOODS Team}}]{Popesso2009}
{Popesso}, P., {Dickinson}, M., {Nonino}, M., {et~al.} 2009, \aap, 494, 443

\bibitem[{{Reddy} \& {Steidel}(2009)}]{ReddySteidel2009}
{Reddy}, N.~A. \& {Steidel}, C.~C. 2009, \apj, 692, 778

\bibitem[{{Richards} {et~al.}(2006){Richards}, {Lacy}, {Storrie-Lombardi},
  {Hall}, {Gallagher}, {Hines}, {Fan}, {Papovich}, {Vanden Berk}, {Trammell},
  {Schneider}, {Vestergaard}, {York}, {Jester}, {Anderson}, {Budav{\'a}ri}, \&
  {Szalay}}]{R06}
{Richards}, G.~T., {Lacy}, M., {Storrie-Lombardi}, L.~J., {et~al.} 2006, \apjs,
  166, 470

\bibitem[{{Rix} {et~al.}(2004){Rix}, {Barden}, {Beckwith}, {Bell}, {Borch},
  {Caldwell}, {H{\"a}ussler}, {Jahnke}, {Jogee}, {McIntosh}, {Meisenheimer},
  {Peng}, {Sanchez}, {Somerville}, {Wisotzki}, \& {Wolf}}]{Rix2004}
{Rix}, H.-W., {Barden}, M., {Beckwith}, S.~V.~W., {et~al.} 2004, \apjs, 152,
  163

\bibitem[{{Robertson} {et~al.}(2015){Robertson}, {Ellis}, {Furlanetto}, \&
  {Dunlop}}]{Robertson2015}
{Robertson}, B.~E., {Ellis}, R.~S., {Furlanetto}, S.~R., \& {Dunlop}, J.~S.
  2015, \apjl, 802, L19

\bibitem[{{Roy} {et~al.}(2015){Roy}, {Nath}, \& {Sharma}}]{Roy2014}
{Roy}, A., {Nath}, B.~B., \& {Sharma}, P. 2015, \mnras, 451, 1939

\bibitem[{{Santini} {et~al.}(2015){Santini}, {Ferguson}, {Fontana}, {Mobasher},
  {Barro}, {Castellano}, {Finkelstein}, {Grazian}, {Hsu}, {Lee}, {Lee},
  {Pforr}, {Salvato}, {Wiklind}, {Wuyts}, {Almaini}, {Cooper}, {Galametz},
  {Weiner}, {Amorin}, {Boutsia}, {Conselice}, {Dahlen}, {Dickinson},
  {Giavalisco}, {Grogin}, {Guo}, {Hathi}, {Kocevski}, {Koekemoer},
  {Kurczynski}, {Merlin}, {Mortlock}, {Newman}, {Paris}, {Pentericci},
  {Simons}, \& {Willner}}]{Santini2015}
{Santini}, P., {Ferguson}, H.~C., {Fontana}, A., {et~al.} 2015, \apj, 801, 97

\bibitem[{{Santini} {et~al.}(2009){Santini}, {Fontana}, {Grazian}, {Salimbeni},
  {Fiore}, {Fontanot}, {Boutsia}, {Castellano}, {Cristiani}, {de Santis},
  {Gallozzi}, {Giallongo}, {Menci}, {Nonino}, {Paris}, {Pentericci}, \&
  {Vanzella}}]{Santini2009}
{Santini}, P., {Fontana}, A., {Grazian}, A., {et~al.} 2009, \aap, 504, 751

\bibitem[{{Schlafly} \& {Finkbeiner}(2011)}]{SF2011}
{Schlafly}, E.~F. \& {Finkbeiner}, D.~P. 2011, \apj, 737, 103

\bibitem[{{Siana} {et~al.}(2015){Siana}, {Shapley}, {Kulas}, {Nestor},
  {Steidel}, {Teplitz}, {Alavi}, {Brown}, {Conselice}, {Ferguson}, {Dickinson},
  {Giavalisco}, {Colbert}, {Bridge}, {Gardner}, \& {de Mello}}]{Siana2015}
{Siana}, B., {Shapley}, A.~E., {Kulas}, K.~R., {et~al.} 2015, \apj, 804, 17

\bibitem[{{Siana} {et~al.}(2010){Siana}, {Teplitz}, {Ferguson}, {Brown},
  {Giavalisco}, {Dickinson}, {Chary}, {de Mello}, {Conselice}, {Bridge},
  {Gardner}, {Colbert}, \& {Scarlata}}]{Siana2010}
{Siana}, B., {Teplitz}, H.~I., {Ferguson}, H.~C., {et~al.} 2010, \apj, 723, 241

\bibitem[{{Stanway} {et~al.}(2016){Stanway}, {Eldridge}, \&
  {Becker}}]{Stanway2015}
{Stanway}, E.~R., {Eldridge}, J.~J., \& {Becker}, G.~D. 2016, \mnras, 456, 485

\bibitem[{{Stevans} {et~al.}(2014){Stevans}, {Shull}, {Danforth}, \&
  {Tilton}}]{Stevans2014}
{Stevans}, M.~L., {Shull}, J.~M., {Danforth}, C.~W., \& {Tilton}, E.~M. 2014,
  \apj, 794, 75

\bibitem[{{Thomas} {et~al.}(2014){Thomas}, {Le F{\`e}vre}, {Cassata},
  {Garilli}, {Lemaux}, {Maccagni}, {Pentericci}, {Tasca}, {Zamorani}, {Zucca},
  {Amorin}, {Bardelli}, {Cassar{\`a}}, {Castellano}, {Cimatti}, {Cucciati},
  {Durkalec}, {Fontana}, {Giavalisco}, {Grazian}, {Hathi}, {Ilbert}, {Paltani},
  {Ribeiro}, {Schaerer}, {Scodeggio}, {Sommariva}, {Talia}, {Tresse},
  {Vanzella}, {Vergani}, {Capak}, {Charlot}, {Contini}, {Cuby}, {de la Torre},
  {Dunlop}, {Fotopoulou}, {Koekemoer}, {L{\'o}pez-Sanjuan}, {Mellier}, {Pforr},
  {Salvato}, {Scoville}, {Taniguchi}, \& {Wang}}]{Thomas2014}
{Thomas}, R., {Le F{\`e}vre}, O., {Cassata}, V.~L.~B.~P., {et~al.} 2014, ArXiv
  1411.5692

\bibitem[{{Tilton} {et~al.}(2015){Tilton}, {Stevans}, {Shull}, \&
  {Danforth}}]{Tilton2015}
{Tilton}, E.~M., {Stevans}, M.~L., {Shull}, J.~M., \& {Danforth}, C.~W. 2015,
  ArXiv 1512.02635

\bibitem[{{Vanzella} {et~al.}(2008){Vanzella}, {Cristiani}, {Dickinson},
  {Giavalisco}, {Kuntschner}, {Haase}, {Nonino}, {Rosati}, {Cesarsky},
  {Ferguson}, {Fosbury}, {Grazian}, {Moustakas}, {Rettura}, {Popesso},
  {Renzini}, {Stern}, \& {GOODS Team}}]{Vanzella2008}
{Vanzella}, E., {Cristiani}, S., {Dickinson}, M., {et~al.} 2008, \aap, 478, 83

\bibitem[{{Vanzella} {et~al.}(2015){Vanzella}, {de Barros}, {Castellano},
  {Grazian}, {Inoue}, {Schaerer}, {Guaita}, {Zamorani}, {Giavalisco}, {Siana},
  {Pentericci}, {Giallongo}, {Fontana}, \& {Vignali}}]{Vanzella2015}
{Vanzella}, E., {de Barros}, S., {Castellano}, M., {et~al.} 2015, \aap, 576,
  A116

\bibitem[{{Vanzella} {et~al.}(2010{\natexlab{a}}){Vanzella}, {Giavalisco},
  {Inoue}, {Nonino}, {Fontanot}, {Cristiani}, {Grazian}, {Dickinson}, {Stern},
  {Tozzi}, {Giallongo}, {Ferguson}, {Spinrad}, {Boutsia}, {Fontana}, {Rosati},
  \& {Pentericci}}]{Vanzella2010c}
{Vanzella}, E., {Giavalisco}, M., {Inoue}, A.~K., {et~al.} 2010{\natexlab{a}},
  \apj, 725, 1011

\bibitem[{{Vanzella} {et~al.}(2012{\natexlab{a}}){Vanzella}, {Guo},
  {Giavalisco}, {Grazian}, {Castellano}, {Cristiani}, {Dickinson}, {Fontana},
  {Nonino}, {Giallongo}, {Pentericci}, {Galametz}, {Faber}, {Ferguson},
  {Grogin}, {Koekemoer}, {Newman}, \& {Siana}}]{Vanzella2012}
{Vanzella}, E., {Guo}, Y., {Giavalisco}, M., {et~al.} 2012{\natexlab{a}}, \apj,
  751, 70

\bibitem[{{Vanzella} {et~al.}(2012{\natexlab{b}}){Vanzella}, {Guo},
  {Giavalisco}, {Grazian}, {Castellano}, {Cristiani}, {Dickinson}, {Fontana},
  {Nonino}, {Giallongo}, {Pentericci}, {Galametz}, {Faber}, {Ferguson},
  {Grogin}, {Koekemoer}, {Newman}, \& {Siana}}]{Vanzella2012a}
{Vanzella}, E., {Guo}, Y., {Giavalisco}, M., {et~al.} 2012{\natexlab{b}}, \apj,
  751, 70

\bibitem[{{Vanzella} {et~al.}(2012{\natexlab{c}}){Vanzella}, {Nonino},
  {Cristiani}, {Rosati}, {Zitrin}, {Bartelmann}, {Grazian}, {Broadhurst},
  {Meneghetti}, \& {Grillo}}]{Vanzella2012b}
{Vanzella}, E., {Nonino}, M., {Cristiani}, S., {et~al.} 2012{\natexlab{c}},
  \mnras, 424, L54

\bibitem[{{Vanzella} {et~al.}(2010{\natexlab{b}}){Vanzella}, {Siana},
  {Cristiani}, \& {Nonino}}]{Vanzella2010a}
{Vanzella}, E., {Siana}, B., {Cristiani}, S., \& {Nonino}, M.
  2010{\natexlab{b}}, \mnras, 404, 1672

\bibitem[{{Westera} {et~al.}(2002){Westera}, {Lejeune}, {Buser}, {Cuisinier},
  \& {Bruzual}}]{Westera2002}
{Westera}, P., {Lejeune}, T., {Buser}, R., {Cuisinier}, F., \& {Bruzual}, G.
  2002, \aap, 381, 524

\bibitem[{{Wise} \& {Cen}(2009)}]{WiseCen2009}
{Wise}, J.~H. \& {Cen}, R. 2009, \apj, 693, 984

\bibitem[{{Wise} {et~al.}(2014){Wise}, {Demchenko}, {Halicek}, {Norman},
  {Turk}, {Abel}, \& {Smith}}]{Wise2014}
{Wise}, J.~H., {Demchenko}, V.~G., {Halicek}, M.~T., {et~al.} 2014, \mnras,
  442, 2560

\bibitem[{{Worseck} {et~al.}(2014){Worseck}, {Prochaska}, {O'Meara}, {Becker},
  {Ellison}, {Lopez}, {Meiksin}, {M{\'e}nard}, {Murphy}, \&
  {Fumagalli}}]{Worseck2014}
{Worseck}, G., {Prochaska}, J.~X., {O'Meara}, J.~M., {et~al.} 2014, \mnras,
  445, 1745

\bibitem[{{Xue} {et~al.}(2011){Xue}, {Luo}, {Brandt}, {Bauer}, {Lehmer},
  {Broos}, {Schneider}, {Alexander}, {Brusa}, {Comastri}, {Fabian}, {Gilli},
  {Hasinger}, {Hornschemeier}, {Koekemoer}, {Liu}, {Mainieri}, {Paolillo},
  {Rafferty}, {Rosati}, {Shemmer}, {Silverman}, {Smail}, {Tozzi}, \&
  {Vignali}}]{Xue2011}
{Xue}, Y.~Q., {Luo}, B., {Brandt}, W.~N., {et~al.} 2011, \apjs, 195, 10

\bibitem[{{Yajima} {et~al.}(2011){Yajima}, {Choi}, \& {Nagamine}}]{Yajima2011}
{Yajima}, H., {Choi}, J.-H., \& {Nagamine}, K. 2011, \mnras, 412, 411

\bibitem[{{Zackrisson} {et~al.}(2013){Zackrisson}, {Inoue}, \&
  {Jensen}}]{Erik2013}
{Zackrisson}, E., {Inoue}, A.~K., \& {Jensen}, H. 2013, \apj, 777, 39

\bibitem[{{Zibetti} {et~al.}(2013){Zibetti}, {Gallazzi}, {Charlot}, {Pierini},
  \& {Pasquali}}]{Zibetti2013}
{Zibetti}, S., {Gallazzi}, A., {Charlot}, S., {Pierini}, D., \& {Pasquali}, A.
  2013, \mnras, 428, 1479

\end{thebibliography}


\appendix
\onecolumn

\section{Third step of the cleaning procedure: SED fitting of individual galaxy clumps}     \label{sec:smart}

During the $\text{cleaning}$ procedure, we inspected the HST images of the multi-component objects to remove sources surrounded by different-redshift neighbours from our
sample. In principle, a star-forming galaxy can be composed of many clumps. Some of them can have different colours that are
caused by stellar populations with different physical properties. Faint and small sources at different redshift 
can mimic these colours and contaminate the estimation of UV emission in ground-based observations. For this reason, we performed
an SED fitting of individual clumps taking advantage of the complete HST photometry in the CANDELS area.

We first inspected the HST/ACS F606W (V606) images and identified eight star-forming galaxies, composed of multiple clumps, on which our analysis could be performed. We ran the SExtractor software on the V606 band to detect each clump separately and perform photometry on them. To identify the clumps, we used a detection THRESHOLD from 3 down to 0.1, MINAREA from 10 down to 1, and DEBLEND\_MINCONT down to 0.0001, depending on the galaxy. We chose AUTO photometry in the (original PSF) V606 band image as the reference total magnitude (V$_{ref}$) of each clump. We then ran SExtractor in dual mode, adopting V606 as the detection image and $B$435, $V$606, $I$775, z814, z850, $Y$105, $J$125, $H$160, all convolved with the $H$160 PSF, as measurement images. We performed aperture photometry on every image within apertures as small as the $H$160 PSF full width at half maximum. The total magnitude and uncertainty of each clump in all the HST bands was then calculated as

\begin{equation} \label{eq:totalmag}
m^{band}_{total} = (m^{band}_{aper}-m^{F606}_{aper}) + V_{ref}
\end{equation}
\begin{equation} \label{eq:totalerr}
\sigma (m^{band}_{total}) = \sqrt{\sigma(m^{band}_{aper}-m^{F606}_{aper})^2 + \sigma(V_{ref})^2}
.\end{equation}

We added VIMOS $U$ and HAWKI $K$ bands (part of CANDELS release) to the SED of the clumps. At the position of each individual clump in the detection image, we applied the TPHOT software \citep{Merlin2015} to obtain the total magnitude in $U$ and $K$, given the prior of the detection in V606 and the image PSFs. The $U$ and $K$ band magnitude are fundamental to constrain the SED of sources at $z\sim3$.  

We used a photometric redshift (zphot) software \citep[][]{Fontana2000} to estimate the best-fit zphot of each clump and the $\chi^2$ of the best fit. We ran the code twice. The first time we left zphot as a free parameter, the second time we fixed the redshifit to the spectroscopic one of the galaxy main component. We then evaluated the quality of the fit. In the following figures we show the BVI colour images of the eight star-forming galaxies presenting multiple clumps and the $\chi^2$ function obtained by running the zphot software on each of the shown clumps.  

\begin{figure*}[h!]
\centering
\includegraphics[width=85mm]{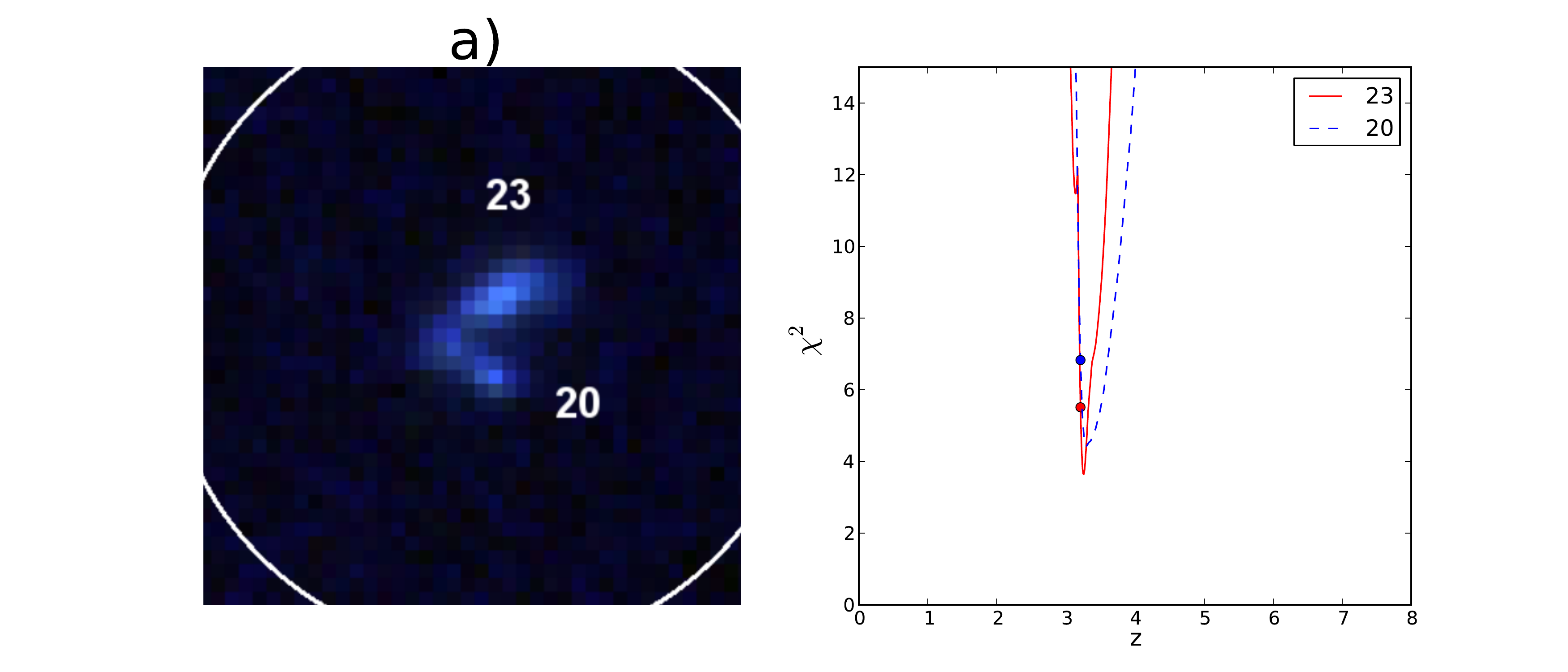}
\includegraphics[width=85mm]{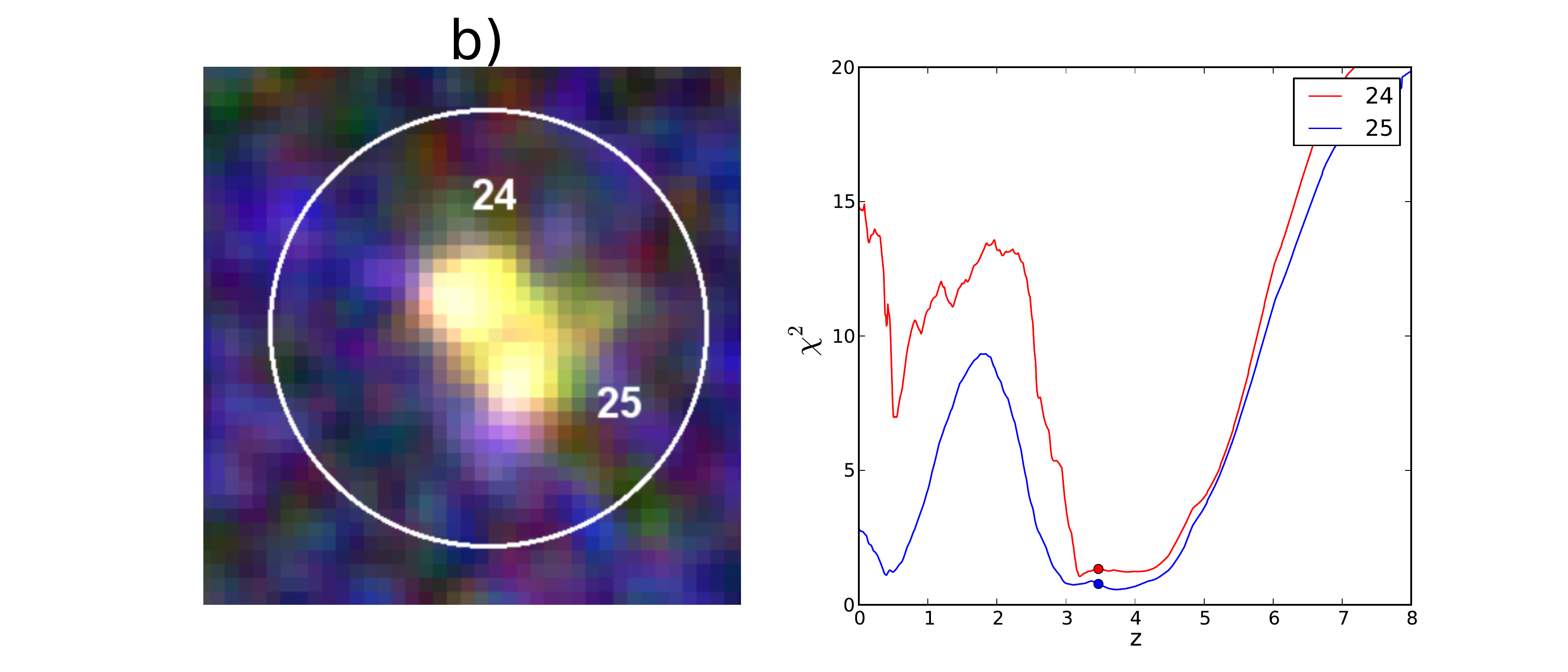}
\includegraphics[width=85mm]{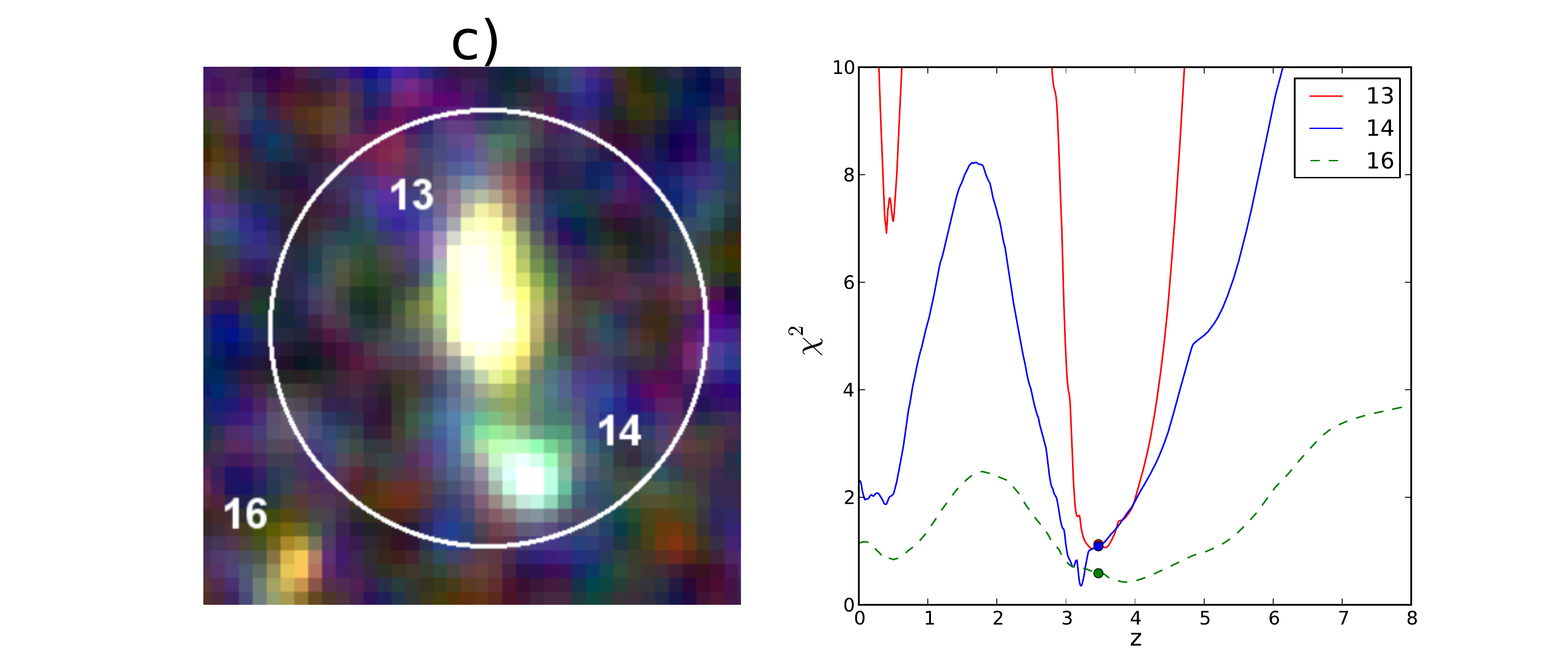}
\includegraphics[width=85mm]{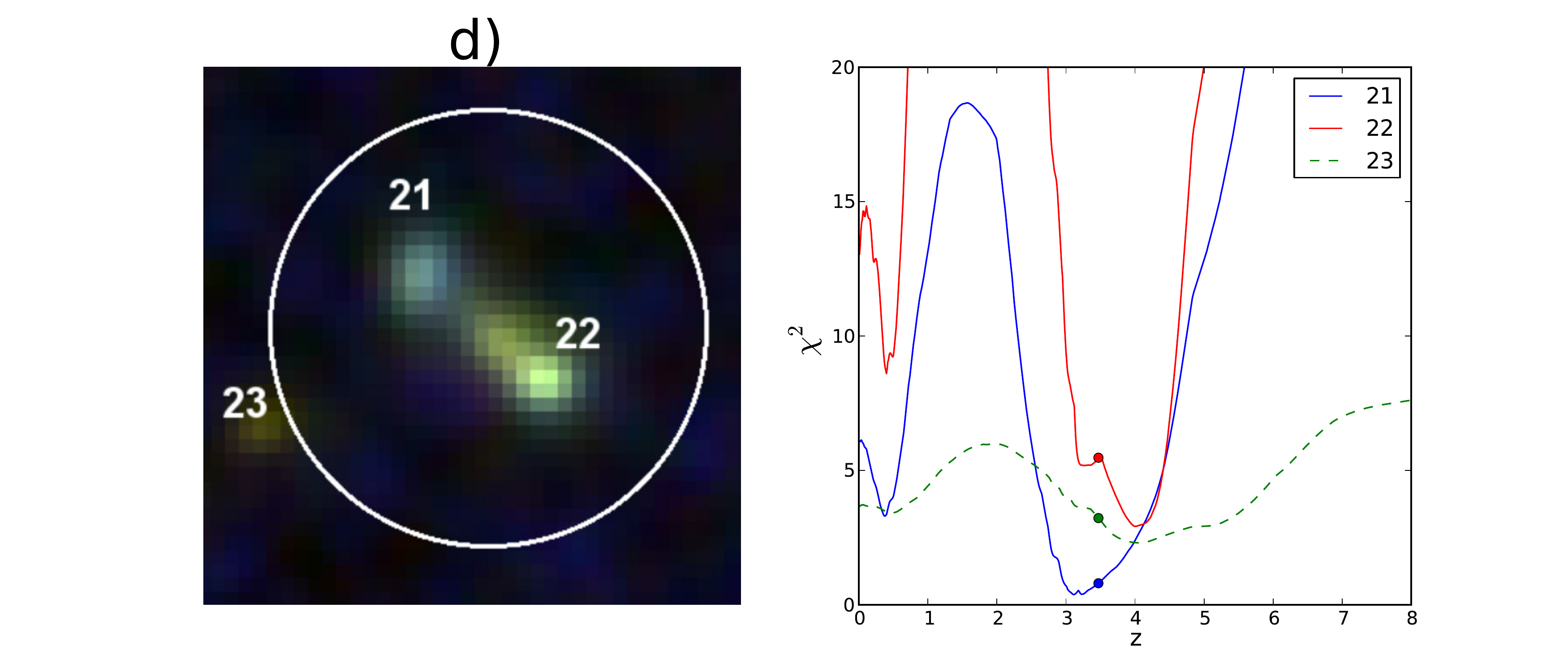}
\includegraphics[width=85mm]{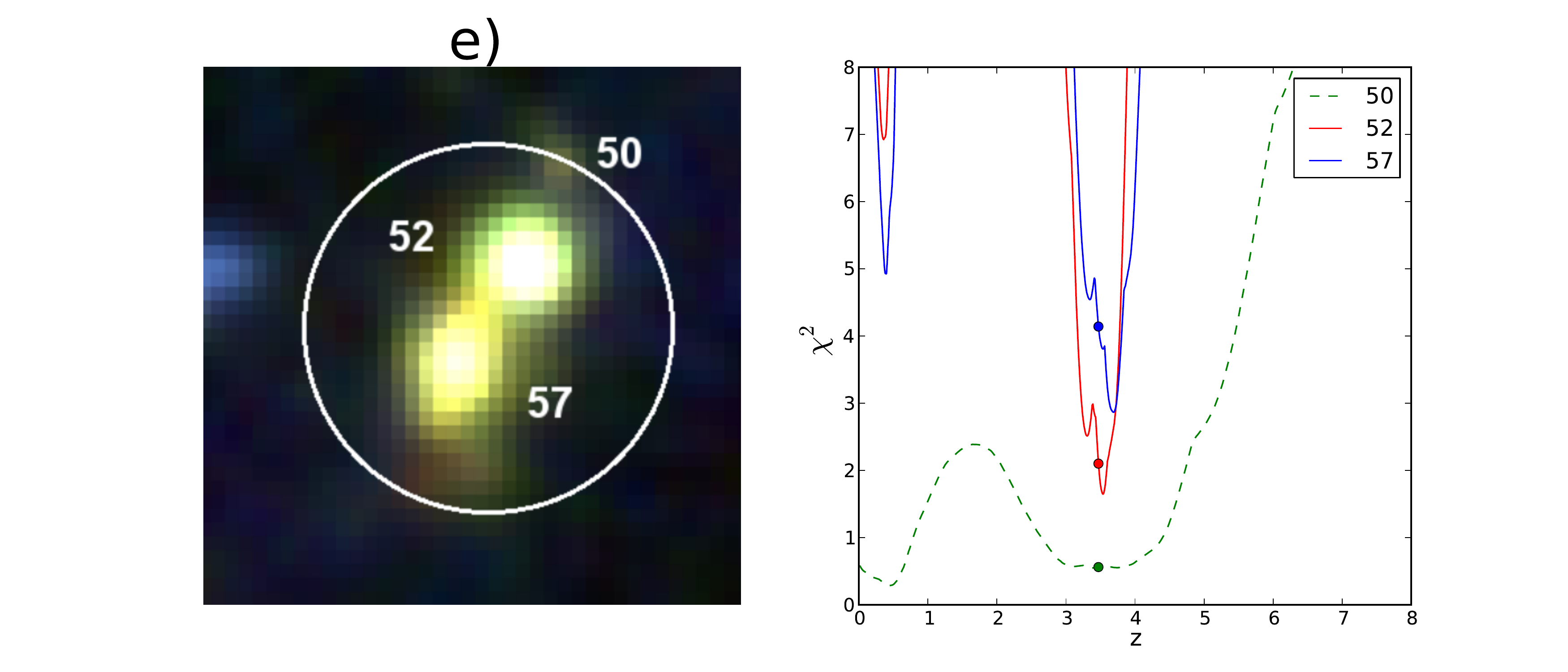}
\includegraphics[width=85mm]{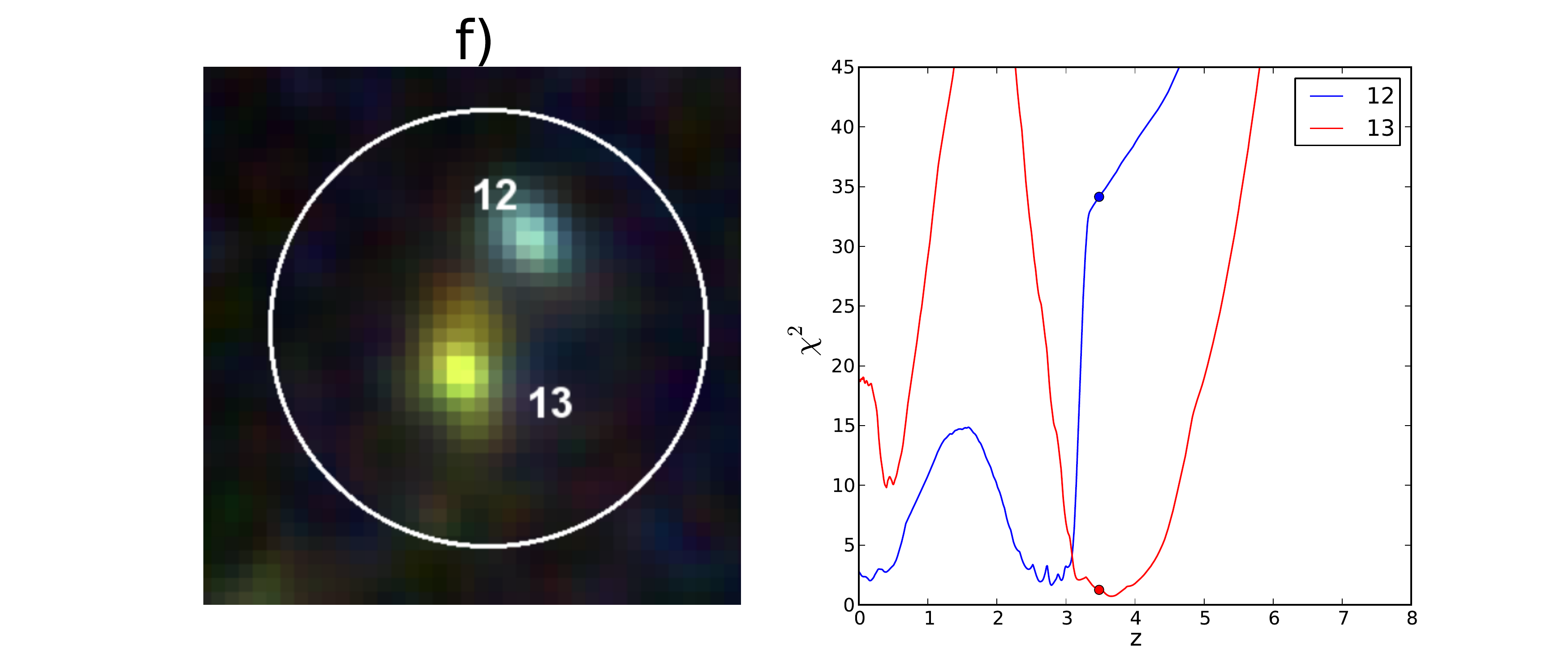}
\includegraphics[width=85mm]{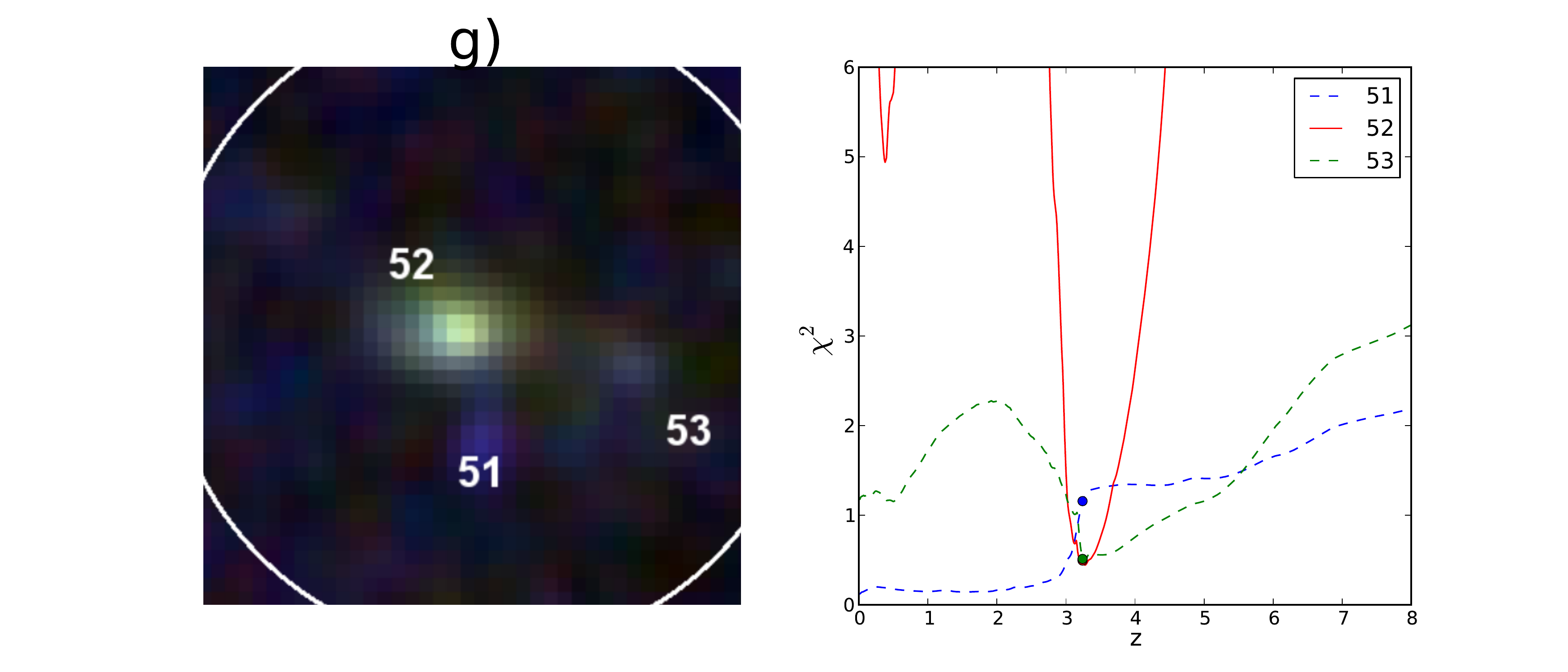}
\includegraphics[width=85mm]{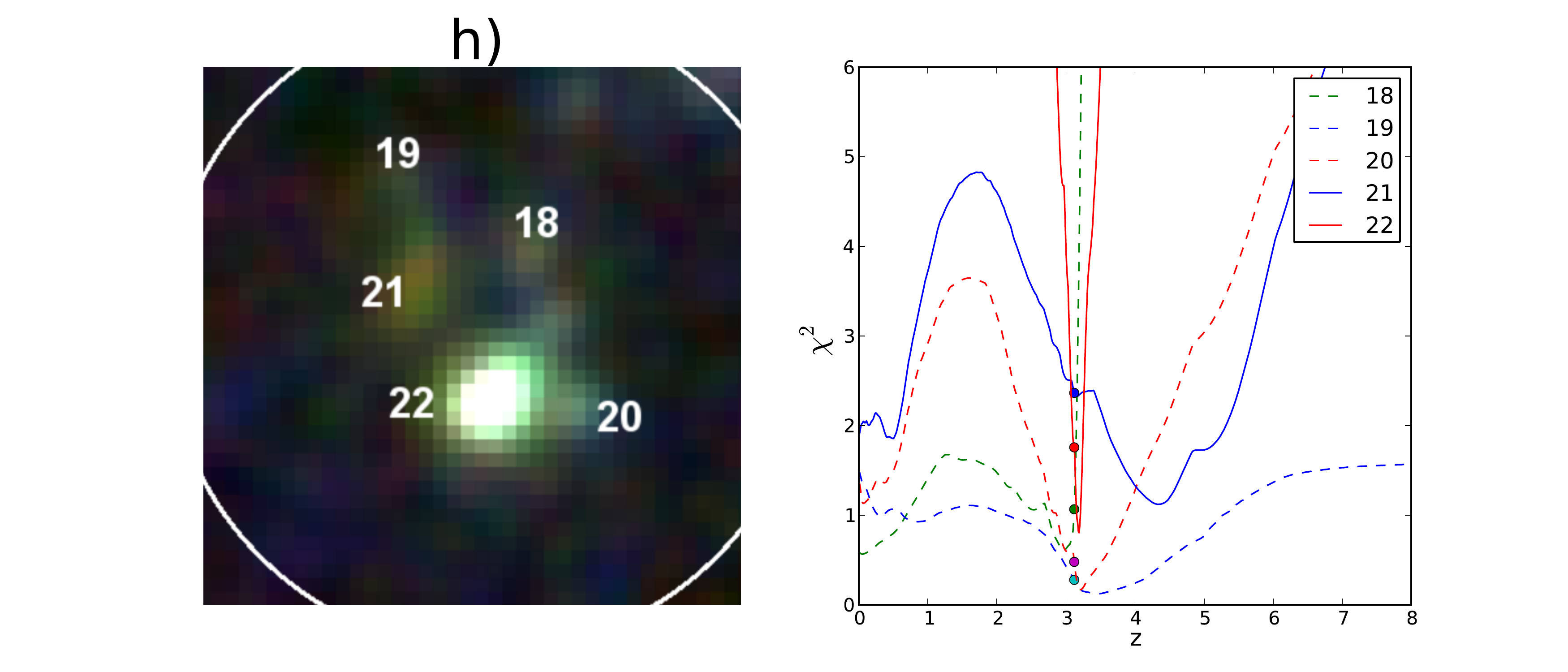}
\caption{Every figure is composed of a left and a right panel, presenting $BVI$ colour images and $\chi^2$ function versus photometric redshift, respectively.
a) Object ID\_CANDELS=6839 ($z=3.208$): $kept$ in the clean sample. It is composed of two clumps (numbers 23 and 20). The upper is brighter than the lower one.  The white circle in the left panel has a radius equal to the PSF full width at half maximum of the ground-based narrow-band image, sampling the LyC at $z=3.208$. 
 The solid (dashed) line in the right panel corresponds to the brighter (fainter) clump, while the blue (red) dot indicates the $\chi^2$ of the SED fit derived by fixing the value of the redshift to the spectroscopic one.
b) Object ID\_CANDELS=4022 ($z=3.466$): $kept$. The two solid lines in the right panel show that the two clumps (numbers 24 and 25) are equally bright.
c) Object ID\_CANDELS=3768 ($z=3.468$): $kept$.  The red and blue dots overlap.
d) Object ID\_CANDELS=3325 ($z=3.473$): $kept$. 
e) Object ID\_CANDELS=15220 ($z=3.469$): $kept$.  The two solid lines in the right panel show that the two clumps are equally bright (numbers 52 and 57).
f) Object ID\_CANDELS=7645 ($z=3.478$): $removed$. 
g) Object ID\_CANDELS=17905 ($z=3.242$): $removed$. In the right panel, the red and green dots overlap.
h) Object ID\_CANDELS=23527 ($z=3.116$): $removed$. It is composed of five clumps. The two reddest ones have a broad $\chi^2$ function. The main component (number 22) is perfectly fitted at the spectroscopic redshift.
}
\label{AppFig}
\end{figure*}

Figure \ref{AppFig}a shows a source composed of two clumps with very similar colours ($\Delta$(B-V)$<0.1$, $\Delta$(V-z) $<0.1$). The $\chi^2$ function versus photometric redshift is narrow for both clumps and has a minimum very close to the spectroscopic redshift. We conclude that the two clumps are part of the same galaxy and kept it in our clean sample. 
For the source presented in Fig. \ref{AppFig}b, we derived the same conclusion, even if in this case the $\chi^2$ function is broader.

The objects shown in Fig. \ref{AppFig}c, Fig. \ref{AppFig}d, and Fig. \ref{AppFig}e are composed of two main clumps, characterized by $\Delta$(B-V)$\sim0.4$ and $\Delta$(V-z) $\sim0.2$. However, the SED fit gives a photometric redshift consistent with the spectroscopic one to both of them. A third faint component is seen at a distance slightly larger than or comparable to (Fig. \ref{AppFig}c, Fig. \ref{AppFig}d, and Fig. \ref{AppFig}e, respectively) to the narrow-band image representing the source LyC rest-frame wavelength. In the former cases the faint component has a photometric redshift still consistent with the spectroscopic one. In the latter it has an inconclusive photometric-redshift solution, but it is far enough from the main clump not to be blended in an image with a PSF comparable with that of the ground-based narrow band, sampling its LyC. Therefore, we kept the objects in our clean sample.

In Fig. \ref{AppFig}f, Fig. \ref{AppFig}g, and Fig. \ref{AppFig}h we present three sources removed from our clean sample, after inspecting their colours and performing the SED fitting of each individual clump. In these cases, one of the clumps is best fitted by a photometric-redshift solution that differs from the spectroscopic value of $\Delta(z_{zphot}-z_{spec})>0.5$. The main component is generally characterized by a satisfactory solution at the spectroscopic redshift.

This procedure, based on the availability of HST band photometry, allowed us to exclude $\sim40$\% of the multi-component sources whose the LyC signal may be contaminated by neighbours with different
redshifts.


\end{document}